\documentclass[]{elsart}
\usepackage{graphicx}
\usepackage{dcolumn}
\usepackage{bm}
\usepackage{epsfig}
\usepackage{float} 
\usepackage{amsmath}
\usepackage{amssymb}
\usepackage{lineno} 

\newcommand \beq{\begin{equation}}
\newcommand \eeq{\end{equation}}
\newcommand \bef{\begin{figure}}
\newcommand \eef{\end{figure}}
\newcommand \bei{\begin{itemize}}
\newcommand \eei{\end{itemize}}
\newcommand \bet{\begin{table}}
\newcommand \eet{\end{table}}

\newcommand \ee {$e^+e^-$ }

\newcommand \mee {$m_{e^{+}e^{-}}$ }

\newcommand \pt {$p_T$ }

\newcommand \sq {$\sqrt{s}$ }

\def\NCA{Nuovo Cimento }

\def\NIMA{{Nucl. Instr. and Meth.}~{\bf A}}
\def\NPA{{Nucl. Phys.}~{\bf A}}

\def\PRL{Phys. Rev. Lett.\ }

\def\PRC{{Phys. Rev.}~{\bf C}}

\def\hypfig#1{\hyperref[#1]{Fig.~\ref*{#1}}}
\def\hyptab#1{\hyperref[#1]{Table~\ref*{#1}}}
\def\hypsec#1{\hyperref[#1]{Section~\ref*{#1}}}
\def\hypapp#1{\hyperref[#1]{Appendix~\ref*{#1}}}
\def\hypeq#1{\hyperref[#1]{Eq.~\ref*{#1}}}

\begin{document}

\begin{frontmatter}

\title{Design, Construction, Operation and Performance of a Hadron Blind Detector for the PHENIX Experiment}

\author[SUNY]{W. Anderson},
\author[BNL]{B. Azmoun},
\author[Weizmann]{A. Cherlin},
\author[Nevis] {C.Y. Chi},
\author[SUNY]{Z. Citron},
\author[SUNY]{M. Connors},
\author[Weizmann]{A. Dubey},
\author[SUNY]{J. M. Durham},
\author[Weizmann,corr1]{Z. Fraenkel},
\author[SUNY]{T. Hemmick},
\author[SUNY]{J. Kamin},
\author[Weizmann]{A. Kozlov},
\author[SUNY]{B. Lewis},
\author[Weizmann]{M. Makek},
\author[Weizmann]{A. Milov},
\author[Weizmann]{M. Naglis},
\author[SUNY,corr2]{V. Pantuev},
\author[BNL]{R. Pisani},
\author[SUNY]{M. Proissl},
\author[Weizmann]{I. Ravinovich},
\author[UCR]{S. Rolnick},
\author[BNL]{T. Sakaguchi},
\author[Weizmann]{D. Sharma},
\author[BNL]{S. Stoll},
\author[SUNY]{J. Sun},
\author[Weizmann,corr3]{I. Tserruya} and
\author[BNL]{C. Woody}
\address[BNL]{Brookhaven National Laboratory, Upton, NY 11973-5000, USA}
\address[Nevis] {Columbia University, New York, NY 10027 and Nevis Laboratories, Irvington, NY 10533, USA}
\address[UCR]{University of California at Riverside, Riverside, CA 92521, USA}
\address[SUNY]{Stony Brook University, SUNY, Stony Brook, NY 11794-3400,USA}
\address[Weizmann]{Weizmann Institute of Science, Rehovot 76100, Israel }
\thanks[corr1]{Deceased.}
\thanks[corr2] {Present address: Institute for Nuclear Research, Russian Academy of Sciences, Moscow, Russia}
\thanks[corr3]{Corresponding author: Itzhak Tserruya
  {\it E-mail}: Itzhak.Tserruya@weizmann.ac.il}


\maketitle

\begin{abstract}
A Hadron Blind Detector (HBD) has been developed, constructed and successfully
operated within the PHENIX detector at RHIC. The HBD is a Cherenkov detector operated
with pure CF$_4$. It has a 50 cm long radiator directly coupled in a windowless
configuration to a readout element consisting of a triple GEM stack, with a CsI photocathode
evaporated on the top surface of the top GEM and pad readout at the bottom of the stack. This paper
gives a comprehensive account of the construction, operation and in-beam performance of the detector.
\end{abstract}


\end{frontmatter}

\section{Introduction}
\label{sec:Intro}  
We have developed a Hadron Blind Detector (HBD) as an upgrade
of the PHENIX experiment at the Relativistic Heavy Ion Collider
(RHIC) for the measurement of electron pairs, particularly in
the low-mass region (m$_{e^+e^-}$ $<$ 1~GeV/c$^2$).
Low-mass dileptons are considered a powerful and unique probe
to diagnose the hot and dense strongly interacting quark gluon
plasma formed in ultra-relativistic heavy ion collisions \cite{ref:white-paper}. They
are sensitive to chiral symmetry restoration effects expected to
take place in these collisions \cite{ref:IT-review}. They can also be used to detect the thermal
radiation emitted by the plasma via virtual photons providing a
direct measurement of the plasma temperature, one of its most
basic properties \cite{ref:ppg086}.

PHENIX is a large multipurpose experiment specially devoted to
the measurement of rare probes, and electromagnetic probes in
particular \cite{ref:PHENIX-det}. At mid-rapidity ($|\eta| < $ 0.35) the detector has
excellent electron identification capabilities based on a RICH
detector and an electromagnetic calorimeter. It also has
 a mass resolution of about 1\% at the $\phi$ mass, which allows
precision spectroscopy measurements of the light vector
mesons $\rho, \omega$ and $\phi$. The observation of spectral
shape modifications of these mesons could provide direct
information on the chiral symmetry restoration.
However, the measurement of low-mass electron pairs in the
original PHENIX detector configuration suffers from a huge
combinatorial background, with  a signal to background ratio
of S/B $\simeq$ 1/200 in the invariant dielectron mass range
of m = 0.3-0.5 GeV/c$^2$ \cite{ref:ppg088}.
The combinatorial background comes from the overwhelming
yield of $\pi^0$ Dalitz decays and $\gamma$ conversions and
originates from the limited geometrical acceptance of the
PHENIX detector (the central arm spectrometers consist of two
arms each one covering the pseudo-rapidity interval
$|\eta| < $ 0.35  and 90$^o$ in azimuthal angle) and the very
strong magnetic field starting at the vertex. Consequently,
very often only one of the two tracks of an \ee
pair is detected in the central arm detectors. The second
track never reaches the detectors (because it falls out of the acceptance or is curled by the magnetic field) or is not detected due to
the inability to reconstruct low-momentum tracks with
\pt $<$ 200 MeV/c. These single tracks, when paired to other
electron tracks in the same event, give rise to the combinatorial
background.

The HBD aims at considerably reducing the combinatorial background
from the two main background sources, $\pi^0$ Dalitz decays and
$\gamma$ conversions. The detector exploits the distinctive feature
of the \ee pairs from these two sources, namely their very small
opening angle. The HBD is therefore located in a field free region
that preserves the original direction of the \ee pair. Electron
tracks identified in the central arm detectors are rejected as likely
partners of a $\pi^0$ Dalitz decay or a  $\gamma$ conversion pair
if the corresponding hit in the HBD has a double amplitude or has a
nearby hit within the typical opening angle of these pairs.

The HBD consists of a Cherenkov radiator that is directly coupled to a
triple Gas Electron Multiplier (GEM)~\cite{ref:Sauli_GEM}
detector with a CsI photocathode. Both the radiator and the GEMs are operated
with pure CF$_4$ in a common gas volume. The
detector was constructed after extensive R\&D to demonstrate
the concept validity (see~\cite{ref:hbd1,ref:hbd2} for
the R\&D results and~\cite{ref:hbd-it,ref:hbd-reports} for other
previous reports related to the HBD).

This paper gives a comprehensive report on the design, construction,
operation and performance of the HBD. The detector was commissioned
in 2007 and has been fully operational since the fall of 2008. It was
used as an integral part of the PHENIX detector in the RHIC runs
of 2009 and 2010 which were devoted to the study of p+p collisions and
Au+Au collisions, respectively. The paper is organized as follows: Section 2
presents the overall detector concept. The realization of the detector,
including design, construction and test, is described in detail in Section 3. The
detector services, including the readout electronics, the gas handling and
monitoring system and the high voltage
system are described in Sections 4, 5 and 6, respectively. The operation and
monitoring of the detector under running conditions are presented in Section 7.
Section 8 gives a comprehensive account of the detector performance. A short summary
is provided in Section 9.

\section{Detector concept}
\label{sec:Concept}  

\begin{figure}[]
 \begin{center}
    \includegraphics[width=80mm]{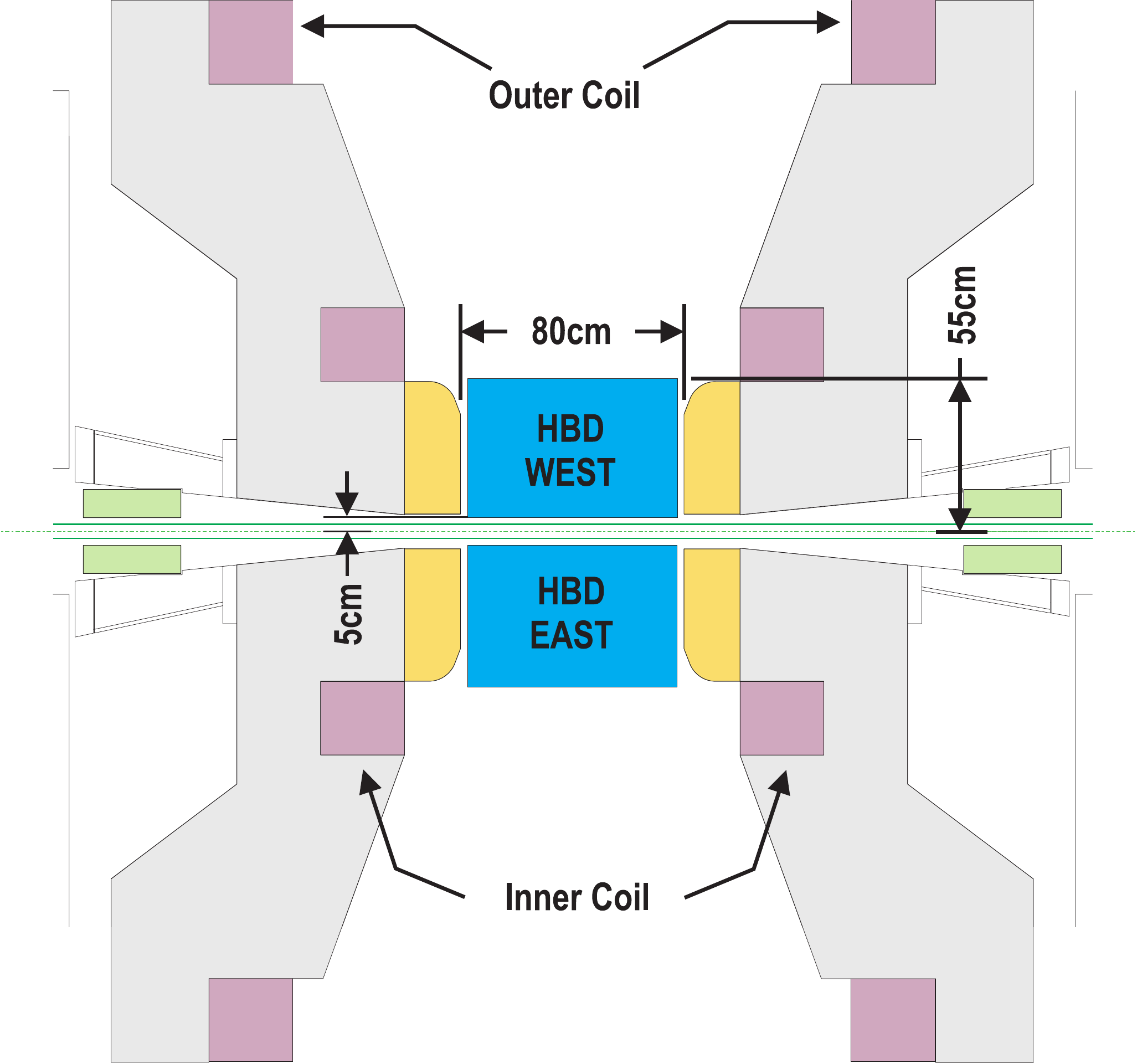}
    \caption{Top layout of the inner part of the PHENIX central arm
    detector showing the location of the HBD and the inner and outer coils.}
  \label{fig:location}
 \end{center}
 \end{figure}

The main task of the HBD is to recognize and reject $\gamma$  conversions and $\pi^o$ Dalitz decays.
The strategy is to exploit the fact that the opening angle of electron pairs from these sources
is very small compared to the pairs from light vector mesons. In a field-free region, this angle is preserved
and by applying an opening angle cut one can reject more than 90\% of the conversions and $\pi^o$ Dalitz decays,
while keeping most of the signal.
The PHENIX central arm magnetic field consists of an inner and outer coil that can be operated
independently. The field free region, necessary for the operation of the HBD, is
generated by allowing the current in these two coils to flow in opposite directions.
In this so-called ``$+-$'' mode, the inner coil located at a radius of $\sim$ 60 cm counteracts
the action of the outer coil resulting in an almost field free region extending out to
$\sim$ 50-60 cm in the radial direction.
The size of the HBD is constrained by the available space in the field free region,
from the beam pipe (at r $\sim$ 5 cm) up to the location of the inner coil. Fig.~\ref{fig:location}
shows the layout of the inner part of the PHENIX detector together with the location of
the coils and the HBD.

The system specifications of the HBD were defined by Monte Carlo
simulations performed at the ideal detector level aiming at reducing
the combinatorial  background originating from conversions and
$\pi^0$ Dalitz decays by two orders of magnitude. At this level of
rejection, the quality of the low-mass $e^+e^-$ pair measurement is
no longer limited by the background originating from these
sources, but rather by the background originating from the semi-leptonic
decay of charmed mesons. The simulations showed that the goal can be
achieved with a detector that provides electron identification with
an efficiency of $\sim$90\%. This also implies a double electron
hit recognition at a comparable level. The separation between single
and double electron hits is one of the main performance parameters
of this detector. On the other hand, a moderate hadron rejection factor
of $\leq$ 50 is sufficient. It is also important to have a larger
acceptance in the HBD compared to the fiducial central arm acceptance
to provide a veto area for the rejection of pairs where only one
partner is inside the fiducial acceptance.

The requirements on electron identification limit the choice to a
Cherenkov-type detector. In order to generate enough UV photons in a
$\sim$50~cm long radiator to ensure good distinction between single and double
hits, we adopted a windowless scheme without mirror and chose
pure CF$_4$ as radiator and detector gas. The use of a UV transparent
window between the radiator and the detector element and of a mirror,
as commonly done in RICH detectors, limits the bandwidth to about 8-9 eV.
The choice of CF$_4$ both as the radiator and detector gas in a windowless
geometry results in a very large bandwidth (from $\sim$6 eV given by the
threshold of CsI to $\sim$11.1~eV given by the CF$_4$ cut-off) and
consequently a very large figure of merit N$_0$. The N$_0$ value is estimated
to be close to 700~cm$^{-1}$ under  ideal conditions with no losses.
The large value of N$_0$  ensures a very high electron efficiency, and
more importantly, is crucial for achieving good double-hit resolution.

\begin{figure}[]
 \begin{center}
    \includegraphics[width=140mm]{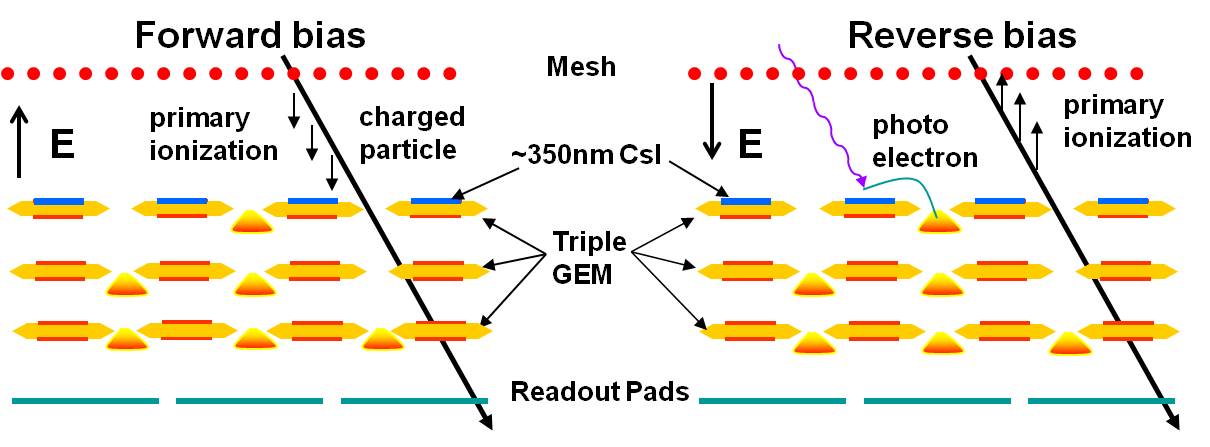}
    \caption{Triple GEM stack operated in the standard forward
    bias mode (left) and in the hadron-blind reverse bias mode (right).}
  \label{fig:gemstack}
 \end{center}
 \end{figure}

In this windowless proximity focus configuration, the Cherenkov
light from particles passing through the radiator is directly collected
on a photosensitive cathode plane, forming an almost circular blob image
rather than a ring as in a conventional RICH detector. After consideration of
relevant options, we chose a triple GEM
detector with a CsI photocathode evaporated on the top surface of the
first GEM foil as the active detector element. The signal is collected by a pad readout at the bottom of the GEM stack (see
Fig.~\ref{fig:gemstack}). In this reflective photocathode scheme, the
photoelectrons are pulled into the holes of the GEM by the strong
electric field inside the holes and the photocathode is
totally screened from photons produced in the avalanche process.

The hadron blindness property of the HBD is achieved by operating
the detector in the so-called reverse bias mode as opposed to the standard forward bias (FB) mode (see Fig.~\ref{fig:gemstack}). In the reverse bias (RB) mode,
the mesh is set at a lower negative voltage with respect to the GEM and
consequently the ionization electrons deposited by a charged
particle in the drift region between the entrance mesh and the
top GEM are mostly repelled towards the mesh (see
Fig.~\ref{fig:gemstack} right panel). Consequently, the signal produced by a
charged particle results only from (i) the collection of ionization charge
from only a thin layer of $\sim 100$ $\mu$m above the top GEM which is
subject to the entire 3-stage amplification, and (ii) the collection
of ionization charge in the first transfer gap (between the top and
the middle GEMs) which is subject to a 2-stage amplification only.
The ionization electrons produced in the second transfer gap and in the induction gap
generate a negligible signal since they experience one and zero stages of amplification, respectively.
For a drift region and a transfer gap of 1.5 mm each and a
total gas gain of 5000, the mean amplitude of a hadron signal
drops to $\sim 10\%$ of its value in the forward bias
mode~\cite{ref:hbd2}.

The readout pad plane consists of hexagonal pads with an area of
6.2 cm$^2$  (hexagon side length a = 1.55 cm) which is comparable to, but smaller than,
the blob size which has a maximum area of 9.9 cm$^2$. Therefore,
the probability of a single-pad hit by an electron entering the
HBD is very small.
On the other hand, a hadron traversing the HBD  will produce a signal
predominantly localized in a single pad. This provides an additional
 strong handle in the hadron rejection of the HBD.

The relatively large pad size also results in a low granularity thereby reducing
the cost of the detector. In addition, since the signal
produced by a single electron is distributed between 2-3 pads, one
expects a primary charge of several photoelectrons per pad,
allowing the operation of the detector at a relatively moderate
gain of a few times 10$^3$. This is a crucial advantage for stable
operation of a UV photon detector.

 \section{HBD design, construction, and testing}
\label{sec:construction}

\subsection{Design overview}
\label{subsec:design_overview}
The detector design derives from the system specifications and the space
constraints discussed in Section~\ref{sec:Concept}. In addition,
special care was taken to minimize (i) the amount of material in
order to reduce as much as possible the number of photon conversions in the central arm acceptance
and (ii) the dead or inactive areas
due to frames or spacing between adjacent detector modules in
order to achieve the highest possible efficiency.
Table~\ref{tab:design-parameters} summarizes the most important
design parameters.

\begin{table*}[h!]
\caption{Design parameters of the HBD.}
\label{tab:design-parameters}
\begin{tabular}[]{l|l}
\hline Acceptance                                &$|\eta| \leq$ 0.45 $\Delta \phi$ =135$^o$     \\
GEM size ($\phi \times z$)                       &  23 $\times$ 27 cm$^2$ \\
GEM supporting frame and cross (w x d)           & frame: 5$\times$1.5 mm$^2$, cross: 0.3$\times$1.5 mm$^2$\\
Hexagonal pad side length                        &  a = 15.5~mm    \\
Number of pads per arm                           &   1152   \\
Dead area within central arm acceptance          &  7\%    \\
Total Radiation length within central arm acceptance   & 2.40\%     \\
Weight per arm (including HV and gas connectors) & $<$10~kg  \\
\hline
\end{tabular}
\end{table*}

\begin{figure}[]
 \begin{center}
    \includegraphics[width=65mm]{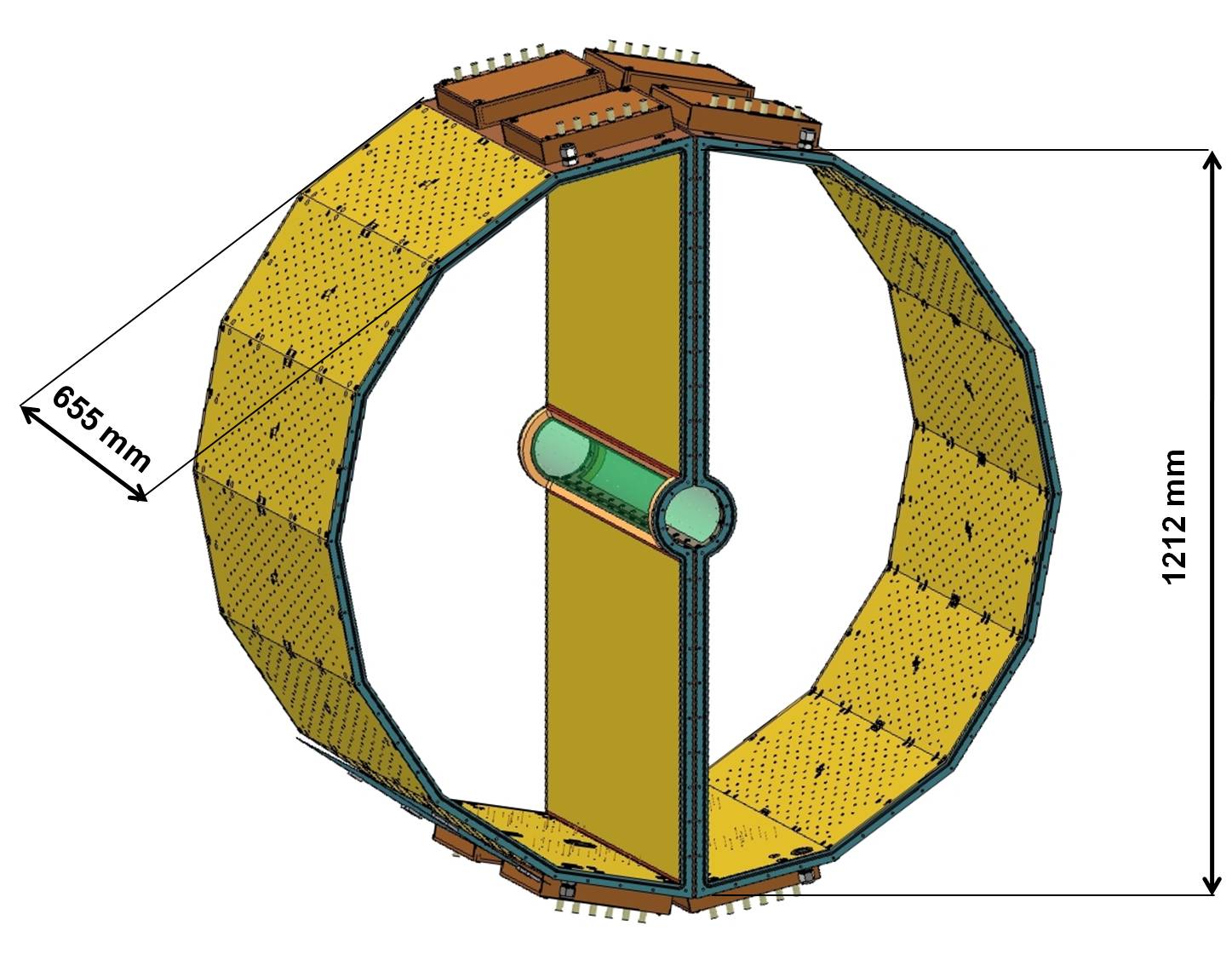}
    \includegraphics[width=65mm]{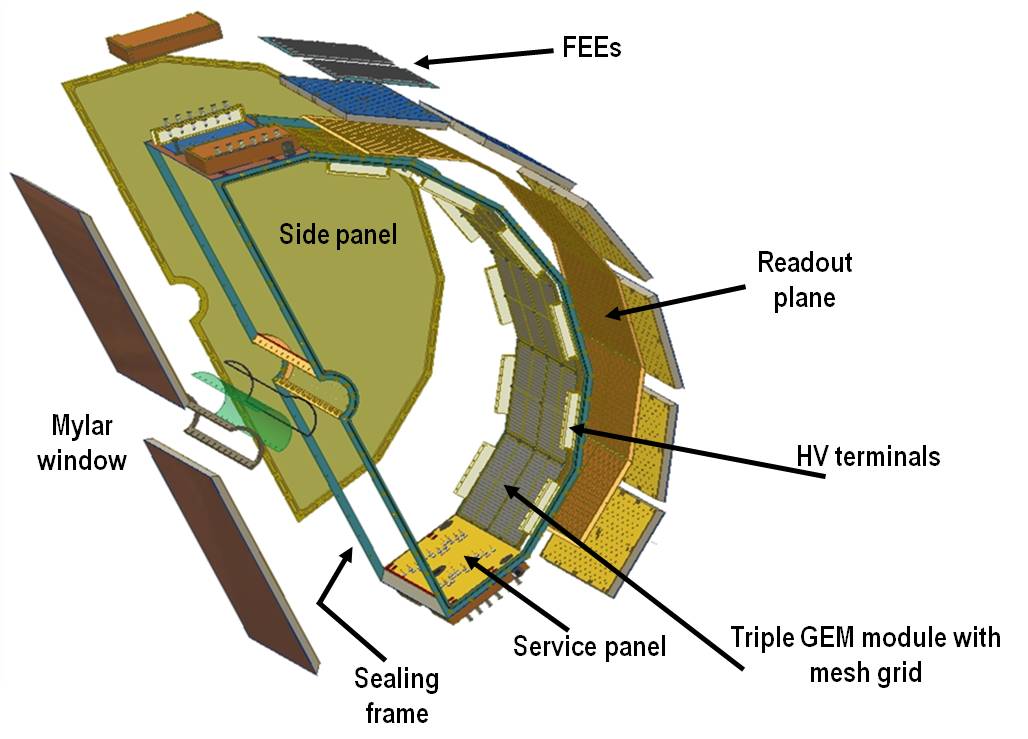}
    \caption{Left panel: 3D view of the two arm HBD. Right panel:
    exploded view of one HBD arm.}
  \label{fig:3d-view}
 \end{center}
 \end{figure}

The HBD is made of two identical arms, located close to the
interaction vertex. The entrance window is located just after the beam
pipe at r$\sim$5 cm. The detector extends to
r$\sim$60 cm in the radial direction and 65.5~cm along the beam axis (see
Fig.~\ref{fig:location} and Fig.~\ref{fig:3d-view} left panel).
Each arm covers 135$^o$ in azimuthal angle $\phi$ and $\pm$0.45
units in pseudorapidity $\eta$. This extended acceptance with
respect to the central arms (which cover 90$^o$ in $\phi$ and
$\pm$0.35 units in $\eta$) provides a very generous veto area
for efficient rejection of close pairs where only one track
falls inside the fiducial acceptance.

The right panel of Fig.~\ref{fig:3d-view} shows an exploded view
of one HBD arm, displaying the various elements of the detector.
Each arm consists of a $\sim$50 cm long radiator directly
coupled to a triple GEM photon detector. The latter is subdivided
in 12 detector modules, 6 along the $\phi$ axis $\times$ 2
along the z axis. With this segmentation, each detector module
is $\sim$23~$\times$~27~cm$^2$ in size. In the 2009 and 2010 RHIC runs,
10 modules were instrumented in each  arm covering an azimuthal range of 112.5$^o$ which is
considerably larger than the azimuthal range of 90$^o$ covered by the central arm detectors.

\subsection{Detector vessel}
\label{subsec:detector_vessel}
The detector vessel has a polygonal shape formed by panels glued
together as shown in Fig.~\ref{fig:3d-view}. Eight panels of
63.0~$\times$~23.7~cm$^2$ and two vertical panels of 63.0~$\times$~54.8~cm$^2$
define the polygonal shape. The panels
consist of a 19~mm thick honeycomb core sandwiched between two
0.25~mm thick FR4 sheets. Six of the eight panels define the HBD
active area. The other two panels, outside the active area,
are service panels. Gas-in and gas-out connections, HV connectors
serving the GEMs, and a small UV-transparent window are located
on these two panels.

Two supporting frames made of FR4, 19~mm thick (dictated by the thickness of the
honeycomb core of the panels) and 7~mm wide, connect all panels
together on each side providing mechanical stability and rigidity
to the entire box. A thin window  around the beam pipe is used to further
reduce the radiation length in the HBD fiducial acceptance. The window is
made of a 50~$\mu$m thick layer of aclar and a 25~$\mu$m thick layer of  black kapton
on the inside to minimize reflections. It is
glued on a semi-cylindrical FR4 frame bolted to the supporting frame along
the beam axis and is therefore easily removable.
The two sides of the box are closed with covers
made of 12.5~mm thick honeycomb core sandwiched between two 0.25~mm
thick FR4 facesheets that are bolted on the supporting frame with an
o-ring seal. Each side panel also has a 12.5~mm thick (dictated by the thickness
of the honeycomb core of the side covers) and
15~mm wide frame around its perimeter to provide rigidity in the bolting
area.

\begin{figure}[]
 \begin{center}
    \includegraphics[width=80mm]{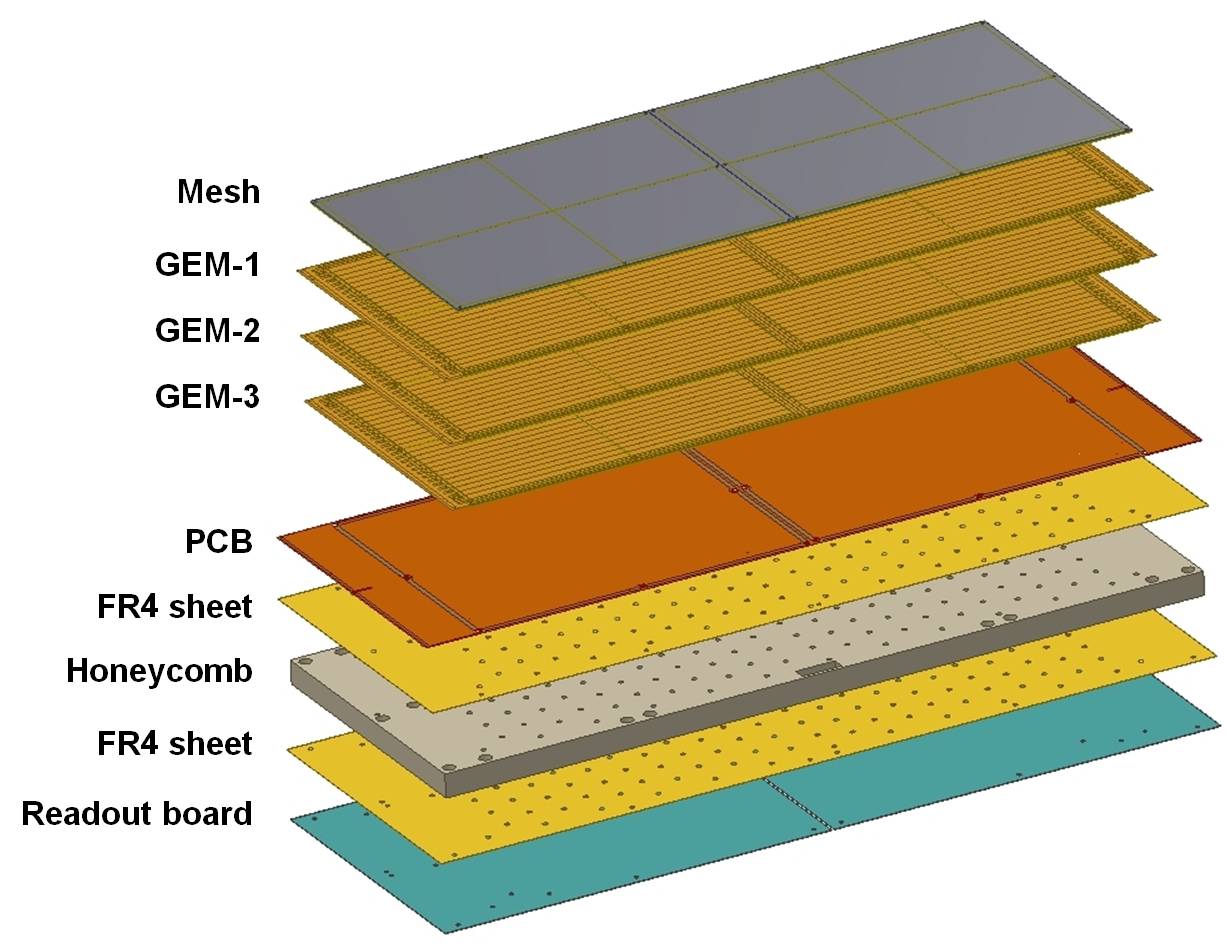}
    \caption{Exploded view of one panel of the HBD vessel and readout board.}
  \label{fig:panel-exploded-view}
 \end{center}
 \end{figure}

Fig.~\ref{fig:panel-exploded-view} presents an exploded view of
a single back-side panel, showing the various components of one
HBD panel (two detector modules) and the readout board attached to it.
The mesh and the three GEM foils are
mounted on FR4 fiberglass frames. The frames have a width of 5~mm
and a thickness of 1.5~mm that defines the intergap distance. To
prevent sagitta of the foils in the electrostatic fields, the
frames have a supporting cross (0.3~mm thick) in the middle. The
three GEM foils and the mesh are stacked together and attached to
the detector vessel by 8 pins. These pins, located at the corners
and the middle of the frame, keep the tension on the GEM foils and the
mesh while maintaining a minimum deformation of the
5~mm wide frames. Special tooling was developed to stretch the
 foils and the mesh and to glue them onto the narrow frames.
 The design allows for only 1~mm clearance between two adjacent
 detectors. With this design, the resulting total dead area
 within the central arm acceptance is calculated to be 7\%.

\begin{figure}[h!]
   \centering
   \includegraphics[keepaspectratio=true, width=8cm]{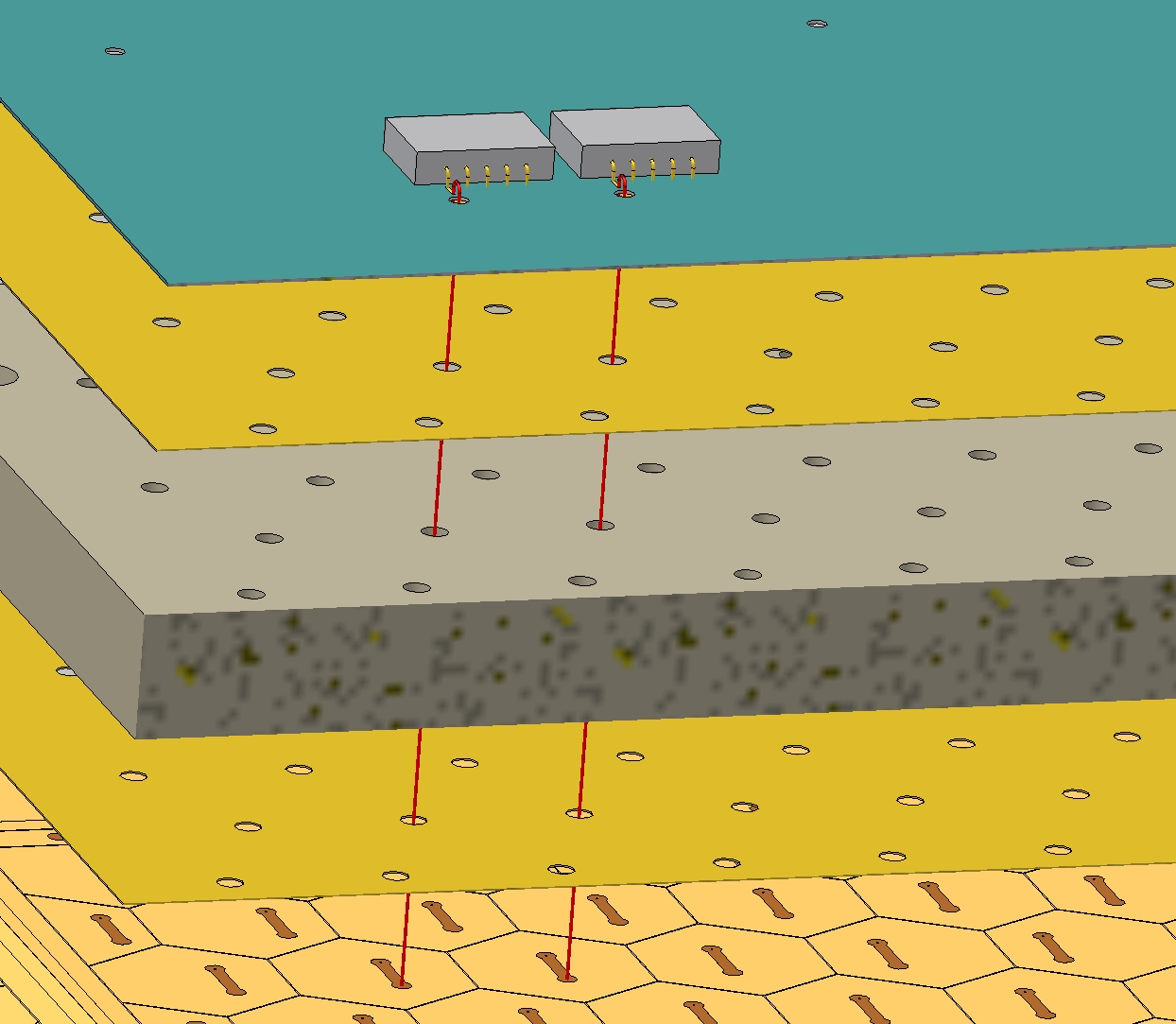}
   \caption{Backplane of detector panel consisting of hexagonal
   pad plane connected by wires to the readout board containing
   the preamplifiers.}
    \label{fig:readout_board}
\end{figure}

The detector anode is a double-sided printed circuit board (PCB)
with a hexagonal pad pattern on the inner side and short signal
traces (seen at the bottom of Fig.~\ref{fig:readout_board}) on the other side.
The side length of the hexagonal pads is a = 15.5~mm resulting in 96 pads in each
detector module and a total of 1152 pads in each arm.
Plated-through holes in the PCB connect
the pads to the signal traces. Short wires are soldered at
the edges of these traces ($\sim$1.5~cm  from the
plated-through holes), passed through small holes in the panels
and soldered to traces on the readout board located outside the detector
that carry the signals to the preamplifiers
(see Fig.~\ref{fig:readout_board}). The PCB is made of 50~$\mu$m
thick kapton foil with 5 $\mu$m copper cladding, in one single
piece ($\sim$140 $\times$ 63~cm$^2$). The PCB is glued onto the
eight panels which define the outside polygonal shape of
the detector vessel (see Fig.~\ref{fig:3d-view}).

Special attention was taken in the design to ensure gas tightness
of the detector vessel. The plated-through holes are effectively
sealed by the panels that are glued on the back side of the PCB. Making
the PCB in one piece and gluing it to the panels behind prevents
any potential leaks at the junctions between adjacent
panels. The junctions between adjacent panels of the vessel are easily
sealed by gluing a 50~$\mu$m thick kapton strip along the inner
side of the junction. The leak rate in each vessel, which has a total volume of 313 l,
was measured to be $\sim$0.12 cc/min.

Each detector vessel alone weighs $\sim$5~kg and  adding all other
components (HV connectors, gas in/out, GEM foils, preamplifier
cards...) results in a total weight of less than 10~kg per arm.
The total radiation length of the detector vessel within the central arm acceptance is
calculated to be 0.82\%. To this, one must add the
contribution  of the readout board and pre-amps that is attached to the vessel (estimated to be 1.03\%) and
the 50~cm CF$_4$ radiator gas (estimated to be 0.56\%)
to give a total of 2.4 \% of a radiation length for the entire detector.  The material budget is itemized in  Table~\ref{tab:material-budget}.

\begin{table*}
 \caption{Material budget of the HBD within the PHENIX central
 arm acceptance. The layout of the readout boards, pre-amps and sockets is rather complex, and for these components the thickness values quoted represent an average over the detector area.}
\label{tab:material-budget}
\begin{tabular}[]{llllcc}
&  &  &  & & \\
Component & Material & X$_0$ (cm) & Thickness (cm) & Area ($\%$) & Rad. Length (\%) \\
\hline
\bf{Vessel}    &            &              &                  &             &                  \\
Window         & Aclar/kapton & 15.8/28.6  & 0.0075/0.0050    & 100         & 0.040 \\
Mesh           & SS         & 1.67         & 0.003            & 11.5        & 0.021 \\
GEM            & Kapton     & 28.6         & 0.005$\times$3   & 64          & 0.034 \\
GEM            & Copper     & 1.43         & 0.0005$\times$6  & 64          & 0.134 \\
GEM frames     & FR4        & 17.1         & 0.15$\times$4    & 6.5         & 0.228 \\
PCB            & Kapton     & 28.6         & 0.005            & 100         & 0.017 \\
PCB            & Copper     & 1.43         & 0.0005           & 80          & 0.028 \\
Facesheet      & FR4        & 17.1         & 0.025$\times$2   & 100         & 0.292 \\
Panel core     & Honeycomb  & 8170         & 1.905            & 100         & 0.023 \\
Total vessel   &            &              &                  &             & 0.82 \\
\hline
\bf{Readout}   &            &              &                  &             &        \\
Readout board   & FR4/copper & 17.1/1.43    & 0.05/0.001       & 100         & 0.367 \\
Preamps + sockets & Copper  & 1.43         & 0.0005           & 100         & 0.66 \\
Total readout  &            &              &                  &             & 1.03\\
\hline
\bf{Gas}       & CF$_4$     & 9240         & 51.5             & 100         & \bf{0.56} \\
\hline
\bf{Total}     &            &              &                  &             & \bf{2.4}  \\
\hline
\end{tabular}
\end{table*}

\subsection{GEM electrodes}\label{sec:GEMs}
\begin{figure}[t]
 \begin{center}
  \includegraphics[keepaspectratio=true, width=14cm]{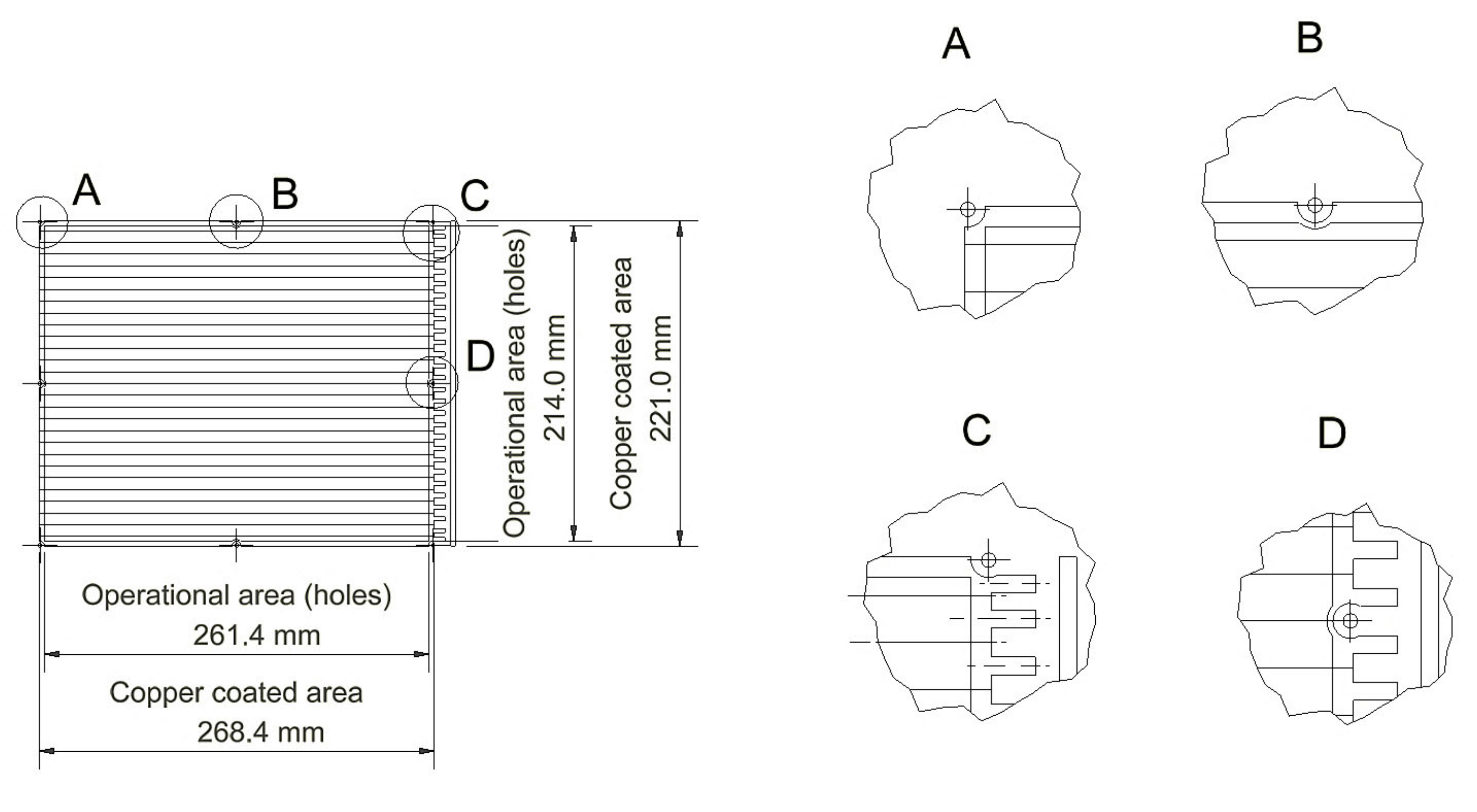}
  \caption{\label{fig:gem_layout}GEM foils used in the HBD showing global dimensions and details of HV strips and copper recesses at the location of the pins.}
 \end{center}
\end{figure}

The GEMs for the HBD were all produced at the Technical Support
Department at CERN. Each GEM is made of a 50~$\mu$m thick, metal-clad
(5~$\mu$m thick copper on each side) polymer foil commonly known as
kapton. It is chemically etched to produce a highly dense pattern of 60-80~$\mu$m
diameter holes with 140~$\mu$m pitch. The copper coated area of the foils is
268.4~$\times$ 221~mm$^{2}$, whereas the operational area i.e. the area with holes
is 261.4~$\times$ 214~mm$^{2}$.
The top face of each GEM foil is divided into 28 high voltage segments:
26 central segments of 8~mm each and a first and last segment of 6.5~mm. The gap
without copper between the segments is 200~$\mu$m. Fig.~\ref{fig:gem_layout}
gives the detailed geometry of the GEM foils, showing the various dimensions.

All GEM foils were electrically tested at CERN prior to shipping to the Weizmann
Institute. The tests were done in a clean room in dry air ($<$30$\%$
humidity). The foils were accepted if the leakage current was below
5~nA at 600~V.

Upon arrival at the Weizmann Institute the foils were again electrically tested
before framing. The test was done in air, either in a clean room or on
a laminar flow table. HV was applied on the top surface of the GEM foil
to each of the 28 strips with the bottom surface grounded. Since the humidity
in the test area was not as low at it was during the initial test at CERN
the voltage was raised up to only 550~V.
However, the criteria for accepting a good foil was the same as it was at CERN,
namely less than $\sim$5~nA of leakage current. A few foils exhibited a high current or even a
short during testing. The foils which
showed high leakage current were gently blown with  clean and dry compressed
nitrogen in order to remove any residual dust that could cause the high
leakage current. If after this cleaning procedure the foil still showed
high leakage current it was rejected. However, this happened very rarely.

\begin{figure}
 \begin{center}
  \includegraphics[keepaspectratio=true, width=12cm]{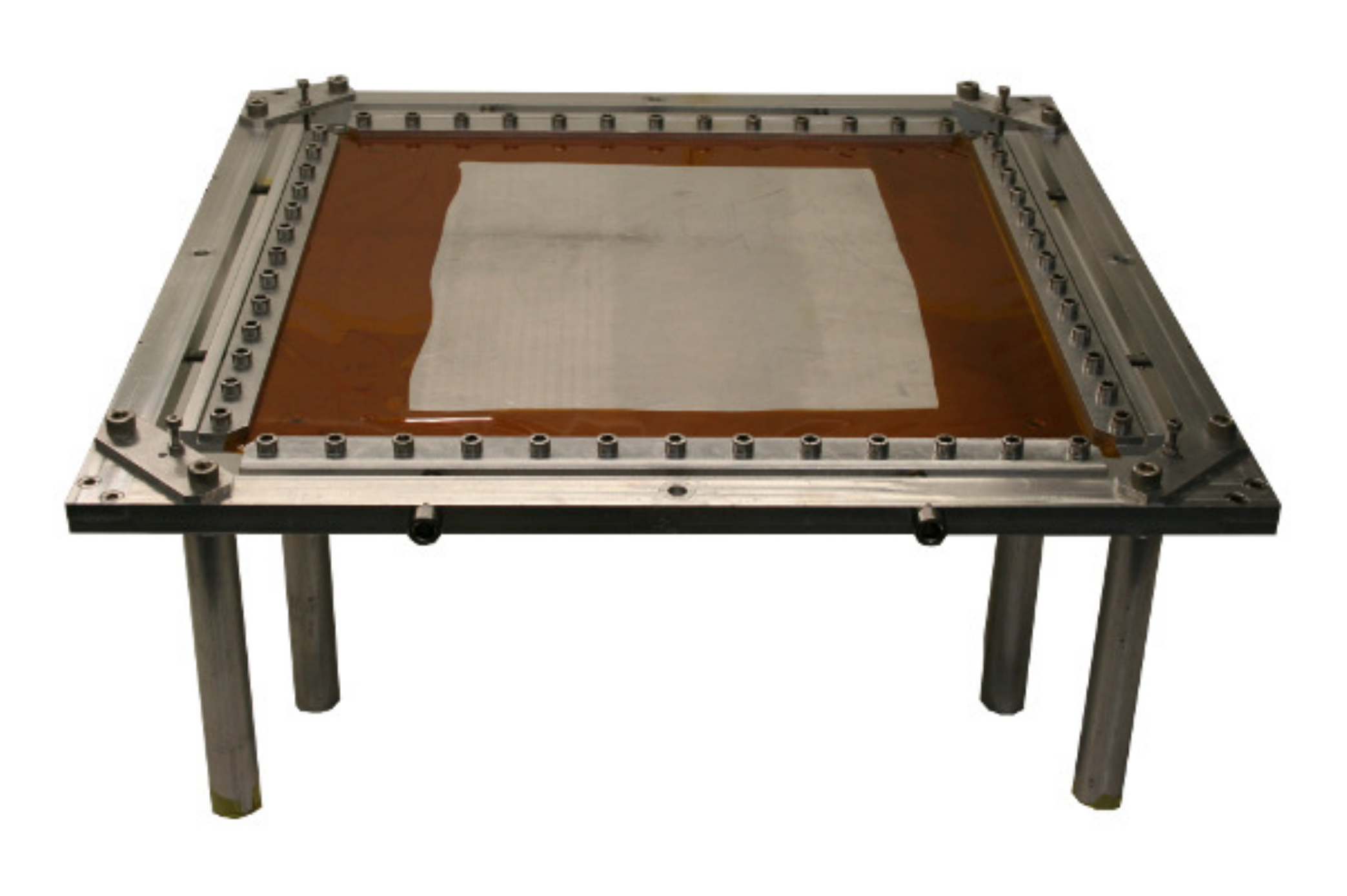}
  \caption{The stretching device used to stretch the GEM foils. See the text for details.}
  \label{fig:stretching-device}
 \end{center}
\end{figure}

After passing the HV test the foils are ready to be glued on FR4 frames.
The frames are 5~mm wide and with inner dimensions of 263.4~$\times$ 216.0~mm$^2$
so that the "dead" area of the foil, i.e. the area without holes, is 1~mm wide along the
inner sides of the frame. This 1~mm  spacing prevents glue
from filling the holes during framing.
First the foils are stretched using a custom designed stretching fixture (see Fig.~\ref{fig:stretching-device}). The  400~$\times$ 400~mm$^{2}$ foil is placed inside
L-shaped frames along each of the four sides, covered with aluminum
bars and tightened down with a set of screws. The foil is then stretched
using two bolts on each of the four sides of the frame to equally distribute the tension
across the foil.

\begin{figure}
 \begin{center}
  \includegraphics[keepaspectratio=true, width=12cm]{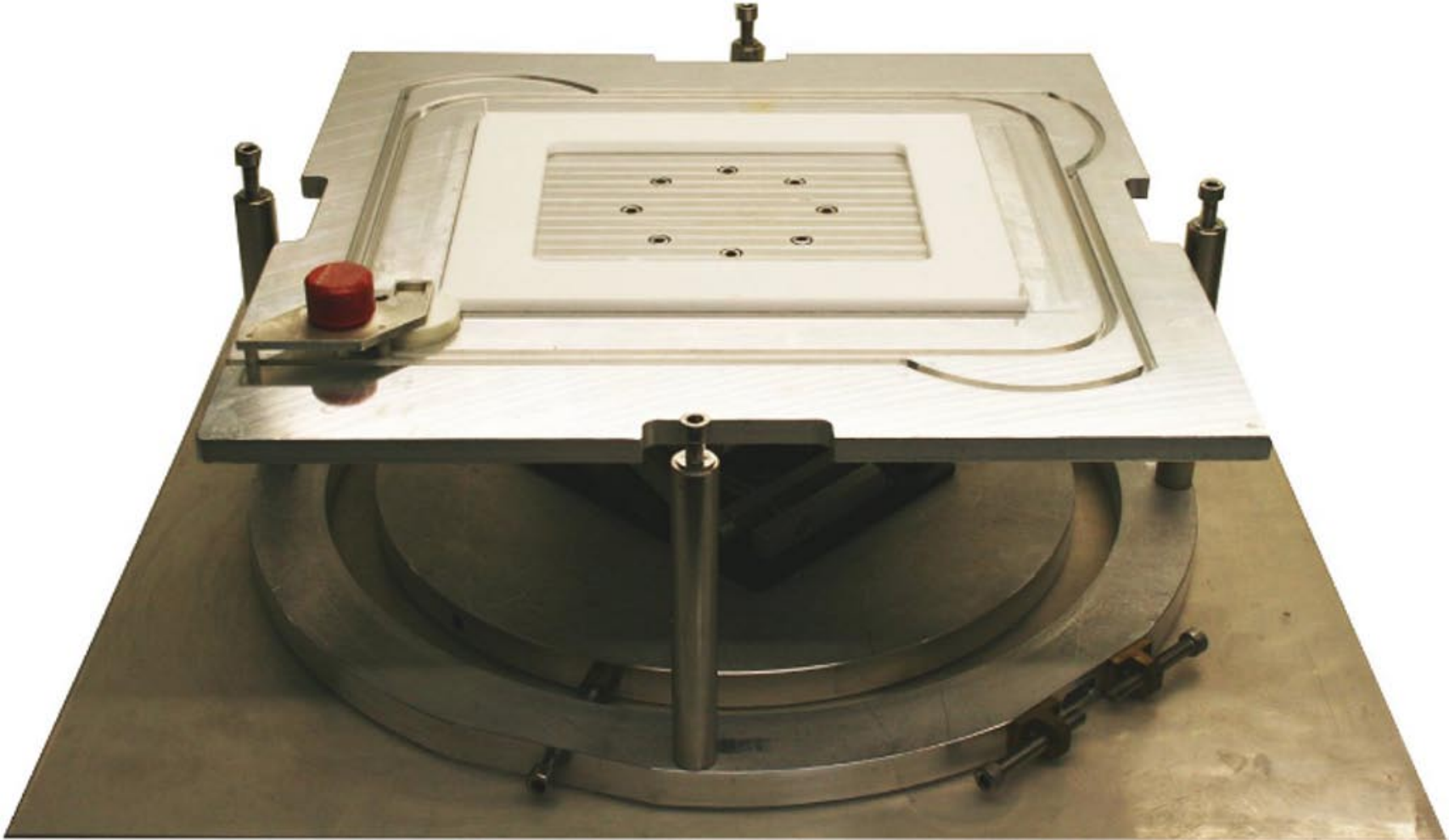}
  \caption{The GEM positioning and glue dispenser device. The GEM is glued on the FR4 frame secured with
  pins on the teflon frame. The glue dispenser, seeing on the lower left corner, applies a uniform glue
  layer of $\sim$100~$\mu$m thickness when moving along the grove on the plate.}
  \label{fig:glue-dispenser}
 \end{center}
\end{figure}

After stretching, the foil is carefully aligned with respect to the FR4 frame in
order to match the eight mounting holes in the foil to those in the frame, using
a positioning device which also serves as glue dispenser (see Fig.~\ref{fig:glue-dispenser}).
The latter consists of a rigid base on which a small lifting table
is mounted that contains a large plate the size of the foil. The FR4 frame is mounted
on top of this plate together with a Teflon frame and eight Teflon pins that
protrude several mm.
The stretching device together with the stretched foil is then mounted a few millimeters above the
positioning device for the hole alignment. The positioning device allows for a fine adjustment of the
position of the large plate. Once the holes are aligned, the stretching device is momentarily
removed to apply glue to the FR4 frame. A custom designed glue dispenser moves along a grove on the
plate and deposits a uniform layer of glue of $\sim$100~$\mu$m thickness onto the FR4 frame.
The glue used is Araldite AY103 epoxy with HY991 hardener mixed in a proportion of 100:40 by weight.
The glue is pumped prior to use in a desiccator in order to remove residual air bubbles and had to be
applied within about one hour after mixing with full polymerization taking approximately 16 hours
at 30-35$^o$C. After applying the glue, the stretching device together with the stretched foil is brought back
to the positioning device.
After verifying the alignment, the lifting table is raised in order to bring the
FR4 frame in contact with the foil. The assembly is then fixed and allowed to cure for typically
$\sim$24~hours.

The foils are then cut and removed from the stretching frame. First, a rough cut is
made around the FR4 frame using a scalpel (Fig.~\ref{fig:stretching-device} shows the
remaining part of the foil after this cut). Then a precise cut, very close to the frame,
is made in order to remove the remaining Kapton. The foil is then cleaned by gently
blowing it with clean dry nitrogen. Another electrical test is performed by raising
each high voltage strip to 550~V in air as had previously been done. In some rare cases,
some foils do not pass the acceptance criteria of less than $\sim$5~nA of leakage current.
These foils are then washed with deionized water, rinsed with alcohol, blown dry and
tested again. This process is sometimes repeated two times. If the leakage current
still exceeds our limit after the second trial, the foil is rejected.

GEMs that pass the high voltage tests then have 20~M$\Omega$ surface mounted resistors
installed on each high voltage segment. When soldering these resistors, the GEM is placed
inside a Plexiglas box such that only the soldering pads for the resistors protrude out for
soldering in order to protect the active area of the GEM foil. After soldering the resistors,
the foils are again tested up to 550~V as a final high voltage test before shipping them
off to Stony Brook University for assembly into the HBD vessel.

\subsection{Assembly and testing}
\label{sec:Assembly}
\subsubsection{GEM storage prior to assembly}
To ensure a dust and water-free environment, GEMs that arrive at
Stony Brook are stored under high vacuum.  A turbo-molecular pump is
used to generate vacuum in the low $10^{-6}$~Torr range.  Prior to
insertion in vacuum, each GEM is further washed and tested.

GEMs are gently sprayed with deionized water for $\sim$30~seconds,
followed by a rinse with clean isopropyl alcohol.  The GEMs are then
blown dry with compressed gas that was passed through a gas ionizer to
facilitate removal of any dust particles.  The GEMs are then placed in
high vacuum for 24~hours to ensure removal of all moisture from the
kapton and FR4 frames.  GEMs that contain moisture are found to
have large leakage currents (on the order of a few $\mu$A at dV =
100~V).  This washing process is repeated on any GEMs which develop
anomalously high leakage current and successfully recovers $\sim$30\%
of these GEMs.

After drying in vacuum, each GEM is moved to a high voltage test
station on a laminar flow table with an ISO Class 1 atmosphere.
Three electrical tests are then performed in air:

\begin{enumerate}
\item The leads of each GEM are checked to have continuity to the
top or bottom of the GEM.  This is most easily tested by confirming
the capacitance of the GEM through the leads with a hand-held multimeter.
\item Each individual strip on the top side of the GEM is tested
for continuity through the resistors to the HV input trace.  With
the bottom of the GEM grounded, the top side of the GEM is raised
to -100~V.  A voltage probe is used to determine that the proper
voltage is present on each of the 28 strips on the top side of the
GEM.  During this process, the leakage current is carefully monitored.
GEMs drawing less than 5~nA are accepted.
\item High voltage is finally applied to the GEM to monitor stability and leakage
current. A current limit of 1~$\mu$A is set on the power supply to limit
damage to the GEM in the event of a discharge.  With the bottom
side of the GEM grounded, the top side is slowly brought to 550~V.
GEMs that are stable and have leakage currents less than 5~nA are
accepted.  GEMs which initially display moderately high leakage
currents ($\sim$10-500~nA), but no discharges, are left at voltage for
up to an hour. Often the current falls back into the
acceptable range.
\end{enumerate}

GEMs that pass these tests are returned to high vacuum for storage,
while those that fail are rewashed and tested again.  GEMs which
continue to fail after two cycles of washing are not used in the HBD.

\subsubsection{Copper GEM assembly}
All GEMs are dust sensitive and must be handled in a clean room or
(preferably) upon a laminar flow table.  Once coated with CsI, the
devices are also water sensitive and will lose their quantum
efficiency if exposed to an atmosphere with high water
concentration for an extended period of time.  For this reason,
CsI-coated photosensitive GEMs are handled in the inert atmosphere
of a glovebox.  Unfortunately, since a glovebox is a closed-loop
system it cannot maintain the level of cleanliness found on the
laminar table.  Because of this, strategies that minimize handling of
the HBD (and GEMs) in the glovebox were found to produce the best
results.  The most successful procedure for HBD assembly involved
assembling the bottom two layers of all GEM stacks in the cleanest
available environment (the laminar flow table), and then adding the
CsI-coated GEMs in the dry glovebox environment.  This procedure
limited exposure to the glovebox environment to 2-3 weeks.

\begin{figure}[ht]
  \begin{center}
    \includegraphics[keepaspectratio=true,
    width=10cm]{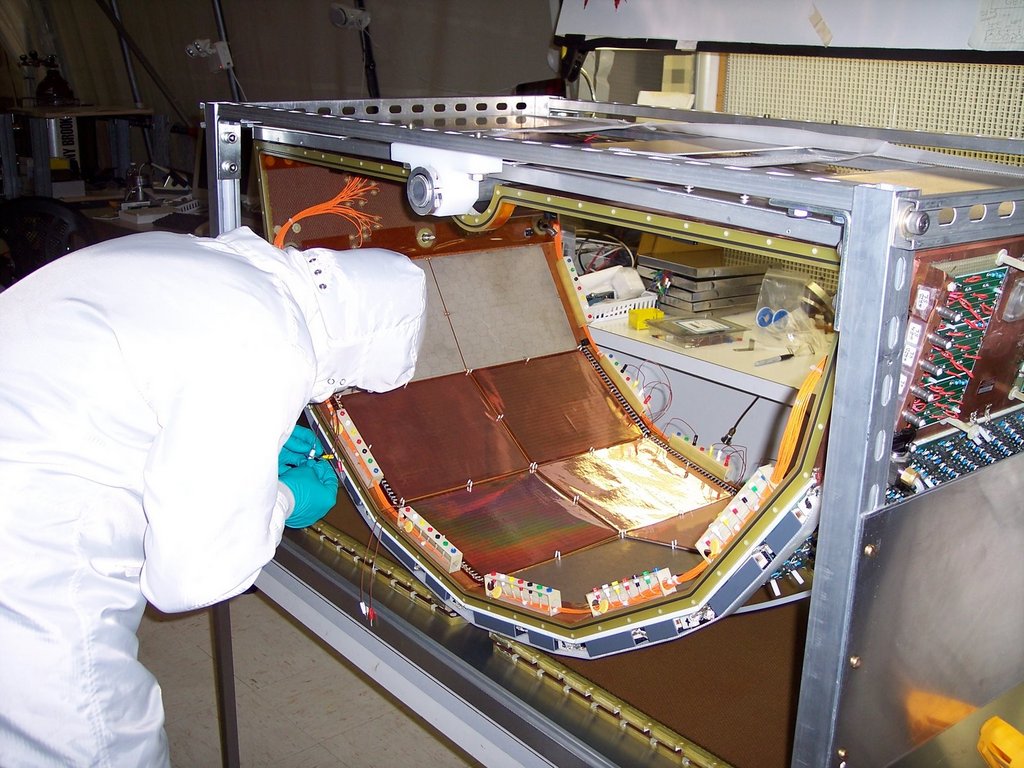}
   \caption{\label{fig:CopperInstallation}Installation of standard
   copper GEMs into the HBD vessel prior to placing the vessel in the
   glovebox.}
 \end{center}
\end{figure}

For installation of the Cu GEMs, the HBD vessel is mounted on a
rotating fixture and placed in front of the laminar flow table.  With
clean air blowing through the interior of the vessel, the standard
copper GEMs (two per module) are mounted in place over the readout
pads as shown in Fig.~\ref{fig:CopperInstallation}.  After mounting,
each GEM is re-tested in situ for continuity and stability (tests
no. 1 and 3 above) to ensure no damage was caused during
installation.

Once all standard GEMs are installed and re-tested, the vessel is
moved into a sealed glovebox to accept the CsI-coated gold GEMs
at the top of the triple-GEM stack.  Once sealed, the
glovebox recirculates nitrogen through a purifier and achieves H$_2$O
concentrations of $<$ 10~ppm.  Regular sweeps of the interior of the
glovebox with an ULPA vacuum cleaner mounted inside the glovebox
atmosphere ensure that particulate contamination is
at an acceptable level.

One critical choice for the glovebox was the selection of the material
of the gloves themselves.  While Butyl gloves provide the best water
barrier, they are not highly rated with regard to generation of
particulate matter.  Hypalon gloves were selected as having the best
rating for particulate matter, and were found to elevate the baseline
water concentration of an empty glovebox from 2-3~ppm to 7-8~ppm, which was still
quite acceptable.

\subsubsection{Evaporation of CsI onto Au plated GEMs}
GEMs are made photosensitive by the evaporation of a thin layer of CsI
on the GEM electrode surface.  This layer is not chemically
stable on a copper substrate since CuI is more tightly bound than CsI.
For this reason, a special subset of the GEM production included GEMs
whose metallic surface was overlayed with Ni (diffusion barrier) and
then Au (chemically inert layer)~\cite{ref:CuNiAu}.
Not surprisingly, these GEMs were seen to have identical gain and
voltage stability characteristics as the standard copper GEMs
and were handled in an identical manner during the testing
and framing stages performed at the Weizmann Institute of Science.

Reflective photocathodes exhibit a quantum efficiency that saturates
as a function of the cathode thickness.  For CsI, this saturation
point is found at $\sim$200~nm thickness.  HBD photocathodes were made to
have 300~nm thickness to ensure full sensitivity in spite of possible
non-uniformities of the coating.

GEM photocathodes are manufactured at Stony Brook by evaporating a
$\sim$300~nm-thick layer of CsI to their top surface using an evaporator that was
on loan from INFN \cite{ref:INFN}. The evaporator was used many times in the past to
evaporate photocathodes for RICH detectors used in CEBAF Hall A kaon
experiments \cite{ref:garibaldi} and is of sufficient size to evaporate 4 HBD photocathodes
simultaneously.

\begin{figure}[ht]
  \begin{center}
    \includegraphics[keepaspectratio=true,
    width=10cm]{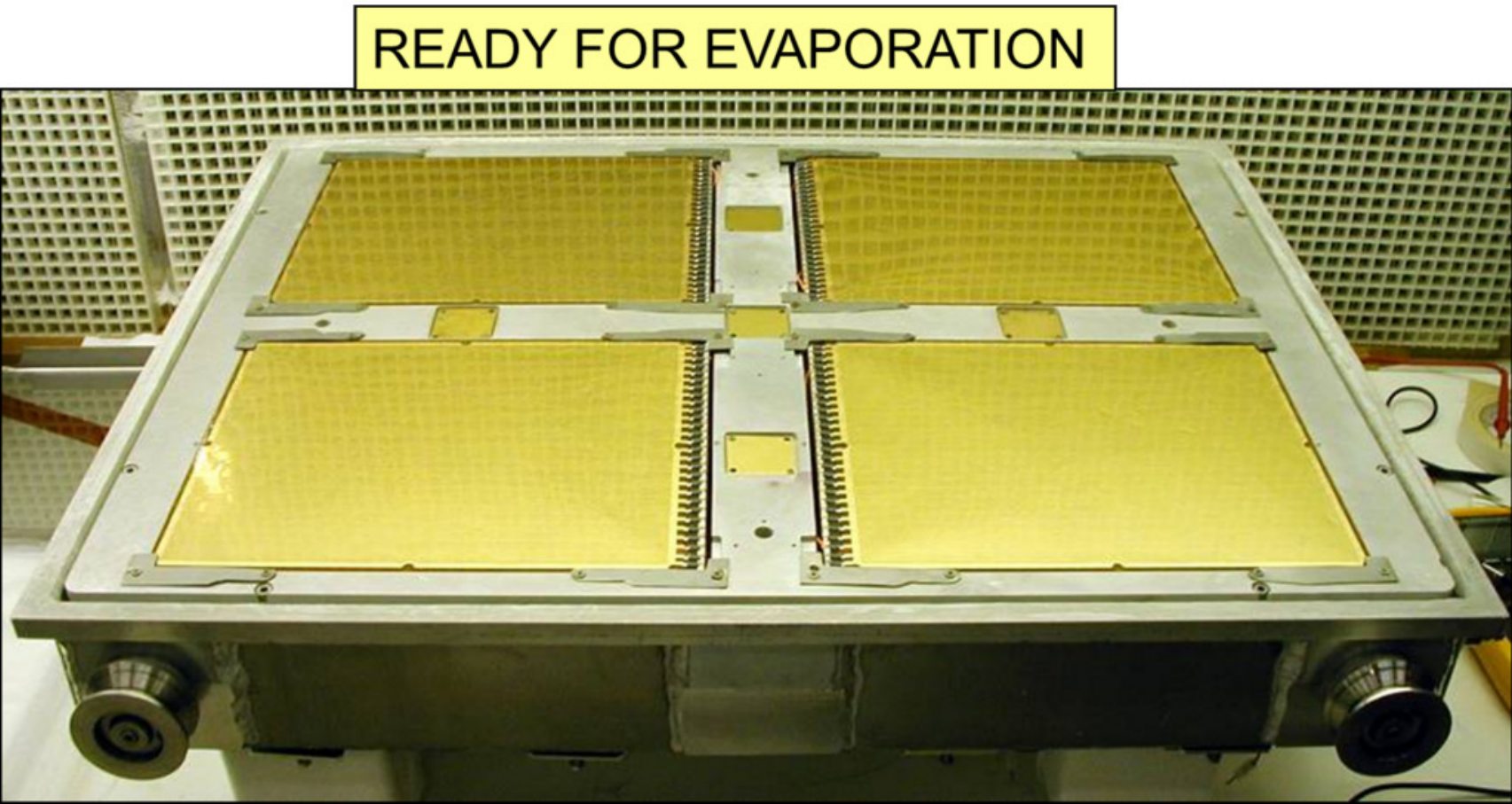}
   \caption{\label{fig:GemsInCart}The evaporation cart holds four
   Au-plated GEMs and five small circuit cards used to monitor the
   quality of each evaporation.  Wheels on the cart allow to be
   moved in vacuum to map the quantum efficiency across the surface of
   each GEM.}
 \end{center}
\end{figure}

Gold GEMs are mounted four at a time into a sealed transfer box and
placed into the evaporator for CsI photocathode deposition.
Additionally, several small (2~cm $\times$ 2~cm) Cu-Ni-Au circuit cards
(called chicklets) are also mounted into the box to be used as a
monitor of the quantum efficiency (QE).  A set of four GEMs and five
chicklets ready for evaporation is shown in
Fig.~\ref{fig:GemsInCart}.  Once in the evaporator, the lid of the
transfer box is removed to expose the GEMs.  The evaporator is pumped
down to a vacuum of $\sim2\times 10^{-8}$~Torr with a combination of a
turbopump and a cryopump.  While pumping, the transfer box containing
the GEMs is heated to 40~$^\circ$C to drive off water and other contaminants.

Facing each GEM is a molybdenum crucible with a single piece of
CsI weighing 0.8~g.  Once ultra high vacuum is achieved, the
crucible is resistively heated to vaporize the CsI.  A quartz
thickness monitor positioned near the GEM surfaces is used to
determine the deposition rate of the CsI.  By varying the
current through the crucibles, the rate is kept near 1~nm/s.
The final thickness of the CsI layer is typically $\sim$300~nm.

After CsI deposition, the transfer box is moved to the QE measurement
section of the evaporator.  It was observed that the QE of newly
deposited photocathodes can change (typically improve) by a factor of
$\sim$2 over a period of $\sim$8~hours, so the measurement is not
performed until this time has passed.  A deuterium lamp shines through
a 160 nm filter, enters the vacuum via a MgF$_2$ window, and shines
onto a movable mirror.  This mirror can be rotated to allow the light
to be directed either onto the GEM surface or onto a reference
phototube of known QE.  Once the light source has been calibrated
using this phototube, the QE of the new CsI photocathode can be
determined relative to the tube.  A mesh with 300~V is used to draw
photoelectrons from the CsI surface, which is measured as current by a
picoammeter.  The transfer box and phototube can both be translated
inside the evaporator, allowing a scan of the entire surface of each
photocathode.  This measurement ensures the photo-sensitivity of each
cathode across its entire surface, but only at a single wavelength.
The small chicklets are later transferred to Brookhaven National
Laboratory, where a scan across the wavelength range 120 nm to 200 nm
is performed using a vacuum photospectrometer.  It was found that
every evaporation during the entire history of the project produced
identical photo-sensitivity and uniformity.

\subsubsection{Installation of GEM photocathodes}
After the QE scan, the evaporator is backfilled with ultra high purity
argon up to atmospheric pressure.  The transfer box containing the gold
GEMs with CsI photocathodes is then sealed in the argon atmosphere
before the evaporator chamber is opened to air.  The sealed transfer
box is put into the glovebox through a load-lock system, which
prevents any room air from entering the glovebox.  The transfer
box is not opened until it is inside the dry nitrogen atmosphere of
the glovebox, ensuring that no humidity affects the photocathodes.

The glove box is set up in three modules, each with a distinct purpose.
The first module has the rail system that accepts the transfer box
from the evaporator, with a winch mounted on the ceiling to lift
the transfer box lid.

The second module serves as the high voltage testing station for
the gold GEMs after CsI deposition.  Here the gold GEMs undergo
all the above mentioned electrical tests, with the exception that
the voltage in step 3 is decreased from 550~V to 500~V in the
nitrogen atmosphere of the glove box.  It is common for a gold GEM
to exhibit several discharges upon the first application of high
voltage after CsI deposition, but stabilize afterwards.  Rarely a
gold GEM exhibits a short or anomalous leakage current after
deposition. If so, it is washed and the testing/deposition
process is repeated.

\begin{figure}[ht]
  \begin{center}
    \includegraphics[keepaspectratio=true,
    width=10cm]{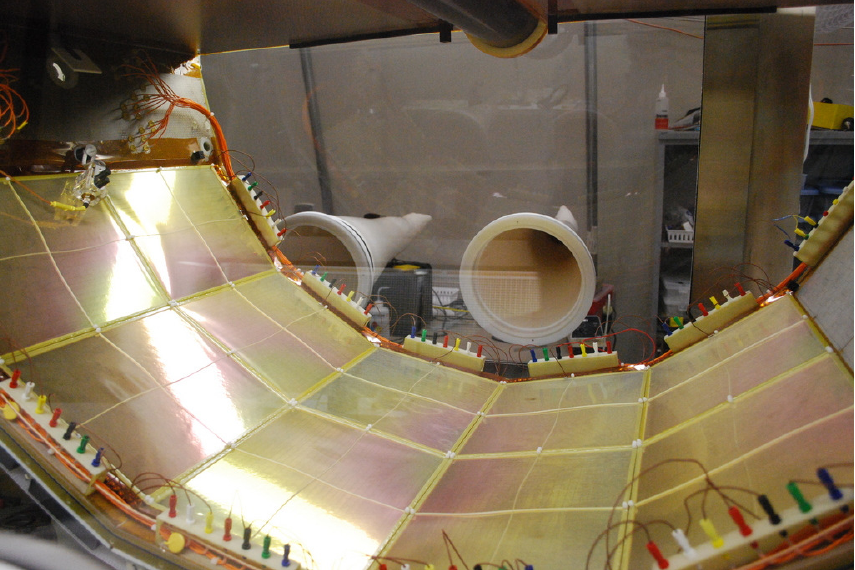}
   \caption{\label{fig:HbdCompleted} A complete HBD in the glovebox
following installation of all photocathodes and meshes.  The lucite
device in the upper left is the scintillation cube, placed outside the
acceptance of PHENIX and used to monitor quantum efficiency in situ
and over time (see Section~\ref{sec:scintillation-cubes}).}
 \end{center}
\end{figure}

The third station houses the HBD vessel.  The vessel is mounted in a
rotating fixture that can be turned to allow access to the edge
modules (normally out of reach of the gloves).  After the gold GEMs
are mounted, all three GEMs in a stack are tested in situ under high
voltage.  The mesh is then installed over the stack, and 500~V is
applied across the drift gap to ensure that there is no electrical
continuity between the GEM and mesh.  A completed HBD vessel is shown in
Fig.~\ref{fig:HbdCompleted}.  The irredescent color of the GEM
surfaces is created by the CsI coating.

Once all the interior components of the HBD are assembled, final
tests are done to ensure that the device is fully functional.  Each
GEM is tested by measuring the capacitance across its HV input leads,
and finally for high voltage stability.  Following these tests, the
sides of the vessel are installed while the vessel is still inside the
glovebox, sealing the dry nitrogen atmosphere inside.  The vessel is
then brought out of the glovebox and onto a test bench, and purged
 with CF$_4$.

\subsection{Cosmic ray tests}
Because many materials of the HBD (FR4, kapton) outgas water, it
is necessary to store the HBD during testing with a high flow rate
recirculating gas system.  This system recycles CF$_4$ gas at a rate of
2.2~slpm, and maintains water levels below 20~ppm, and oxygen below 2~ppm.
Measurements of higher water contamination than oxygen are indicative
of a system that is well sealed against leaks, but outgasses water from
its interior surfaces.

\begin{figure}[ht]
  \begin{center}
    \includegraphics[keepaspectratio=true,
    width=10cm]{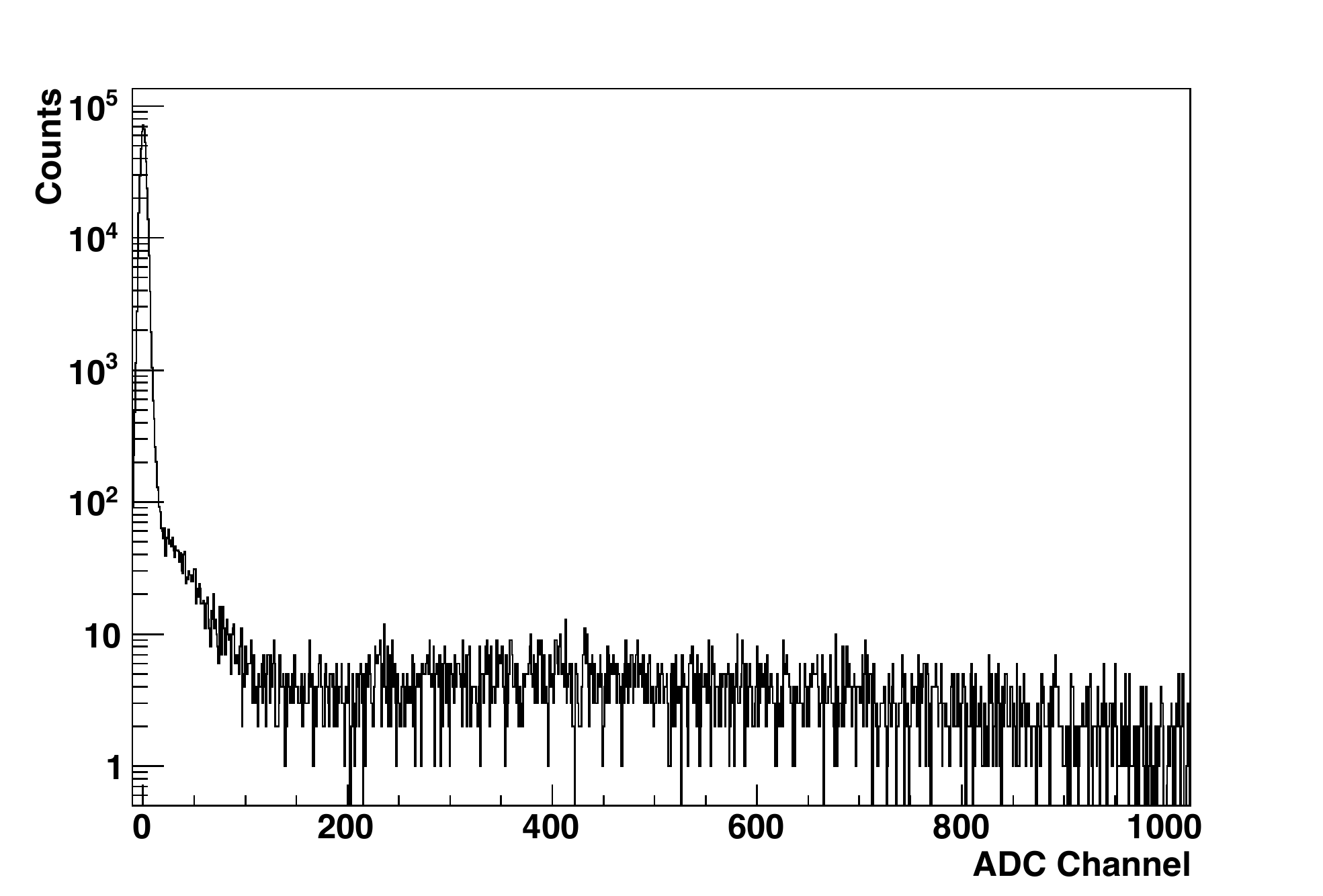}
   \caption{\label{fig:Cosmics} A typical spectrum from the HBD being
   tested with a cosmic ray trigger.  In addition to the response to
   cosmics and pedestal, an exponential distribution due to single
   photo-electrons from scintillation is seen and used for absolute
   gain determination.}
 \end{center}
\end{figure}

While under flow of CF$_4$, the HBD modules are turned on one-by-one
with the mesh high voltage set in the forward bias mode.
This polarity of mesh voltage makes the HBD sensitive to ionizing
particles and thereby cosmic ray muons.  A small scintillator is
placed above the sector under test and used as a trigger for
pulse-height spectra taken by a CAMAC-based DAQ system.  Eight
hours of tests not only provide  an excellent stability measurement,
but also sufficient statistics from 14 pads (those shadowed by the
trigger scintillator) to measure the gain of each module.  A typical
measurement from the cosmic ray data is shown in
Fig.~\ref{fig:Cosmics}.  The spectrum contains three components: (i)
Pedestal (fit with a Gaussian function), (ii) Single photoelectrons from scintillation
(Exponential), and (iii) Ionization from ionizing tracks (Landau).  Of
these three components, the response to the single photoelectrons from
scintillation proved to be the most stable in establishing each
module's gain.  These data were later used to set the operating
voltage for each module.

\subsection{Scintillaton Cubes}
\label{sec:scintillation-cubes}
  The photocathodes which were produced during the evaporation, installation and assembly were
required to perform at maximum efficiency for the duration of the entire experiment, which spanned
a period of almost two years. As described in Section~\ref{sec:Gas}, the HBD gas system was designed to
maintain the detector gas at extremely low levels of water and oxygen, and performed extensive monitoring
of these levels throughout the run. However, in order to be sure that the photocathodes did not deteriorate
over time, an independent means of monitoring their quantum efficiency was developed.
\begin{figure}[h!]
  \begin{center}
  \vspace{-0.3cm}
    \includegraphics[keepaspectratio=true, width=10cm]{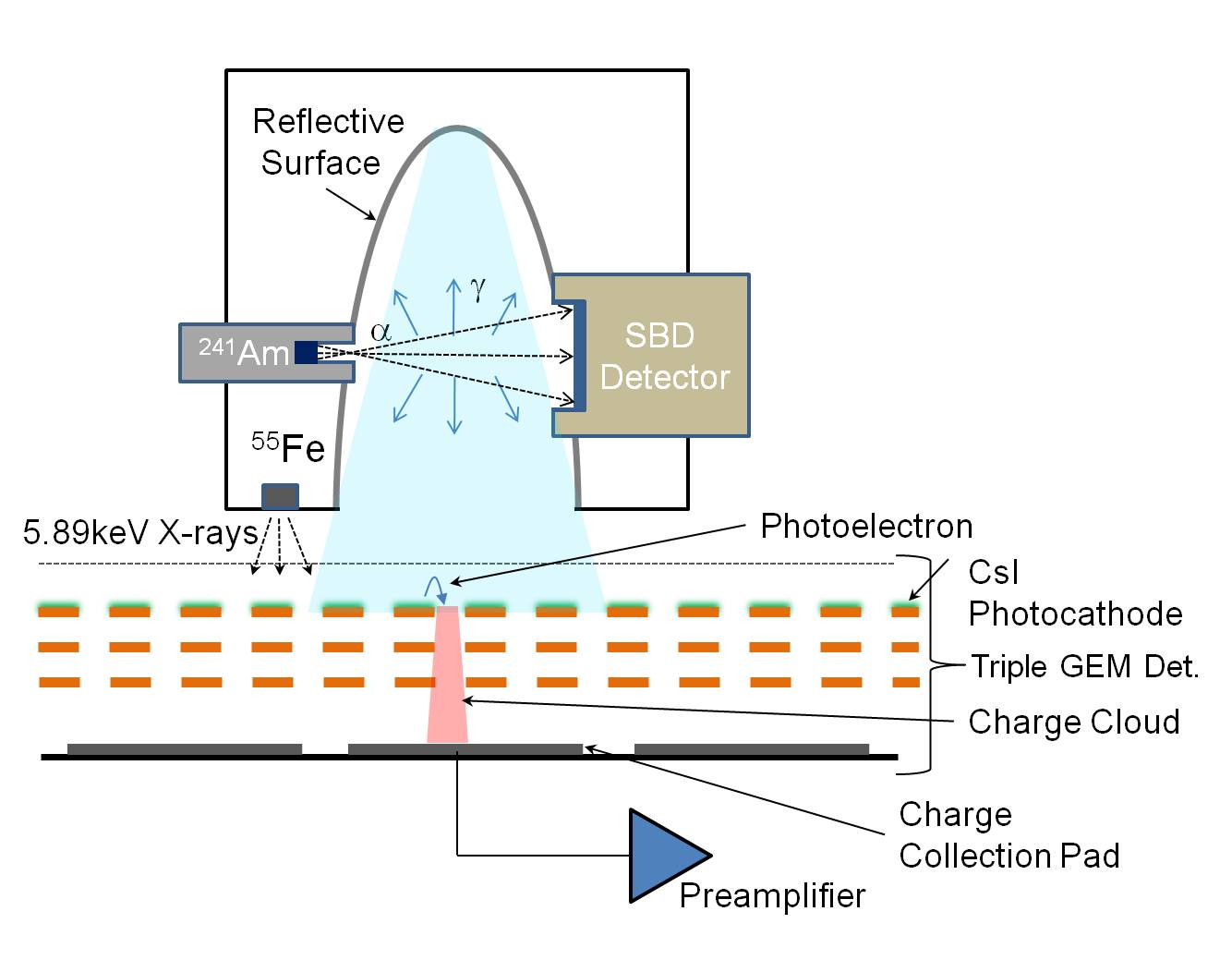}
   \caption{\label{fig:scint-cube-sketch}Schematic representation of the scintillation cube inside the HBD
    detector. Alpha particles from the $^{241}$Am source produce scintillation light which is focused onto
    the photocathode in one area of the detector. An $^{55}$Fe source measures the gas gain in the same region,
    allowing a determination of the photoelectron yield from the photocathode.}
 \end{center}
\end{figure}

  The method utilizes the scintillation light produced in CF$_4$
   to illuminate a small region of the
photocathode which could be used to determine the photoelectron yield. It is then assumed that the rest
of the photocathode is the same, since all photocathodes are inside the same gas volume for each half of
the detector. Fig.~\ref{fig:scint-cube-sketch} shows a schematic representation of this device. It consists
of a lucite cube, which we refer to as a ``scintillation cube'', in which a small (1~$\mu$Ci) $^{241}$Am source is
mounted. Approximately 1~cm away, a silicon surface barrier detector (SBD) is mounted to detect alpha particles
which pass  through the gas. The alpha particles deposit  $\sim$ 4~MeV in the gas, and still have enough
residual energy to trigger the SBD, which is used to externally trigger on the source. The scintillation light
produced in the gas is focused onto the photocathode illuminating essentially one pad and producing
$\sim$ 4-5 photoelectrons. An $^{55}$Fe source is also mounted in the cube which produces 5.9~keV X-rays that
convert in the gas and allow a determination of the gas gain in the same region of the detector. Using the
gas gain and the peak of the photoelectron distribution, the photoelectron yield can be determined. While this
does not give us a measure of the absolute quantum efficiency, it can be used to monitor any changes in the
quantum efficiency over time. Measurements of the photoelectron yield were performed several times throughout
the experiment, and, as discussed in Section~\ref{subsec:photocathode-monitoring}, no change in the quantum efficiency
was observed.

\section{Readout electronics}
\label{sec:Elec}   
   As shown in Fig.~\ref{fig:readout_board}, the readout pads are connected by wires passing through
the honeycomb back panels to individual hybrid preamplifiers located on readout boards mounted on the back of the detector.
These wires were kept very short ($\sim$2~cm) in order to minimize pickup and noise. The preamps plug into sockets on the
readout boards, which also contain power and ground planes that form part of the overall ground plane
of the detector. The entire detector is shielded by virtue of the ground planes on the readout boards
and by copper cladding on all of the other FR4 structural panels. This essentially forms a complete Faraday shield for the detector,
which is very important and effective in minimizing the noise on the front end readout electronics.

\subsection{Hybrid preamps}
\begin{figure}
 \begin{center}
    \includegraphics[width=12cm]{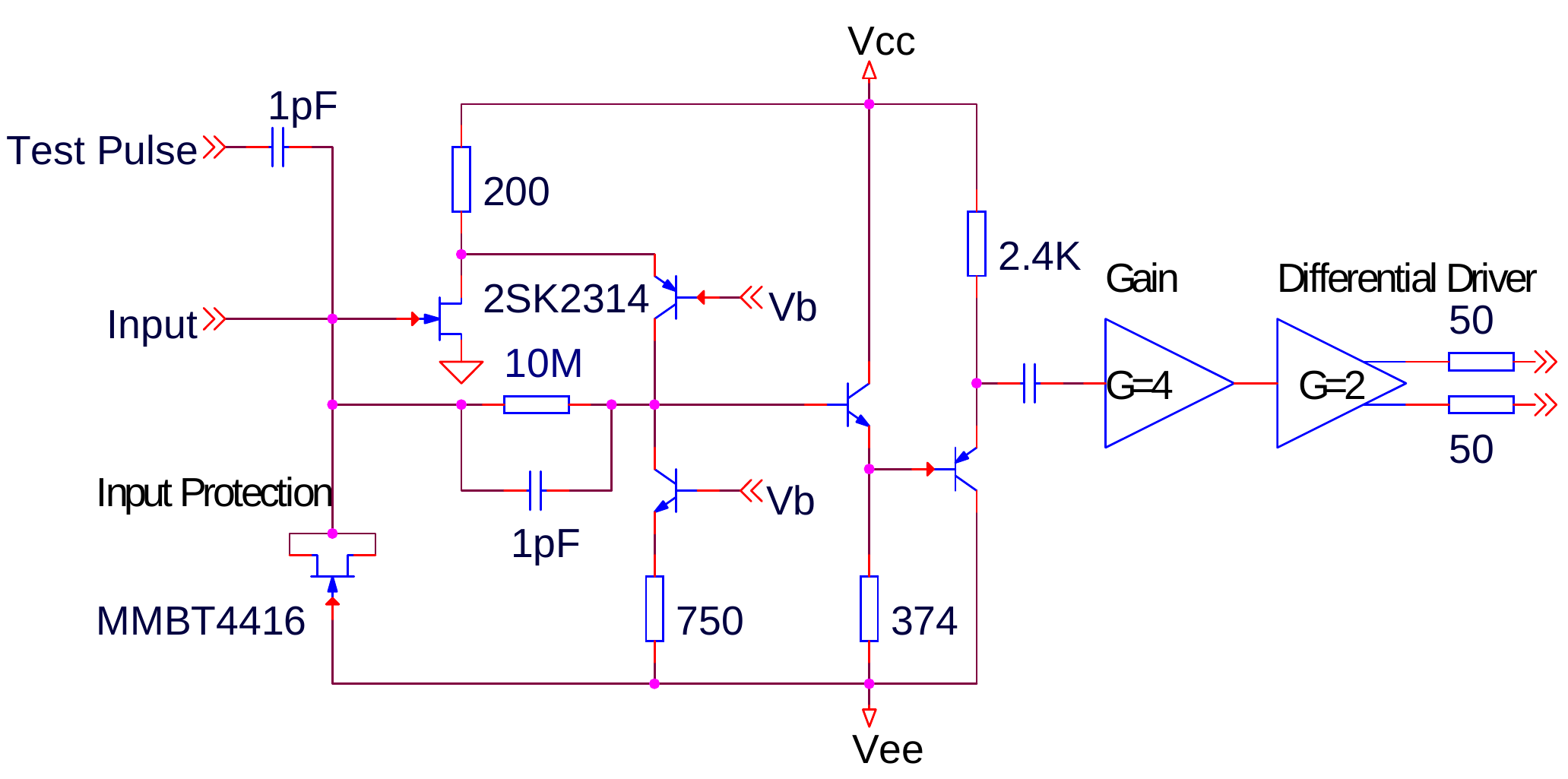}
	\caption{\label{fig:IO1195-2}Simplified schematic of the IO1195-1 hybrid preamp.}
 \end{center}
\end{figure}

   The charge signal from each readout pad of the GEMs is amplified by a custom designed hybrid preamplifier (IO1195-1) developed
by the Instrumentation Division at BNL. A simplified schematic circuit diagram is shown in Fig.~\ref{fig:IO1195-2}.
A detailed schematic can be obtained from BNL's Instrumentation Division ~\cite{ref:BNL_Inst_Div}. It uses an FET input
(2SK2314) that is capable of accepting bipolar signals, but is used with only negative inputs with the HBD. The gain was set to give an output of $\pm$ 100~mV for an input signal of 16~fC (100,000 e), which corresponds to an average primary charge of 20 photoelectrons at
a gas gain of $5 \cdot 10^3$. It has a peaking time of 70~ns and a decay time of 1~$\mu$s. When connected to the GEM pad, the noise was measured to be $\sim$ 1000 e, which is equivalent to 0.2 photoelectrons (p.e.) at a gain of $5 \cdot 10^3$. The input to the preamp incorporates an additional FET that is used for discharge protection, which adds slightly to the noise, but helps to protect against damage
from sparks. It should be noted that throughout the entire operation of the detector, not a single preamp was ever damaged due to sparking.

The output has a differential driver which has a maximum dynamic range of $\pm$ 1.5~V, corresponding to an
input signal of up to 300~p.e., and delivers its differential output signal to a shaper located in the Front End Module (FEM). The cable
used to connect the preamp to the shaper (Meritec Hard Metric 700319-01) is 10 m long and consists of 2 isolated 26 gauge parallel wires per channel
with an overall shield and has an impedance of 100~$\Omega$. The preamp operates with $\pm$5~V and draws 165~mW per channel. The preamps
were designed with a small overall profile (19~mm~$\times$~15~mm) in order to minimize their mass (see Table~\ref{tab:material-budget}).

The preamp also includes a 1~pF capacitor that allows injecting a known amount of charge into the input stage. A test pulse
of known amplitude is generated externally and brought in on the readout boards where it is distributed in groups of 8-12 channels
to the individual preamps. The test pulse provides a convenient way to test and calibrate all the preamps on the detector and
to monitor the electronic gain of the system.

\subsection{Front End Module}
\begin{figure}
 \begin{center}
    \includegraphics[width=10cm]{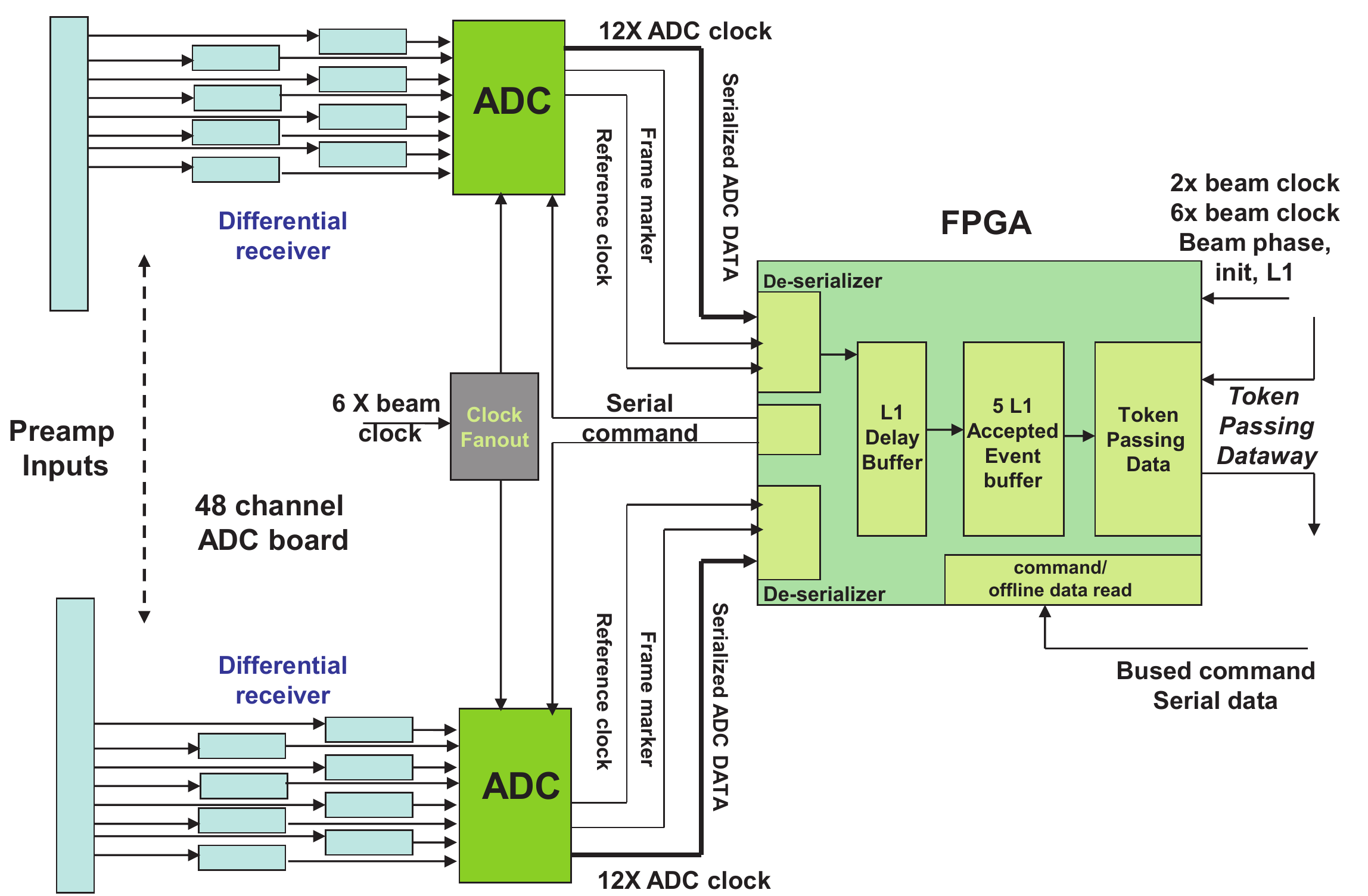}
	\caption{\label{fig:HBD_FEM}Overall layout of HBD Front End Module.}
 \end{center}
\end{figure}

The overall layout for the HBD FEM is shown in Fig.~\ref{fig:HBD_FEM}, and further details are given
in~\cite{ref:ChiIEEE2007}. The FEM receives the preamp signals using a differential receiver, which
provides some additional shaping, and then digitizes them using a 65 MHz 8 channel 12 bit flash ADC (Texas Instrument ADS5272).
Each FEM contains six 8 channel ADCs for a total of 48 channels per module. The gain of the shaper was set such that a preamp
input charge of 16 fC (10$^5$ electrons) produces a signal of 160 ADC channels (0.1~fC/ch). The preamp signal was set in the range
of $\pm$ 1.5~V around the ADC common mode voltage, which effectively uses only half of the full dynamic range of the ADC.
The ADC clock is derived from the 9.6~MHz RHIC clock at six times its frequency, i.e. 57.6~MHz, in normal data taking mode.
The output of the ADC is serialized at 12 times the ADC clock and sent along with additional reference signals to an ALTERA Straitx III 60 FPGA where the data is de-serialized.
The FPGA has 8 independent serializer/de-serializer blocks which are used for the eight channels of each ADC. It also provides a de-serialized function that outputs 6 bit wide data at 120~MHz, which are then regrouped into 12 bit wide ADC data within the FPGA. The PHENIX DAQ requires event
buffering for up to 5 events and 40 beam crossings for its Level~1 trigger, which is also accomplished within the FPGA. The FPGA could
also provide a Level~1 trigger output for the PHENIX DAQ, but this feature was never implemented during data taking.

\begin{figure}[h!]
 \begin{center}
    \includegraphics[width=10cm]{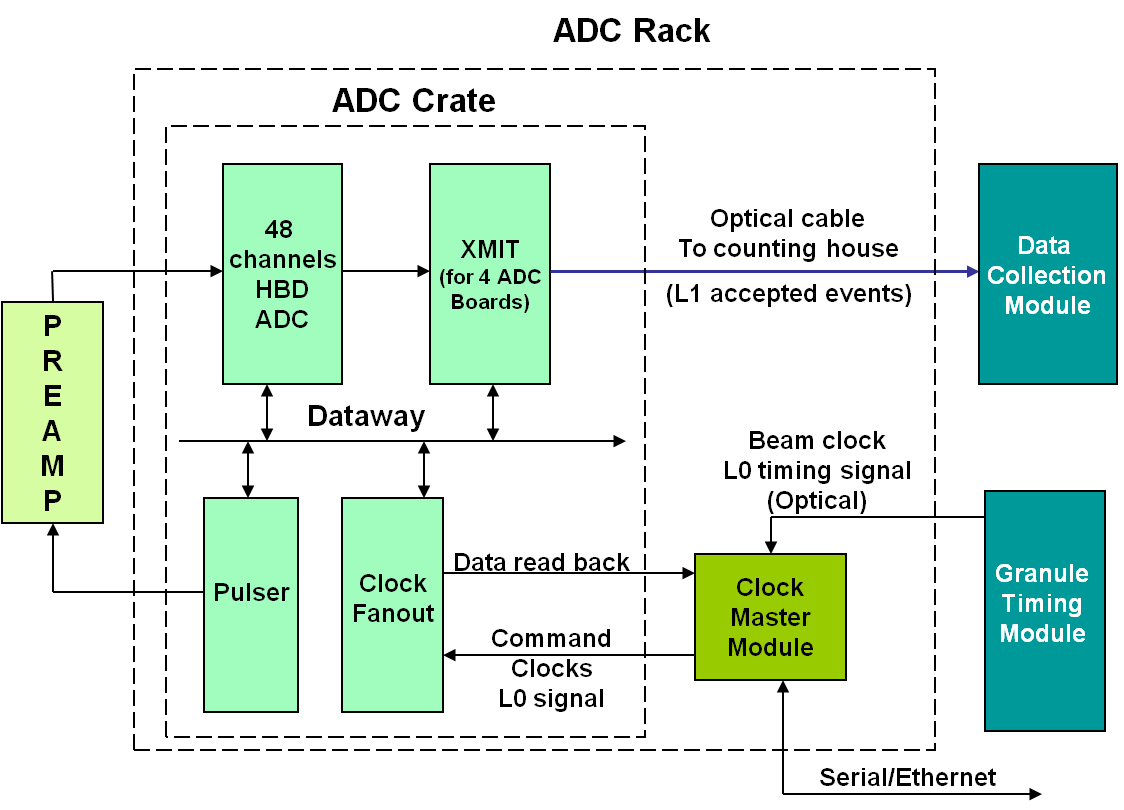}
	\caption{\label{fig:HBD_FEM_System} HBD FEM system and interface to the PHENIX Data Acquisition System.}
 \end{center}
\end{figure}

The FEM system and its interface with the overall PHENIX DAQ are shown in Fig.~\ref{fig:HBD_FEM_System} and
are further described in~\cite{ref:ChiIEEE2007}. The FEMs are hosted in a set of 6U VME crates located in a single rack close
to the detector. A custom dataway is used as a bus for serial data transmission between adjacent modules, and a Clock Master Module
is used to interface with the PHENIX Granule Timing Module that provides synchronization with the PHENIX timing and control
system, and also provides a separate interface for slow downloads. The Level~0 (L0) and Level~1 (L1) timing signals and serial data
are generated within the Clock Master Module and sent to the FEMs. Upon receiving an L1 trigger, 12 samples of data per ADC channel
are sent by optical fiber at six times the beam crossing frequency to a set of Data Collection Modules located in the PHENIX
Control Room with an average transfer time of $\sim$ 40~$\mu$s.

\section{Gas system and monitoring}
\label{sec:Gas}   
 
\begin{figure}[h!]
  \vspace{0.2cm}
 \begin{center}
  \includegraphics[width=10cm]{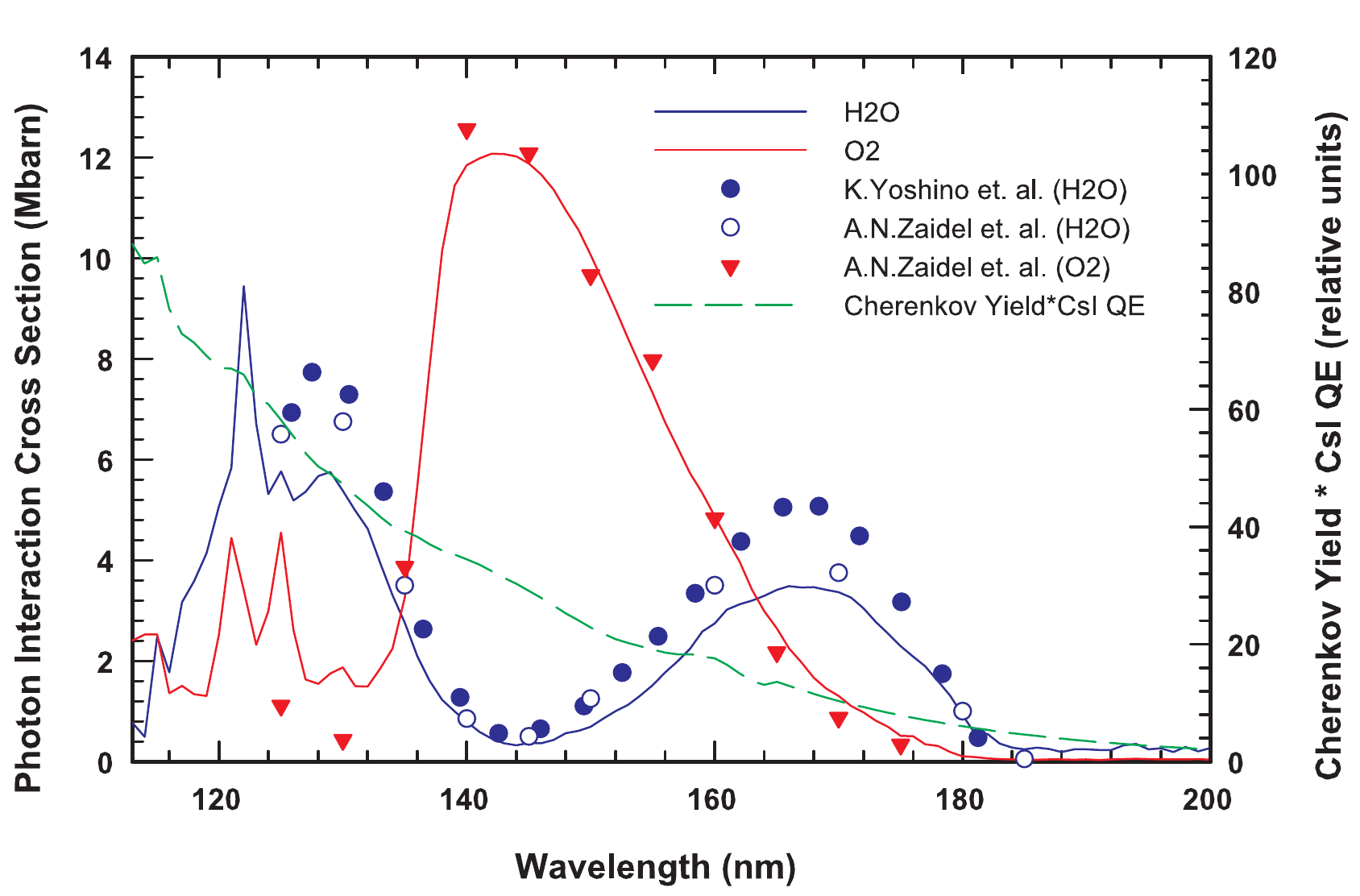}
  \caption{\label{fig:H2O_O2Xsect_RelCherenkovYield}Photon
  absorption cross sections for water and oxygen over the wavelength
  range of sensitivity of CsI to Cherenkov light. Solid curves are
  measurements made on a spectrometer at BNL. Other water and oxygen
  measurements are from \cite{ref:Zaidel,ref:Yoshino}. The Cherenkov
  yield times quantum efficiency is shown by the dashed curve and given in relative units on the right ordinate.}
 \end{center}
\end{figure}

\begin{figure}[h]
 \begin{center}
  \vspace{0.2cm}
  \includegraphics[width=10cm]{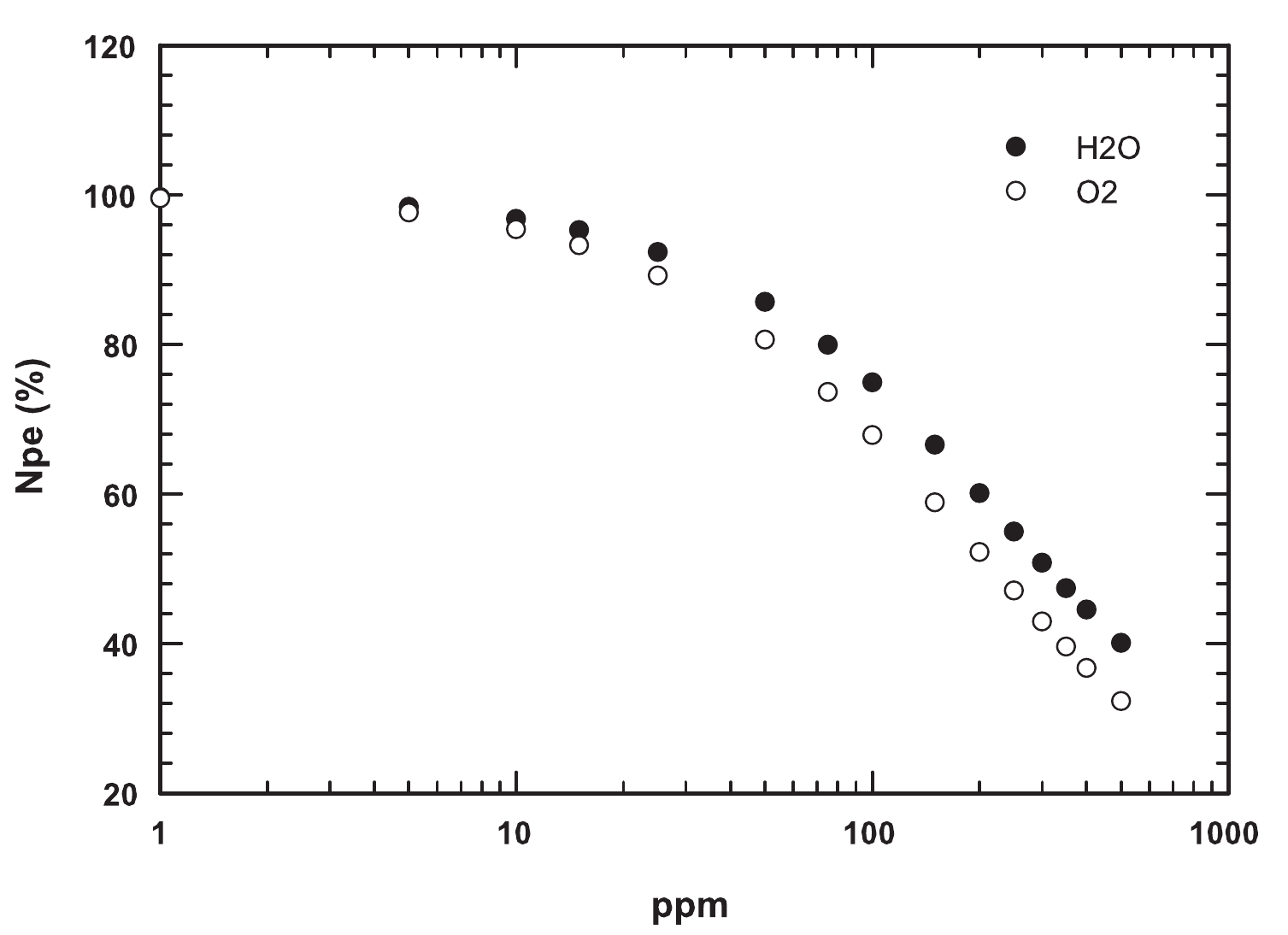}
  \caption{\label{fig:NpeVsPPMs}Relative number of photoelectrons, Npe,
  produced in 50 cm CF$_4$ as a function of the number of ppms of
  water and oxygen contamination in the gas.}
  \end{center}
\end{figure}

Maintaining high gas purity is a critical factor for the overall
performance and operation of the HBD. In particular, impurities
such as water and oxygen adversely affect the performance
in several ways. Both water and oxygen have strong absorption
peaks for Cherenkov light in the spectral range of sensitivity of the
CsI photocathodes, and even small levels of either of these
contaminants can produce a significant loss of photoelectrons.
This is illustrated in Fig.~\ref{fig:H2O_O2Xsect_RelCherenkovYield},
which shows the photon interaction cross sections for water and
oxygen in the wavelength range of interest. Fig.~\ref{fig:NpeVsPPMs} shows how this translates into
a loss in the number of photoelectrons in a 50 cm long radiator.
The main source of oxygen contamination is  from leaks in the
detector vessel. While these leaks are very small ($<$ 0.12 cc/min),
they nevertheless allow  some diffusion of oxygen into the detector.
The main source of water is from outgassing within the detector
itself, particularly from the GEM foils and pad readout plane, which
are made of kapton, and from the walls of the vessel, which are
made of FR4. These components produce a constant source of water
inside the detector which could only effectively be reduced by
maintaining a high gas flow rate.

In addition, prolonged exposure to water can cause a deterioration
in the CsI quantum efficiency \cite{ref:Breskin}, and can also
affect the gain stability of the GEMs \cite{ref:Azmoun_Tech_Etch_Study},
although these effects were not problems at the levels at which
the HBD was operated in PHENIX.
Finally, CF$_4$ is an aggressive gas requiring that all
gas system components be resistant to chemical corrosion, and,
in addition, CF$_4$ is also very expensive.

\begin{figure}[h]
 \begin{center}
  \includegraphics[width=10cm]{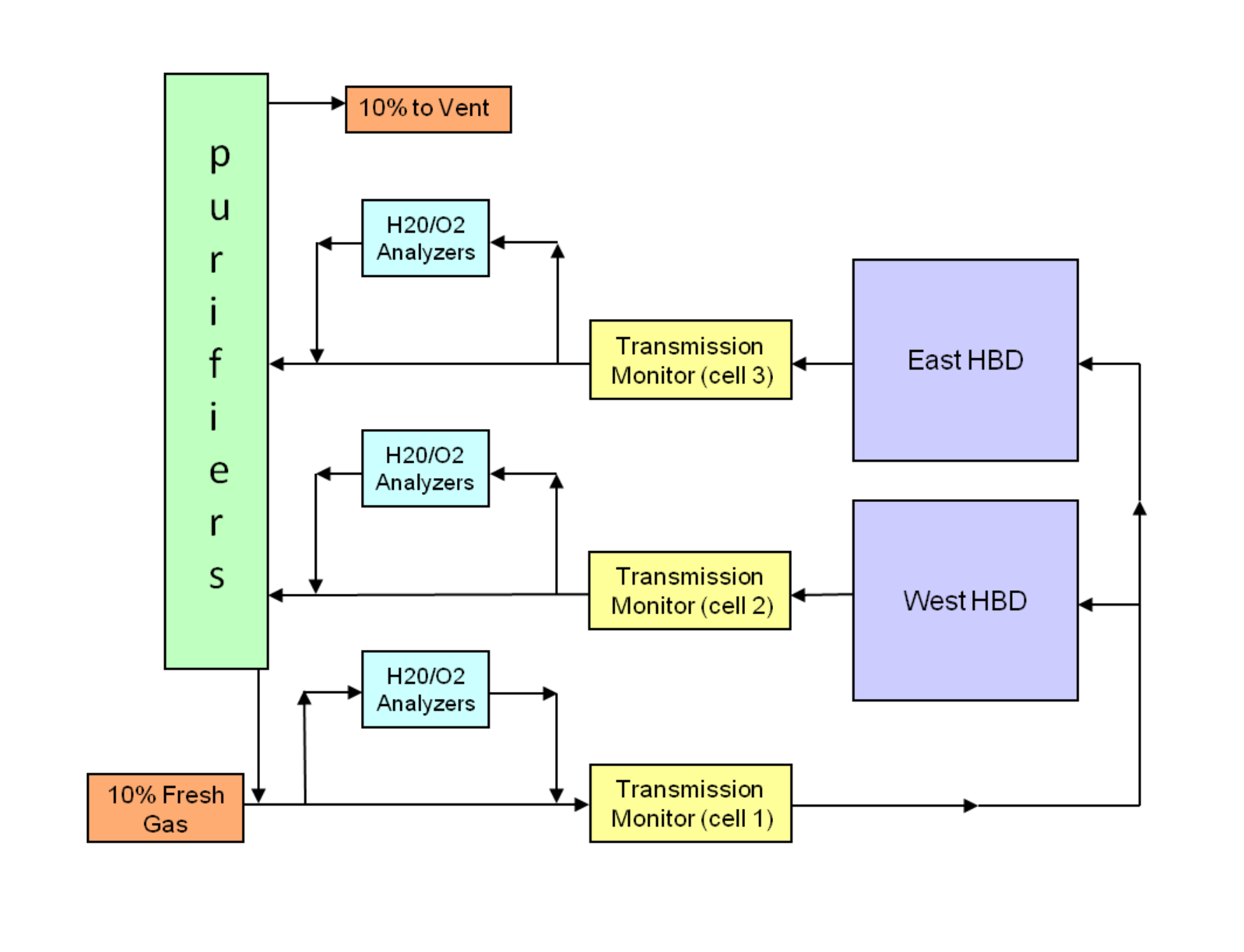}
  \caption{\label{fig:HBD-gas-sys}Recirculating gas system
  used to supply and monitor pure CF$_4$ gas to the HBD detector.}
 \end{center}
\end{figure}

These stringent demands require a sophisticated gas system that
 not only delivers high quality gas to the detector, but also
monitors the levels of the critical contaminants to high precision.
A recirculating gas system, shown in Fig.~\ref{fig:HBD-gas-sys}, is
implemented in order to
limit gas consumption and reduce operating costs \cite{ref:Stoll}.
The system delivers clean gas ($\le$ 1 ppm of H$_2$O and O$_2$)
to the detector. The output gas is repurified by a set of
filters and scrubbers before being reused. In addition, a percentage
of fresh gas (typically $\sim$ 10\%) is introduced during recirculation
in order to help maintain the required level of purity. The main
components of the system are housed in the PHENIX Gas Mixing House,
which is located approximately 100 meters from the detector, and
is connected to the detector though stainless steel pipes.
The input gas is common to both detector arms up to a gas monitoring
station  located just outside the PHENIX
intersection region, where it is split into separate inputs lines
to the East and West detectors. Each detector has separate gas returns to the
monitoring station. The flow rate was typically
$\sim$ 3.75 l/min to each detector, and the operating pressure was
$\sim$1.4 Torr above atmospheric pressure.

In order to measure and control the high gas purity required,
extensive monitoring of the oxygen and water levels is implemented
throughout the system. Oxygen and water sensors (GE Panamatrics O2X1
and Kahn Cermet II) are installed in the Gas Mixing House to
measure the levels of the input gas at the source and the return
gas before repurification. The Kahn Cermet II water sensors were
chosen for their compatibility with CF$_4$. Additional sensors are
also installed for the input and return gases in the monitoring
station, which measure the water and oxygen levels approximately
50 meters from the detector.

These sensors provide continuous information on the contaminant levels.
However, their sensitivity at the low ppm levels achieved during normal operation
is not sufficient to determine the true level of impurities in
the gas. Therefore, a completely independent monitoring system
was built to measure the actual UV transmission of both the input
gas and the output gas of each detector.

\begin{figure}[h]
 \begin{center}
  \includegraphics[width=10cm]{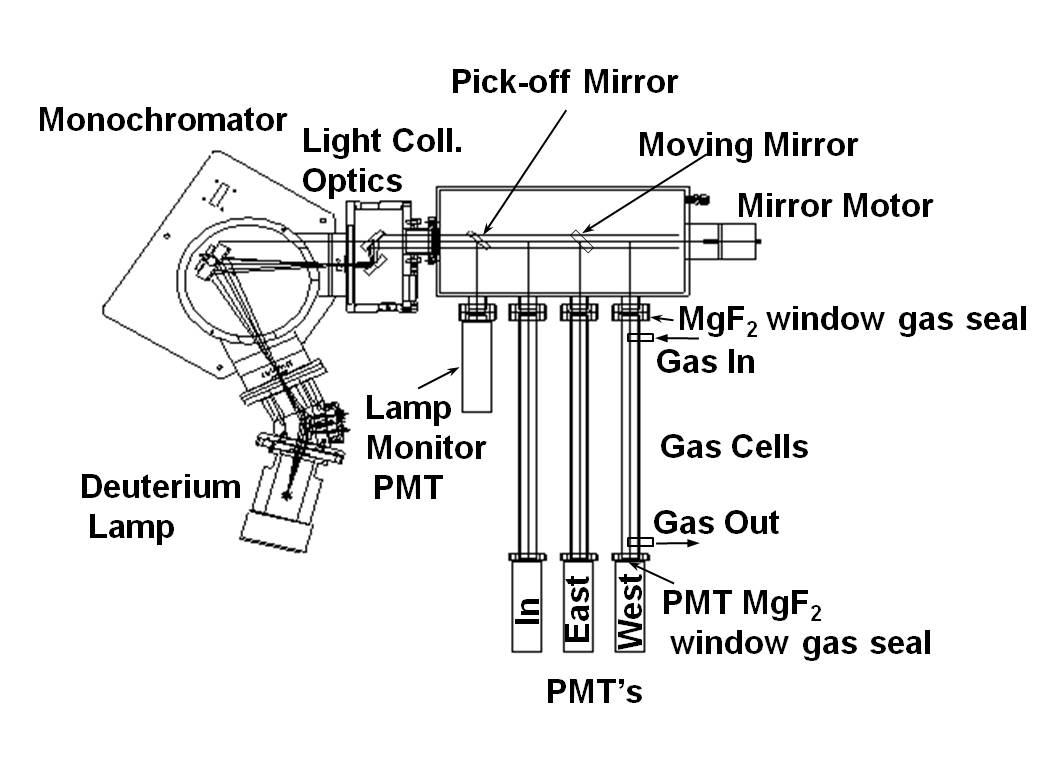}
  \vspace{-0.6cm}
  \caption{\label{fig:VUVspect_sketch_labels} The gas transmission monitor system.}
 \end{center}
\end{figure}

The gas transmission monitor is shown in
Fig.~\ref{fig:VUVspect_sketch_labels} and also further described
in \cite{ref:Stoll,ref:Azmoun}. It consists of a vacuum ultraviolet spectrometer
(McPherson Model 234/302) and a set of three transmission cells that
are used to measure the transmission of the common input gas and
separate output gases of each detector. A scanning monochromator
combined with a translatable mirror delivers a collimated beam
of monochromatic light from a deuterium lamp down one of three
51~cm long transmission cells equipped with a CsI photocathode
phototube (Hamamatsu R6835) on the end. One transmission cell is
used for measuring the input gas and two for the output gas. A
portion of the beam is also measured by a separate phototube in
order to monitor the lamp intensity. The phototubes are operated
in a photodiode mode and their photocathode currents are measured
using two Keithley 6487 picoammeters. The monochromator,
translatable mirror, and readout of the picoammeters are controlled
by a Labview program running on a PC in the monitoring station.

The entire spectrometer and each transmission cell can be pumped
and also purged with pure argon. After almost two years of operation
of the spectrometer under vacuum, it was found that there was a
significant loss in light intensity due to the buildup of deposits on the
beam optics. All of the beam optics were subsequently replaced and,
during the last year of operation, the main
compartment of the spectrometer containing the beam optics was
operated at atmospheric pressure under a flow of pure argon. However, the transmission cells
were still evacuated for reference scans in order to determine the
CF$_4$ transmission relative to vacuum.

The gas transmission is defined as the double ratio of the currents, I, in the cell and in the monitor PMTs. :

$T = (I_{CF_4}(Cell)/I_{vac}(Cell))/(I_{CF_4}(Mon)/I_{vac}(Mon)) $

\begin{figure}[h!]
 \begin{center}
 \vspace{-0.3cm}
 \includegraphics[width=6.5cm]{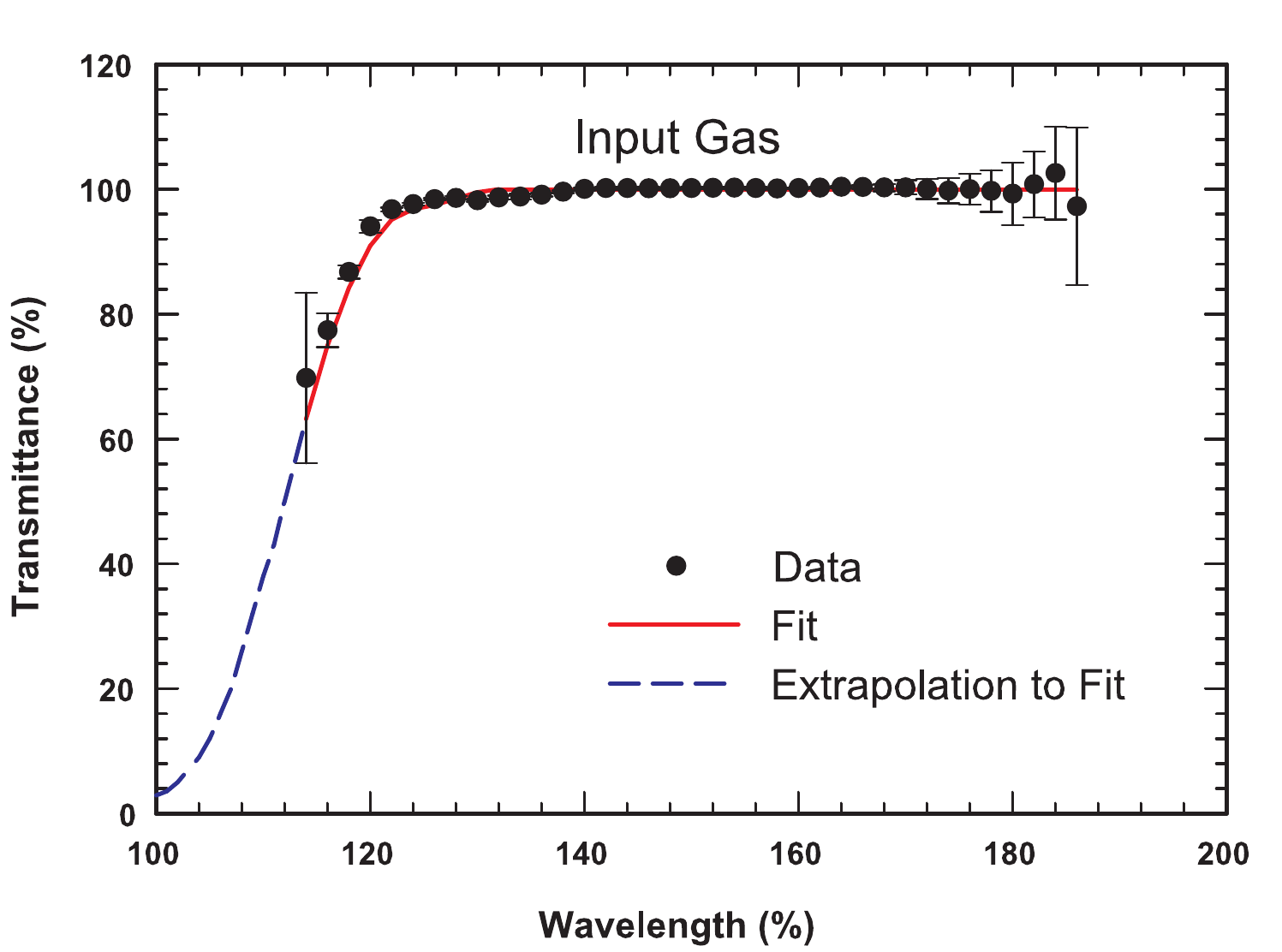}
 \end{center}
 \vspace{0.3cm}
 \begin{center}
 \includegraphics[width=6.5cm]{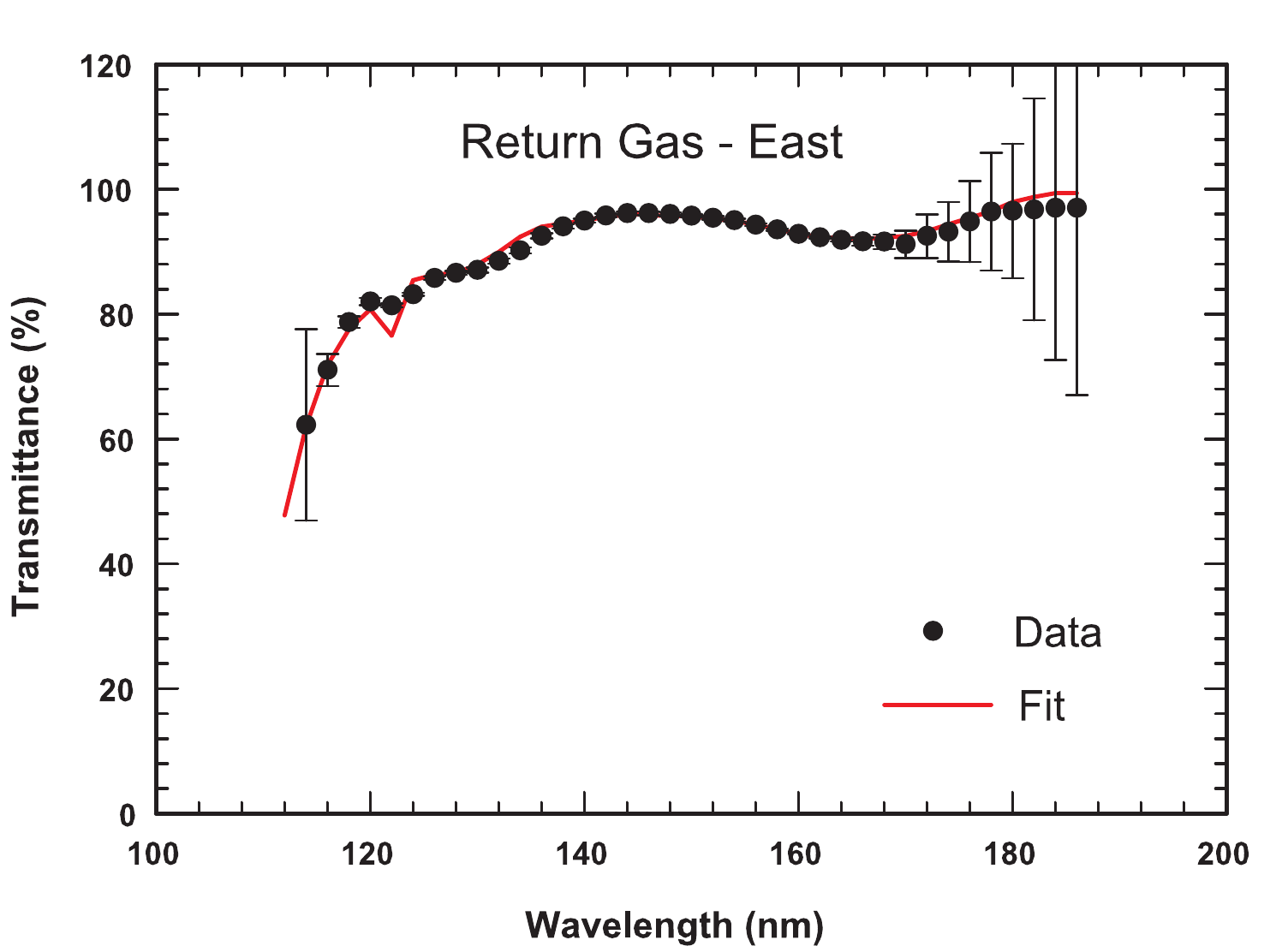}
 \includegraphics[width=6.5cm]{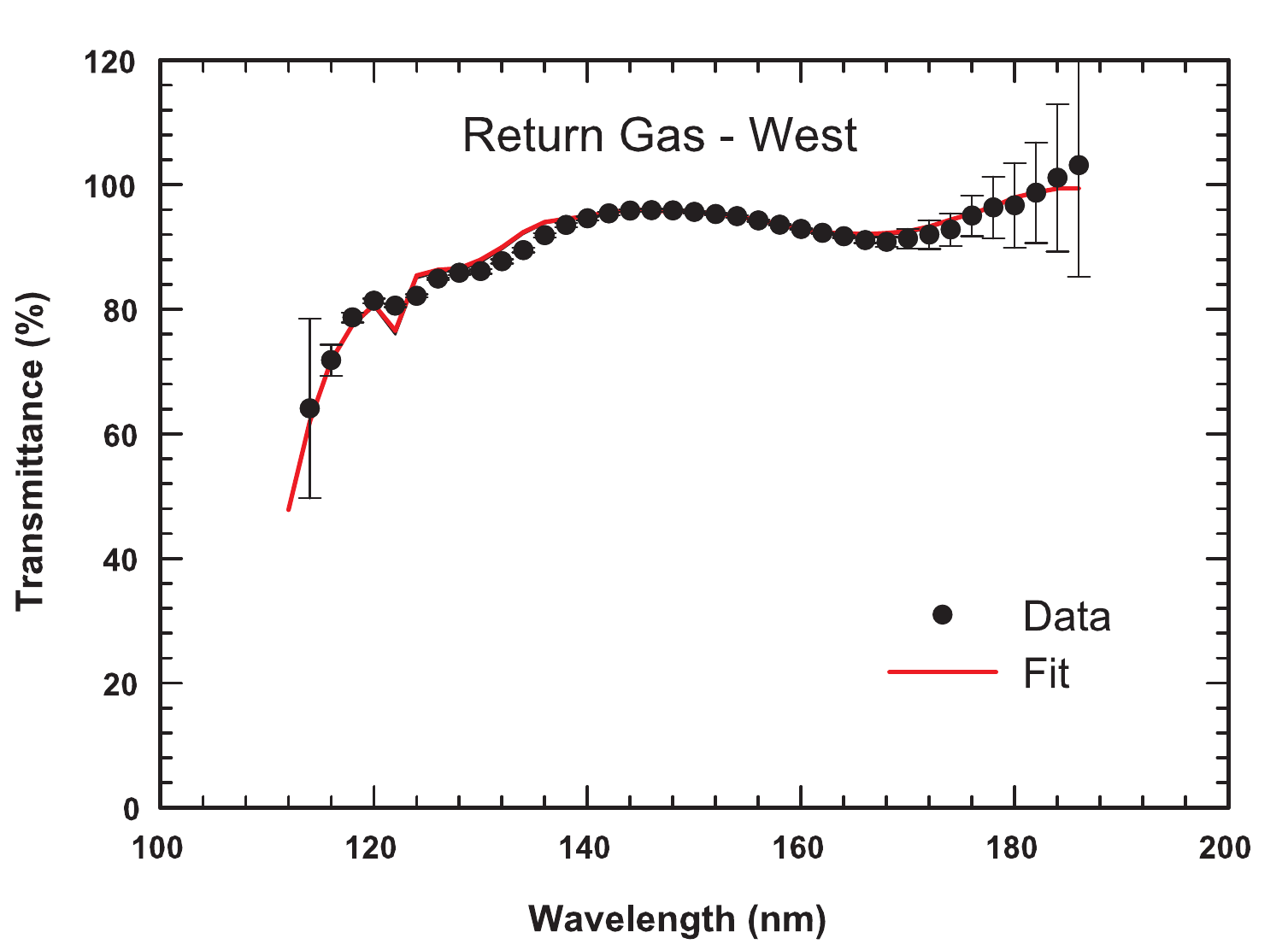}
 \caption{\label{fig:Gas_Transmission_Spectra}UV transmission spectra for
  (top panel) the common input gas, (left panel) the output gas from the East detector and (right panel) the output gas from the West detector.}
 \end{center}
\end{figure}

\begin{figure}[h]
 \begin{center}
  \includegraphics[width=10cm]{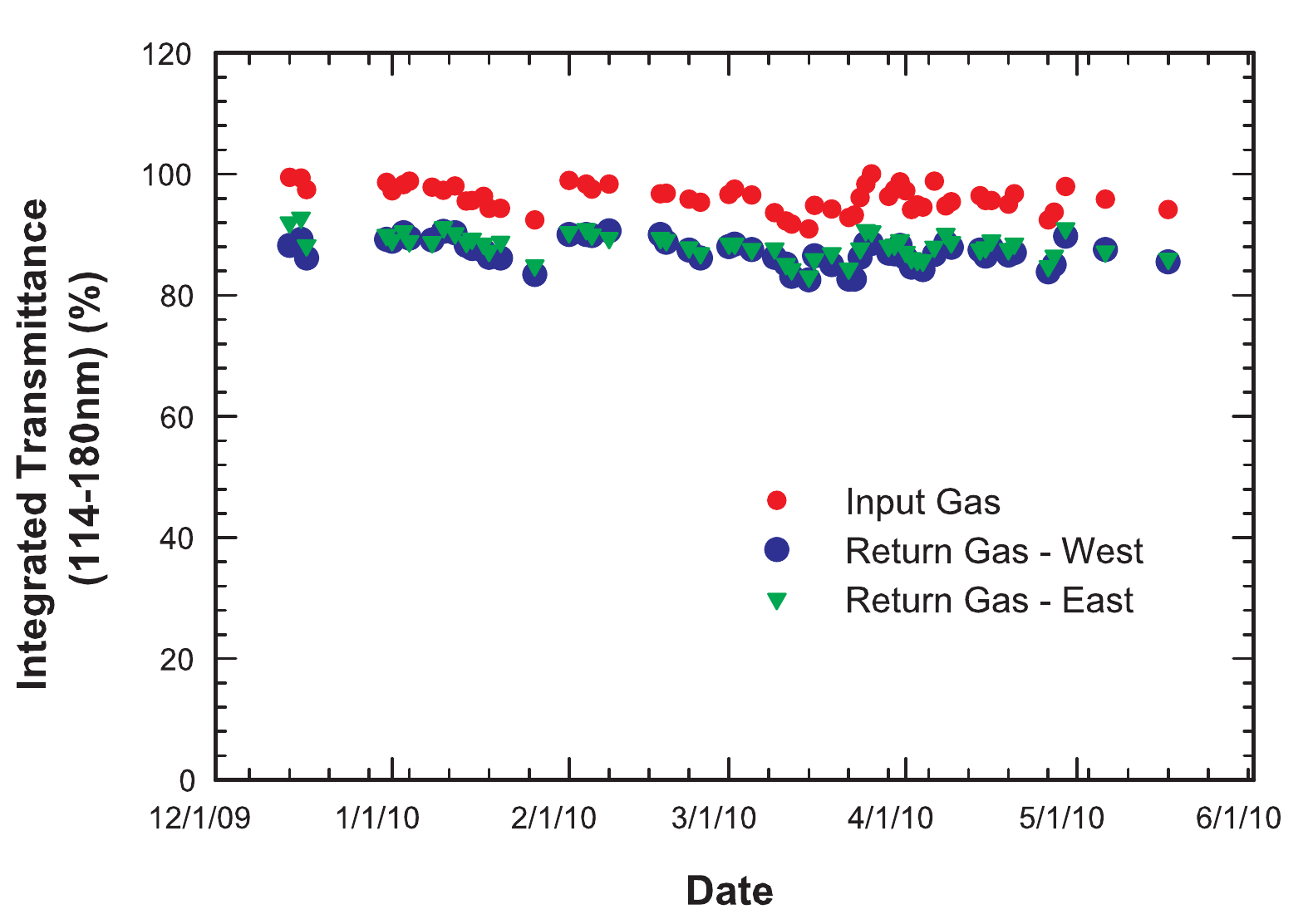}
  \caption{\label{fig:Integrated_Gas_Transmission}Integrated gas
  transmission over the range from 114-180 nm for the input gas and
  east and west output gas for a five month time period during
  Run 10 at RHIC.}
 \end{center}
\end{figure}

The gas transmittance was measured continuously throughout each
run, with scans being done sometimes several times a day
to once a week or more as necessary. A typical set of input and
output transmission spectra is shown in
Fig.~\ref{fig:Gas_Transmission_Spectra}. Using the measured transmission values and the known absorption
cross sections for water and oxygen (see Fig.~\ref{fig:H2O_O2Xsect_RelCherenkovYield}),
a fit is performed to
determine the true impurity levels in the gas.
The spectra are fit over a wavelength range from 111-200 nm including a component
due to the intrinsic absorption in CF$_4$ and components due to water and and oxygen
impurities. This method not
only allows a more precise determination of the actual impurity levels, but gives a
direct indication of the total integrated absorbance in the gas.
The input spectrum shows virtually no additional absorbance other than the intrinsic absorbance
of the gas. The two output spectra exhibit a shape that is dominated by water absorption with a
small component of oxygen (see again Fig.~\ref{fig:H2O_O2Xsect_RelCherenkovYield}).
As determined by these fits, the impurity levels of the input gas were typically less than
2 ppm of both water and oxygen,
and the output gas typically showed 20-30 ppm of water and
$\sim$ 2-3 ppm of oxygen. Fig.~\ref{fig:Integrated_Gas_Transmission}
shows the integrated transmittance over the range from 114-180 nm for
a roughly five month time period during Run 10 at RHIC, and shows
that the input gas was in the range from 90-100\%, and the output
gas was in the range from 80-90\%. This was sufficient to keep the
transmission loss of photoelectrons to $\lesssim$ 5\% throughout the
entire run.

\section{High voltage system}
\label{sec:HV}   

\begin{figure}[h]
 \begin{center}
  \includegraphics[width=10cm]{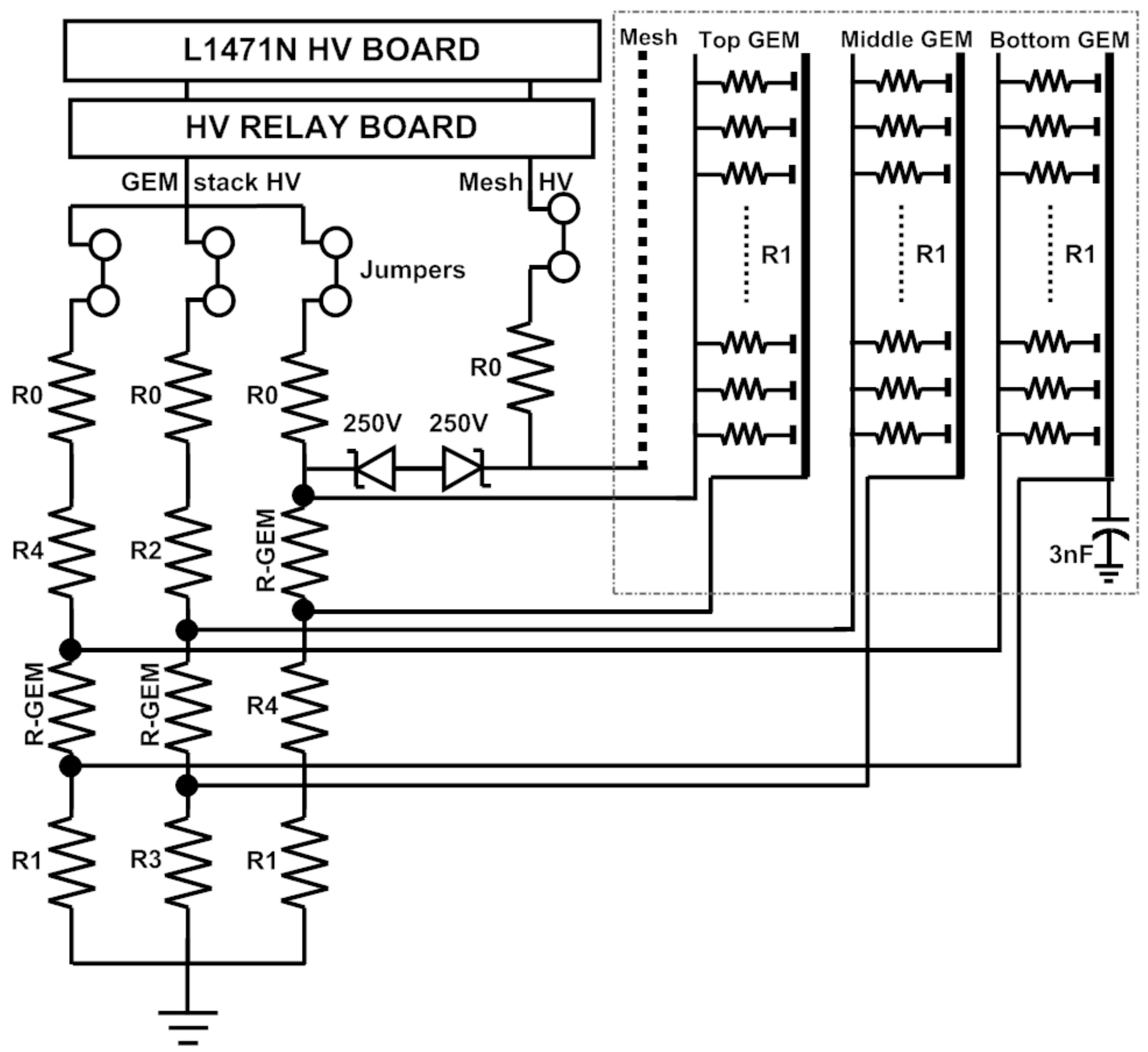}
  \caption{\label{fig:HBD_Voltage_Divider}Voltage divider and power
  supply configuration used to supply high voltage  to the GEM
  stack and mesh of each HBD module. Values of the resistors are:
  R$_0$ = 2~M$\Omega$, R$_1$ = 20~M$\Omega$, R$_2$ = 26~M$\Omega$,
  R$_3$ = 46~M$\Omega$, R$_4$ = 52~M$\Omega$ and R-GEM~=~10~M$\Omega$.
  The 20~M$\Omega$
  resistors on the GEM modules are mounted on the GEM foils inside
  the detector.}
 \end{center}
\end{figure}

The high voltage supply for the HBD is based on the LeCroy 1450
high voltage system, but modified with a number of additional
features to improve calibration, enhance voltage and current monitoring,
and provide additional trip protection capabilities. The main high
voltage supply is a LeCroy 1458 mainframe equipped with a set of
six LeCroy 1471N high voltage modules capable of supplying up to 200~$\mu$A
at 6~kV. Two channels of a 1471N unit are used for each HBD module,
one to power the voltage divider for the GEMs and another to supply
high voltage to the mesh. The configuration of the voltage divider and
power supply to the GEM stack and the mesh is shown in
Fig.~\ref{fig:HBD_Voltage_Divider}. The 20 M$\Omega$ resistors in series with
each of the 28 segmented HV strips of the GEM are mounted on the GEM foils,
while all other resistors are outside the HBD vessel. The internal resistors
are used to prevent a short in one strip from affecting the high voltage
to other strips in the same GEM. As described below, if a short
occurs in one strip, the external divider resistor can be easily
changed to compensate for the short, keeping the total effective
resistance across the GEM constant.

Each of the three GEMs in a stack is powered by its own separate divider.
The values for the divider resistors are chosen to provide equal voltages
across each of the GEMs, relatively high fields in  the two transfer
gaps, and an even larger field in the induction gap,
in order to improve the electron collection efficiency in the three gaps
\cite{ref:hbd1}. This effectively results in higher
overall gain of the triple GEM stack without increasing the voltage
across the GEMs. The total resistance of each leg of the divider is
84~M$\Omega$, giving a total resistance of 28~M$\Omega$ for the three
legs in parallel. In order to achieve a typical operating gain of
4000, the total applied voltage to the top of the divider chain is
$\sim$4000~V, resulting in a total current of $\sim$143~$\mu$A, and an
applied voltage across each of the GEMs of $\sim$476~V, which is
comfortably below the breakdown voltage for the GEMs in CF$_4$ \cite{ref:hbd1}.

A number of additional features are built into the high voltage
system  to prevent excessive voltage from being
applied to the GEMs or between the top GEM and the mesh. A pair
of back-to-back Zener diodes installed between the mesh and
the upper electrode of the top GEM limit the voltage to
less than 250 V to ensure that no voltage capable of causing
breakdown in this gap could be be applied. In addition, the
1471Ns HV modules were modified to incorporate a trip detection
circuit that is able to recognize a trip in one channel and initiate
a trip in another. This is used to trip the mesh of a given module
if the GEM tripped or vice versa. The output of this circuit is
connected to a custom designed ``Relay Board'' that utilizes high
voltage relays to disconnect a tripped channel from the high voltage
power supply and quickly discharge the stored energy in both the GEM
stack and mesh.

\begin{figure}[h]
 \begin{center}
\vspace{-0.3cm}
  \includegraphics[width=10cm]{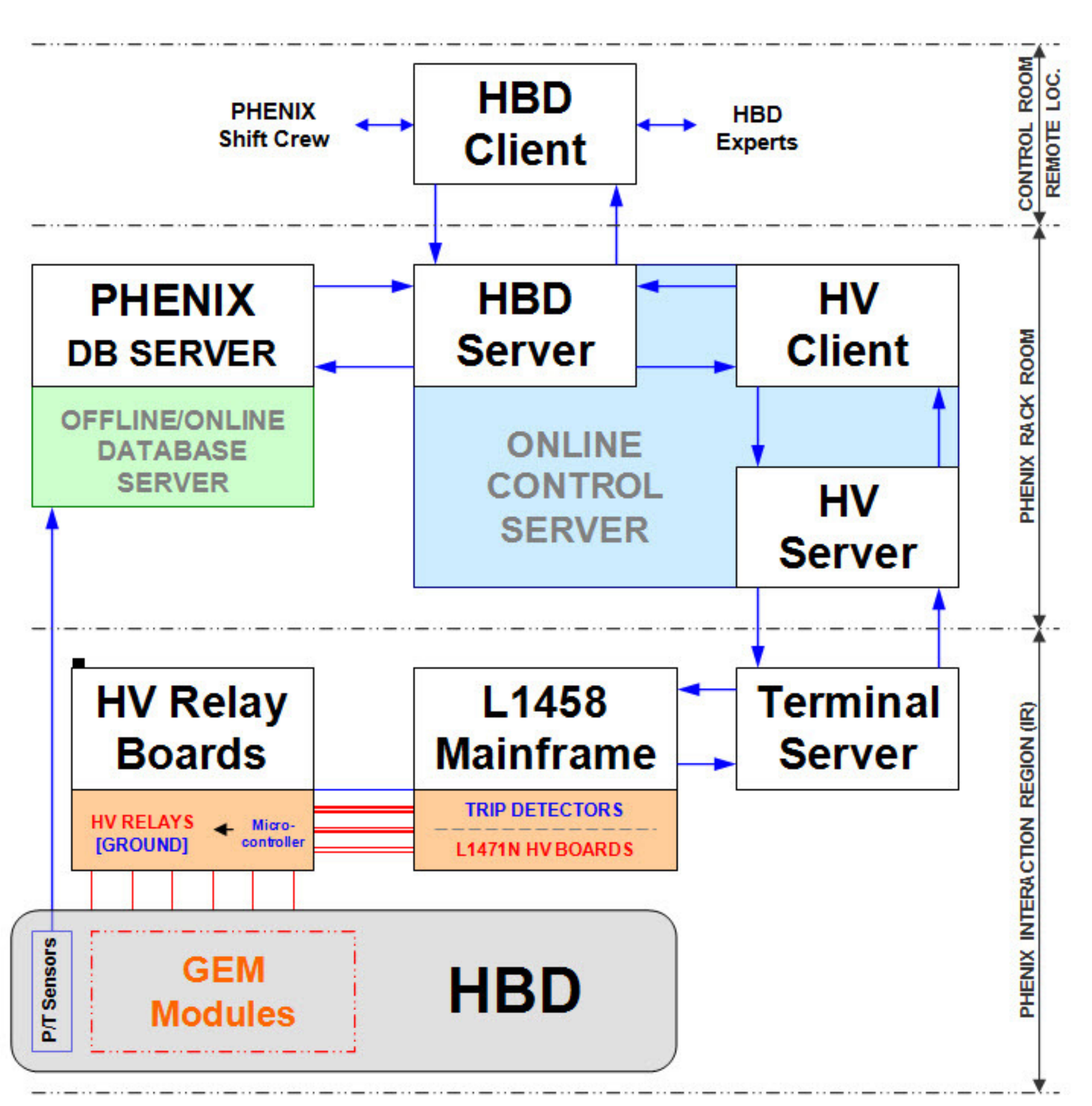}
  \caption{\label{fig:HBD_HV_Control_System}HBD High Voltage Control
  System with its Client/Server configuration. The HBD HV server
  runs on one of the online computers in the PHENIX control room,
  while the client can be either run locally or remotely.}
 \end{center}
\end{figure}

The high voltage is controlled by a High Voltage Control and
Monitoring System (HVC)~\cite{ref:Proissl} that is designed to
provide precision control of the high voltage to the detector,
allowing programmable ramps, setting of standby and operational
voltages for different gain settings, modification of the high
voltage to compensate for pressure and temperature (P/T) changes,
monitoring and recording of all voltages and currents, including
storage of all values into a data base, and several modes of
trip detection and recovery. The system is based on modern
Optimal Control Theory and is implemented in a client-server
environment using mainly Java. The overall system is shown in
Fig.~\ref{fig:HBD_HV_Control_System}. The HBD HV Client/Server
interacts through the main PHENIX HV Client/Server to communicate
with the LeCroy mainframe and perform the basic HV control functions,
and also provides a GUI to interact with the operators in the PHENIX
Control room, or with other operators connected remotely.

In addition to the factory calibration, an in situ calibration
of each 1471 module was done using the actual divider chain of each
GEM. This allows precise monitoring of the
current in each divider and enables detection of any additional
current drawn by the GEM to a level of $\sim$100~nA, which would be
an indication of a short, or partial short, in any of the GEM modules.
If such an increase in current is detected, the HVC  initiates
a ``virtual trip'' and notifies the operator that intervention is
required. This feature is in addition to the normal fast trip
detection provided by the 1471 which causes a trip if the peak
current exceeds a certain value for a short time, as in the case
of a discharge. During normal operation, the fast trip thresholds are
set at 20~$\mu$A, and the slow or virtual trip thresholds are set to
$\sim 1.5\%$ or $\sim$2~$\mu$A, above the standing current value.

It should be noted that if a partial short does occur, it does not
necessarily trip the power supply, but results in a small
increase in current drawn by the divider chain due to the appearance
of an additional resistance in parallel with the nominal 10~M$\Omega$
resistor across each of the GEMs. However, the internal resistors in
series with each of the segmented HV strips limit this parallel
resistance to 20~M$\Omega$, even in the case of a dead short. In
either case, a partial or complete short lowers the resistance,
and hence the voltage across the affected GEM, thus lowering its gain.
These types of shorts can be compensated for by measuring the total
internal resistance of the individual GEM (20~M$\Omega$ plus the
resistance of the short) and changing the external 10~M$\Omega$ resistor
to an appropriate value to restore the total resistance to
10~M$\Omega$. Up to two complete shorts can be compensated for in this
way without having to change any other resistors in the divider chain.
In practice, during the 2010 RHIC run, only two modules showed two
complete shorts and two showed partial shorts, all of which were
compensated for by simply changing a single external resistor.

\section{HBD operation and monitoring}
\label{sec:operation-monitoring}

A number of parameters need to be carefully adjusted and monitored
to ensure good performance of the detector over a run which is
typically 4-6 months long. These include setting the HV for each
detector module to reach the desired operating gain, optimizing the
reverse bias field in the gap between the mesh and the top GEM
to achieve maximum hadron rejection while keeping maximum
photoelectron collection efficiency, monitoring the gas gain variations due
to pressure and temperature changes, and monitoring the gas quality and photocathode
sensitivity. The monitoring of the gas quality was already discussed in Section~\ref{sec:Gas}.
The other parameters are discussed in the present section.

\subsection{Gain determination}
\label{subsec:gain}
The excellent noise performance of the readout electronics
(see Section~\ref{subsec:Noise})
allows the detector to be operated at a relatively low gain. During the RHIC runs of 2009 an 2010 the HBD was operated at a gain of $\sim$4,000.
This gain is achieved with a voltage across the GEM
$\Delta V_{GEM} \sim$470 V, which is
comfortably below the breakdown voltage
of the GEMs (typically around 550~V in CF$_4$ \cite{ref:hbd1}).

\begin{figure}[ht]
 \begin{center}
    \includegraphics[width=10cm]{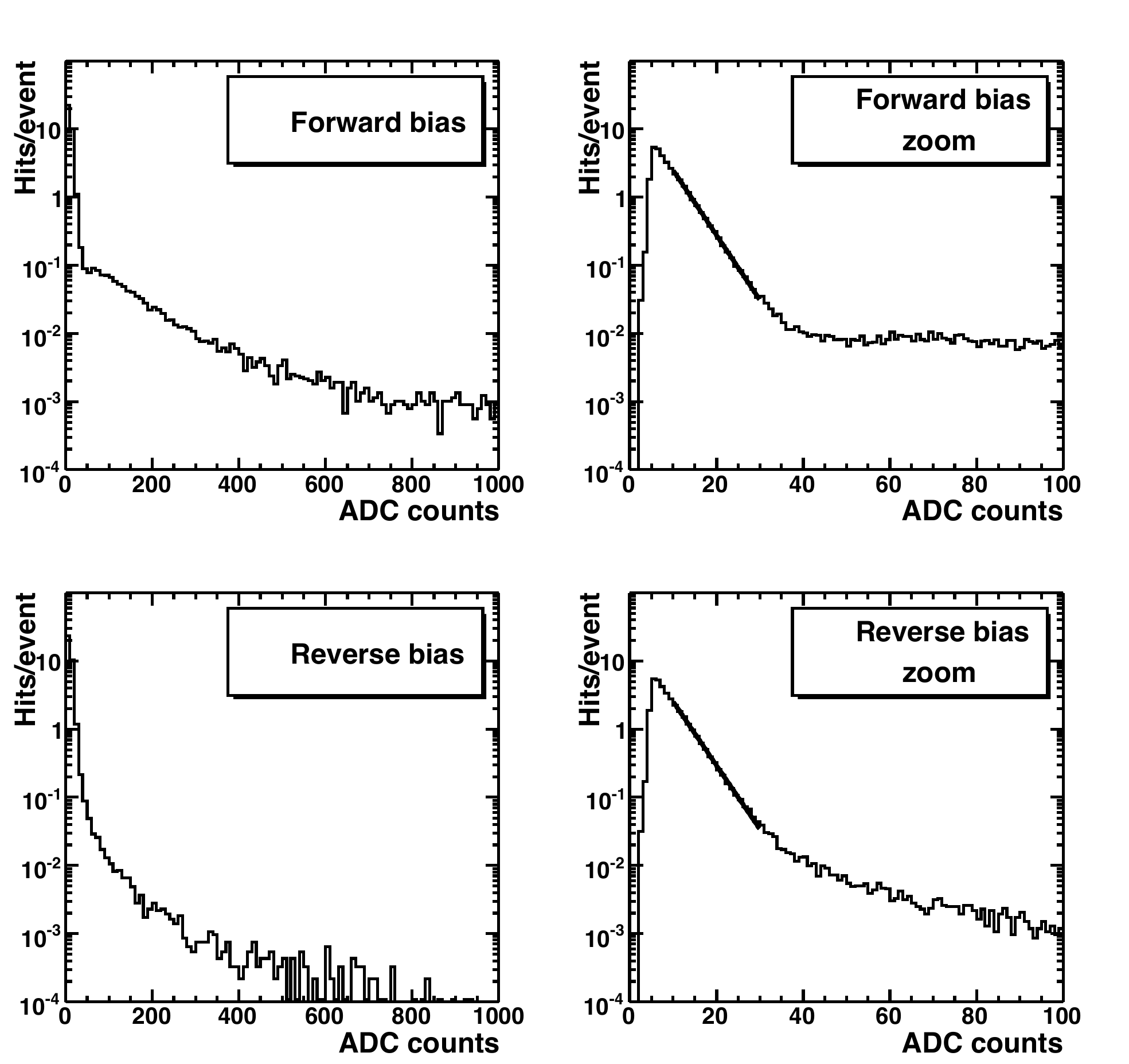}
    \caption{Pulse height distribution in one detector module in
    FB (upper panels) and RB (lower panels) modes. The ordinate is
    normalized to represent the number of hits per event.}
  \label{fig:scintillation}
 \end{center}
 \end{figure}

The gain of each detector module is conveniently and precisely determined by
exploiting the scintillation light produced by charged particles
traversing the CF$_4$ radiator. The scintillation signal is easily
identified in the low amplitude part of the pulse height
distribution as illustrated in Fig.~\ref{fig:scintillation}. In
the FB mode (upper panels) the spectrum has two clear components:
a steep exponential distribution at very low amplitudes attributed to
scintillation photons and a longer tail at higher
amplitudes which is due to ionization of the gas in the drift gap.
When the detector is operated in RB mode (lower panels), the latter is
largely suppressed as the ionization charges get repelled to the mesh,
whereas the exponential part due to scintillation remains unaffected.
The scintillation signal has another characteristic feature: it produces
single pad hits which are not associated to any charged track identified
in the outer PHENIX detectors.

\begin{figure}[ht]
 \begin{center}
    \includegraphics[width=65mm]{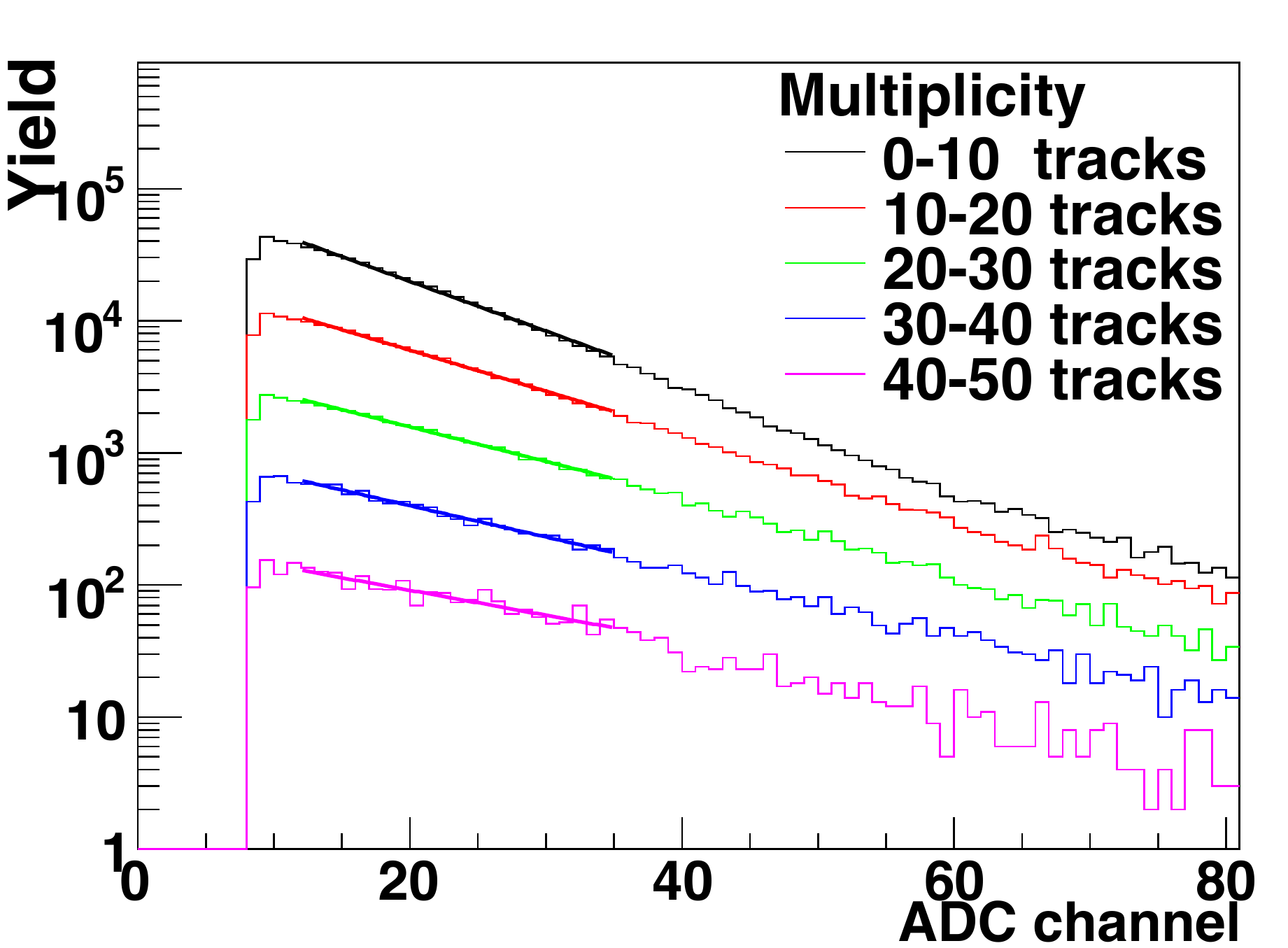}
 \end{center}
 \begin{center}
    \includegraphics[width=65mm]{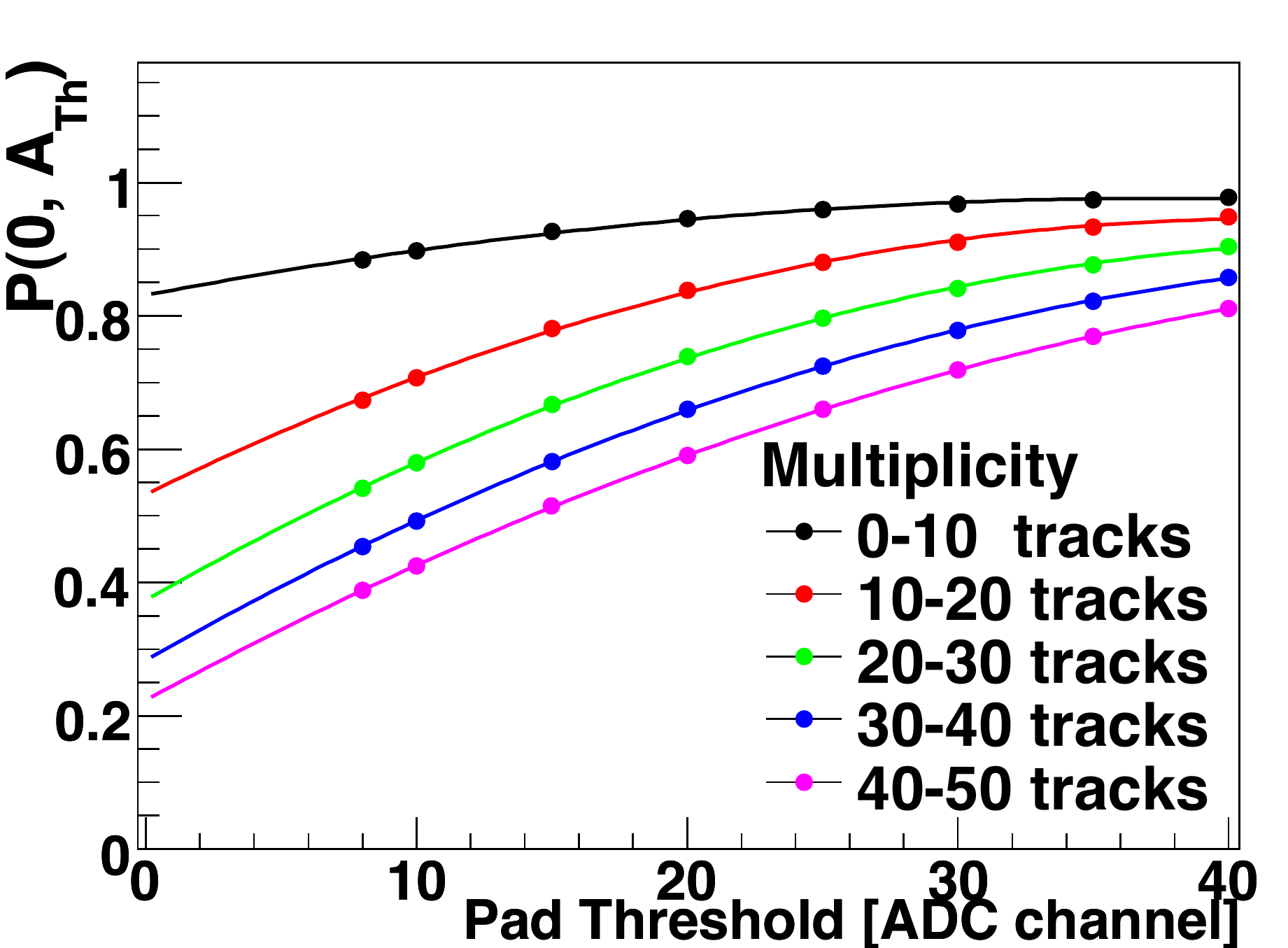}
    \includegraphics[width=65mm]{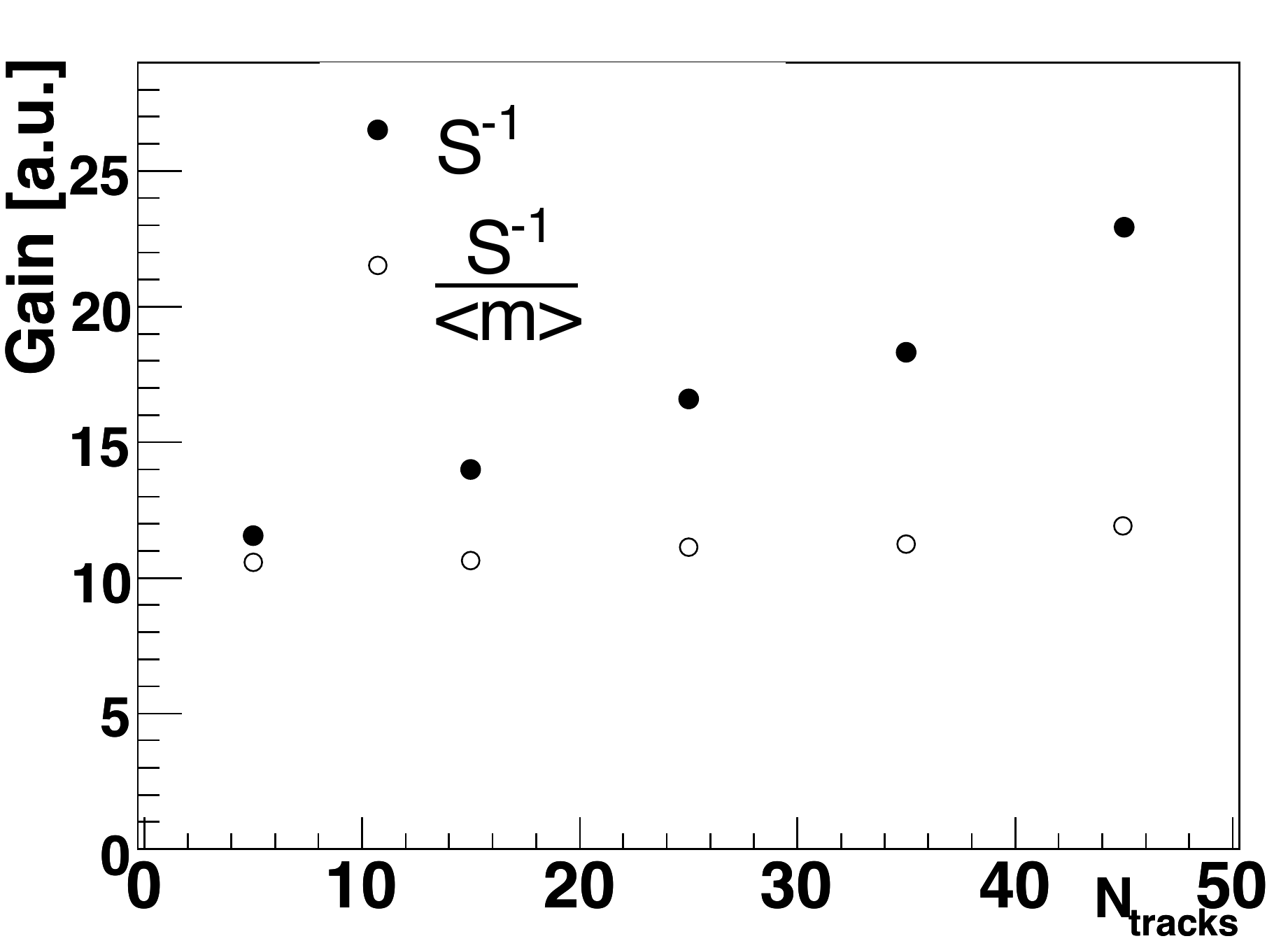}
    \caption{Top panel: Pulse height distribution in one detector
    module measured in Au+Au collisions for different event
    centralities  characterized by the number of hadron tracks
    reconstructed in one central arm of the PHENIX detector. The lines
    represent fits with an exponential function. Bottom left panel:
    Probability to have no pad fired with an amplitude larger than
    A$_{th}$ vs. the amplitude pad threshold. The lines represent the
    fit with an arbitrary function. Bottom right panel: Inverse slopes
    (solid circles) derived from the fits in the left panel and detector
    gain (open circles) obtained using Eq.~\ref{eq:avrg-m}.}
  \label{fig:gain}
 \end{center}
 \end{figure}

The gain G of the detector is obtained from:
\beq
   G =  \frac{S^{-1}} {<m>}
\eeq
where S is the slope of the exponential shape at low amplitudes
and $<m>$ is the average number of scintillation photons in a fired
pad. In p+p collisions $<m>$ is very close to 1 and the gain is readily given by
the inverse slope of the exponential distribution: G $\sim$ S$^{-1}$. In Au+Au
collisions however, the inverse slope increases with the number of charged particles
traversing the detector as shown in the top and bottom right panels of Fig.~\ref{fig:gain}.
Due to the large scintillation yield of CF$_4$, as the number of tracks
increases, the probability of scintillation pile up increases and the
primary charge in the scintillation signal $<m>$ corresponds on
the average to more than one photoelectron. Assuming that the number n
of scintillation photons per pad follows
a Poisson distribution P(n) with an average $\mu$, then $<m>$ is given by:
\beq
    <m> = \frac{\sum_{n \geq 1}nP(n)} {\sum_{n \geq 1}P(n)} = \frac{\mu} {1 - P(0)}
\eeq
where P(0) is the probability to have no hit in a pad: P(0) = e$^{-\mu}$. Therefore:
\beq
    <m> = \frac {\mu}{1 - e^{-\mu}} \simeq 1 + \mu/2 = 1 - ln[P(0)]/2
\label{eq:avrg-m}
\eeq
P(0) is not directly accessible since the data are always collected
with an amplitude larger than some threshold value A$_{th}$ and
what is really measured is P(0, A$_{th}$), i.e. the probability of no hit
with an amplitude larger than A$_{th}$. P(0) is determined
by fitting the variation of P(0, A$_{th}$) vs A$_{th}$ with some
arbitrary function and extrapolating it to A$_{th}$= 0 as illustrated
in the bottom left panel of Fig.~\ref{fig:gain}. The gain derived using
Eq.~\ref{eq:avrg-m} is independent of the collision centrality as
shown in the bottom right panel of Fig.~\ref{fig:gain}.

The online gain was determined by the inverse slope of the exponent in p+p collisions.
The same procedure was also used for the online determination of the gain in Au+Au
collisions but selecting only very peripheral events. In both cases,
the real gain calculated offline with the extrapolation procedure
outlined above is only a few percent lower.

    \subsection{Gain equilibration}
The detector gain is not uniform over its entire active area.
There are two types of gain variations. There are spatial gain
variations across each GEM's area that arise from small changes in
the size of the holes and from the mechanical tolerances of the
various gaps. The
second type of gain variations is global gain variations as a
function of time which are due to changes of the atmospheric pressure
and the temperature.

\begin{figure}[h!]
   \begin{center}
   \vspace{0.2cm}
   \includegraphics[keepaspectratio=true, width=8.5cm]{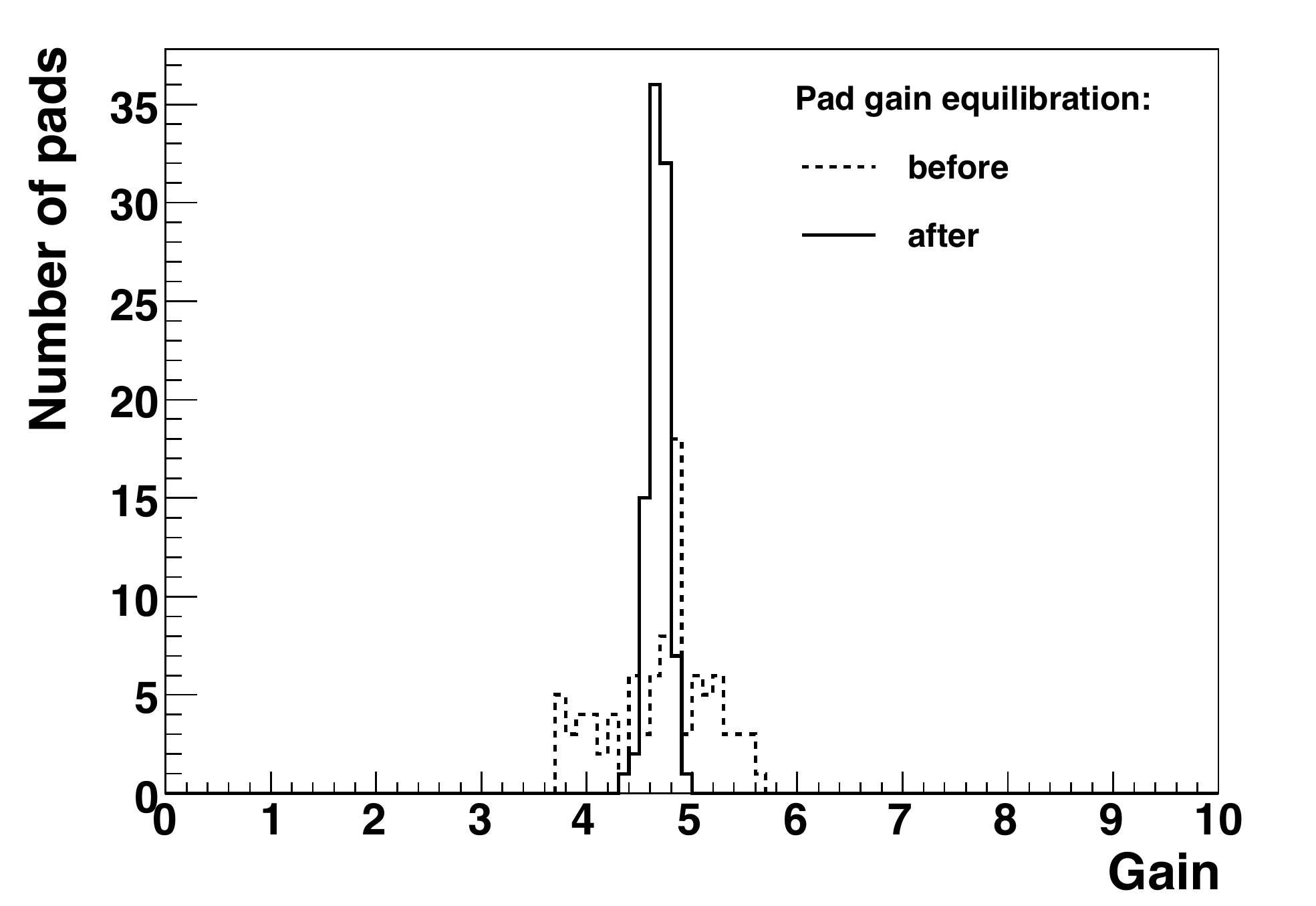}
   \caption{Gain distribution of all pads in one module before and after equilibration.
   The correction factors $<$G$>$/G$_i$ were derived from a different run.}
    \label{fig:gain_equilibration}
\end{center}
\end{figure}

Gain uniformity, both in space and time, is essential for the HBD
performance since its analog response is used to distinguish single
from double electron hits. In order to correct for the spatial gain
variations we use a gain equilibration procedure which brings all
the pads in a given module to the average gain $<$G$>$ in that module.
Using a large statistics run, the average gain $<$G$>$ is calculated
from the gain G$_i$ of each pad {\it i} determined using the procedure
outlined in the previous section. The signal a$_i$ in a given pad
is then corrected according to the expression:
\beq
             a_i \rightarrow a_i \frac{<G>}{G_i}
\eeq
An example showing the spread of gain values across the pads of one
module, before and after pad gain equilibration, is shown in
Fig.~\ref{fig:gain_equilibration}.
One can see that before equilibration
the spread of gains is quite large with an rms of 0.45 which gets reduced to 0.10
after equilibration.

    \subsection{Gain variations with P/T}
    \label{subsec:gain-variations}
\begin{figure}[ht]
   \centering
\includegraphics[keepaspectratio=true, width=8.5cm]{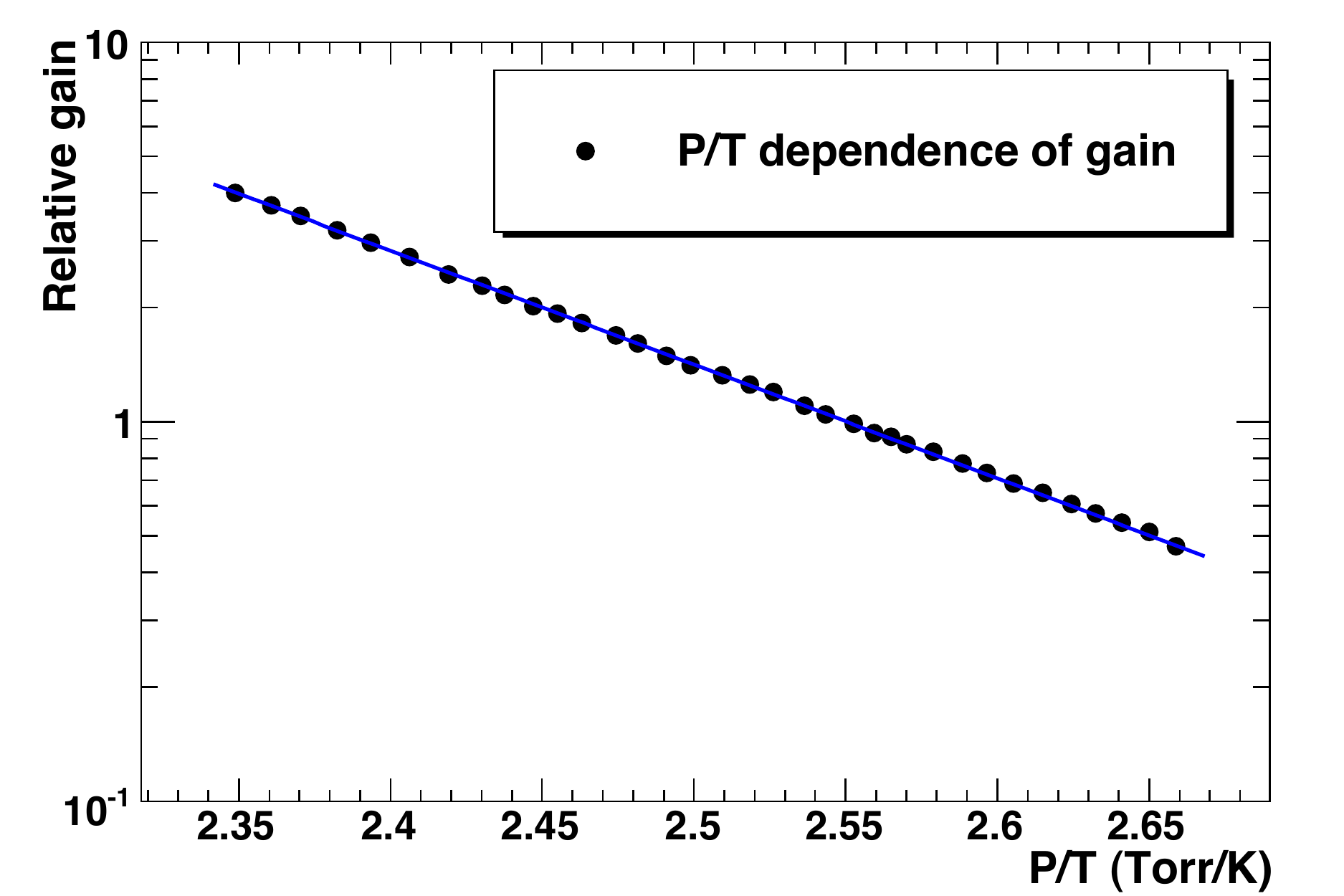}
   \caption{Relative gain variations in CF$_4$ due to P/T changes. The gain
   is normalized to 1 at P/T = 2.55 (Torr/K). The line represents a fit of the data points with
   an exponential function.}
    \label{fig:p-over-t}
\end{figure}

The gas gain in CF$_4$ is very sensitive to variations of the gas
density i.e. changes in P/T. The temperature of the detector is
maintained fairly constant at 21$^o$C by the temperature control in the
experimental hall. However, the gas pressure varies since the detector
is kept at a constant overpressure of 1.4 Torr above the atmospheric
pressure and the atmospheric pressure varies greatly according to weather
conditions. The detector is quite sensitive to these changes as can be seen
in Fig.~\ref{fig:p-over-t}. A change of P/T by $\sim$6\% induces a factor of 2 change
in the gas gain. The solid line represents a fit of the data with
an exponential function demonstrating that the gain varies exponentially with
P/T.

\begin{figure}[ht]
 \begin{center}
   \includegraphics[height=50mm,width=65mm]{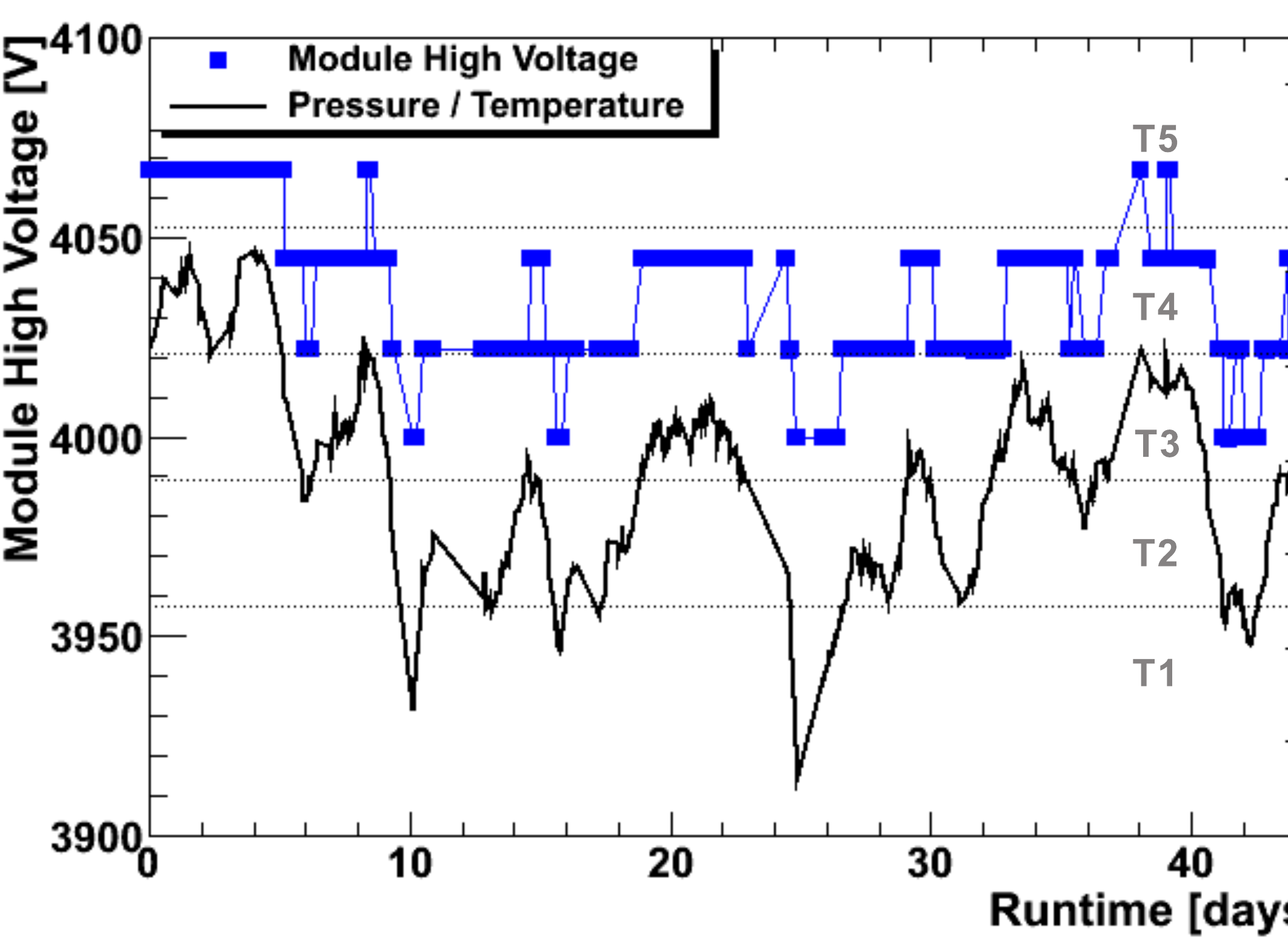}
   \includegraphics[height=50mm,width=65mm]{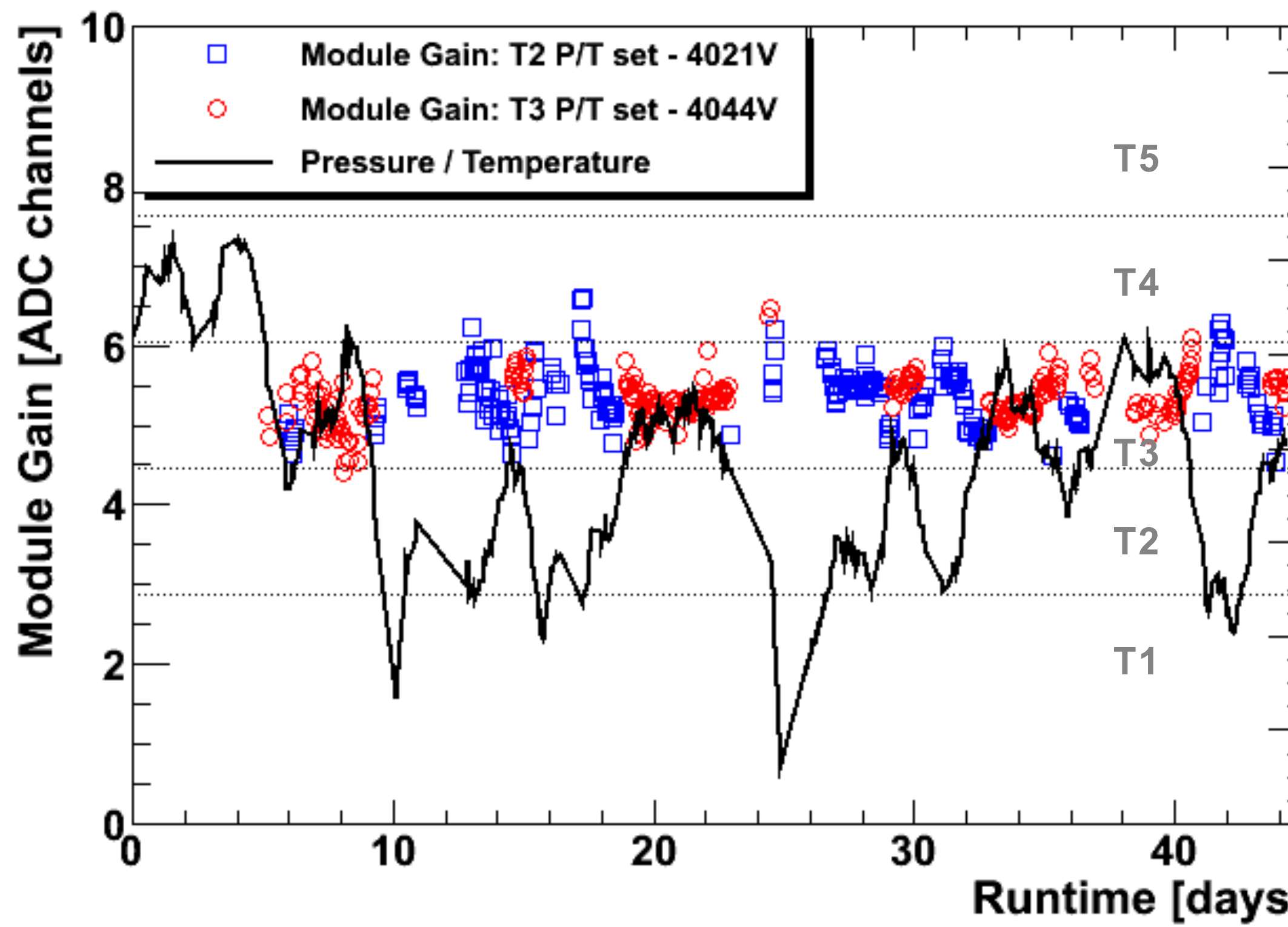}
  \caption{\label{fig:run10_200gev_es5_gain}Left panel: HV applied to one HBD module and P/T
  values during a period of 45 d  of the 200 GeV section of the 2010 RHIC run. Five distinct windows (T1 to T5) in
   P/T  are defined  and the high voltage control program applies a custom set of voltages to the detector
   for each of these windows. Right panel: Measured gain of the same module during
   the same period of time.}
 \end{center}
\end{figure}

To avoid these large excursions of the gain, we defined 5 pre-determined
P/T windows such that over each of them the gain varies by not more than 20\%.
We compensated the gain variations by automatically varying the
operating HV whenever the P/T values crossed the window boundaries.
The left panel of Fig.~\ref{fig:run10_200gev_es5_gain} shows the
applied voltage to a given detector module and the P/T values over
an extended period of 45 days during Run-10. The measured gain of the
same module during the same time period is shown in the right panel,
demonstrating that the gain is kept constant within $\pm$10\%.

\subsection{Reverse bias }
    \label{subsec:reverse-bias}
\begin{figure}[ht]
  \begin{center}
    \includegraphics[keepaspectratio=true, width=8.5cm]{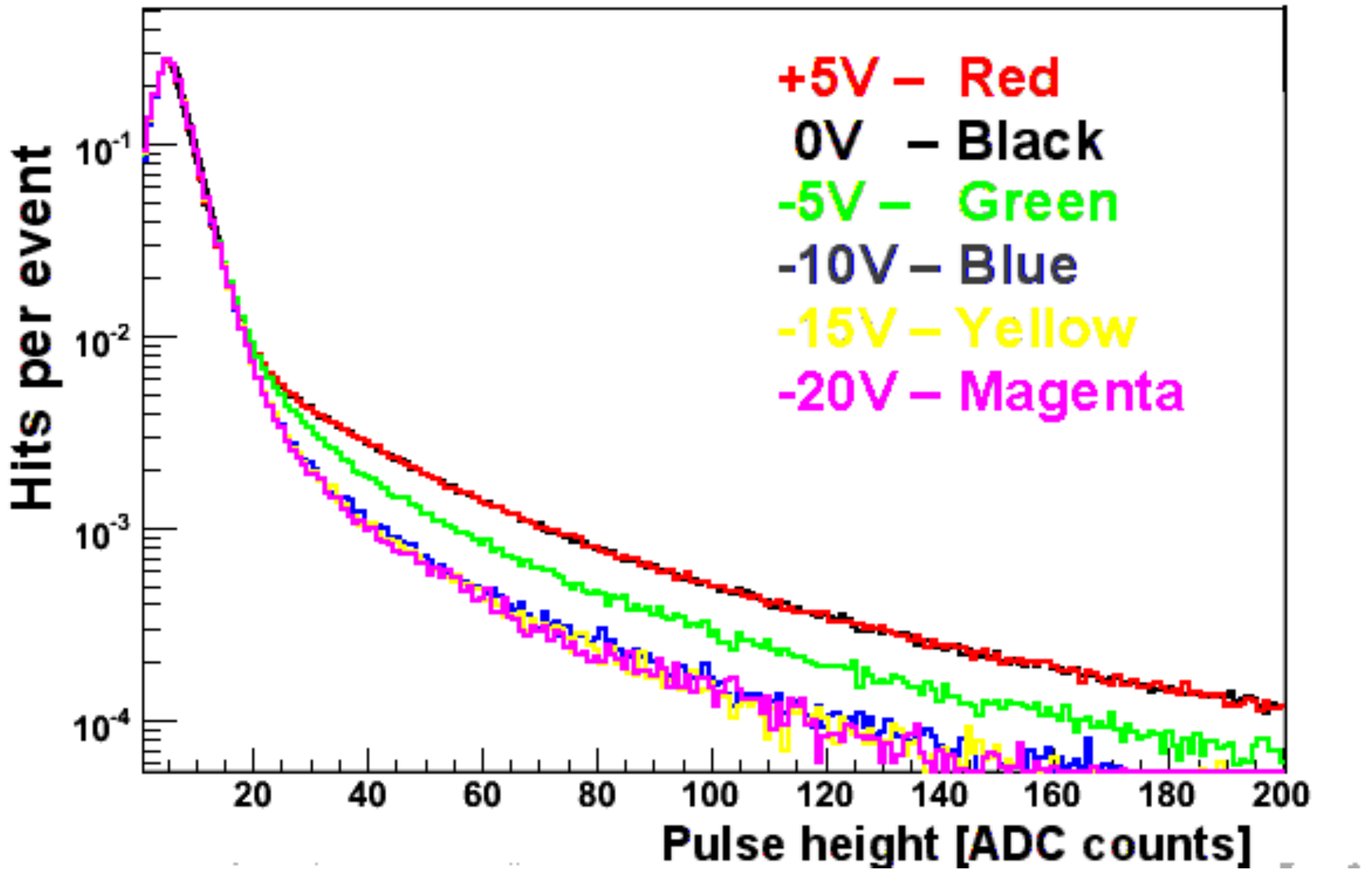}
   \caption{\label{fig:reverse-bias}Pulse height spectra of one detector module in a
   scan from +5 to -20 V of the relative voltage between the mesh and
   the top GEM.}
 \end{center}
\end{figure}

Optimizing the reverse bias mode of operation is of prime importance
for the performance of the detector. As shown in~\cite{ref:hbd2},
the ionization signal in the gap between the mesh and the top GEM
drops sharply as the field is reversed and the primary charges get
repelled towards the mesh. The signal drops quickly by almost a
factor of 10 within $\sim$10~V, while  the photoelectron
collection efficiency drops much more slowly.
Achieving maximum hadron rejection while keeping maximum photoelectron
collection requires setting the relative voltage between the mesh
and the top GEM very close to 0 with a precision of a few volts out of
$\sim$4000 V applied to the voltage divider, which is far beyond the precision
of the absolute high voltage values of the LeCroy 1471N power supplies.
An accurate and fast method to adjust the mesh HV with respect to the
GEM divider voltage was developed that exploits the scintillation signal.
For each  module, a series of short measurements were done where the gain was
kept constant and the voltage across the gap between the mesh and the top
GEM was varied in steps of 5 V. An example of such a voltage scan is
shown in Fig.~\ref{fig:reverse-bias} for one particular module. For a
meaningful comparison among the different spectra, the ordinate is normalized to
represent the number of hits per event. One sees that the yield of the
scintillation signal remains unaffected when the voltage of the mesh
varies from +5 to -20~V with respect to the top GEM,   whereas the ionization
signal sharply drops within this voltage scan.  For this module the optimal
mesh voltage is -10~V with respect to the top GEM, i.e. the minimal voltage
needed that produces the maximal reduction of the ionization tail.

\subsection{Monitoring photocathode sensitivity}
\label{subsec:photocathode-monitoring}
  Maintaining high quantum efficiency of the CsI photocathodes was crucial for achieving
maximum photoelecton yield and it was therefore important to monitor their stability and performance
throughout the run. Monitoring was accomplished using two scintillation cubes mounted inside the two
halves of the detector described in Section~\ref{sec:scintillation-cubes}. Alpha particles from the
$^{241}$Am source mounted inside the cube produce scintillation light which is focused on the CsI
photocathode in one location inside each detector. The amount of light is sufficient to produce
$\sim$ 4-5 photoelectrons, which have a Poisson distribution that can be used to determine the photoelectron
\begin{figure}[h!]
  \begin{center}
    \includegraphics[keepaspectratio=true, width=8.0cm]{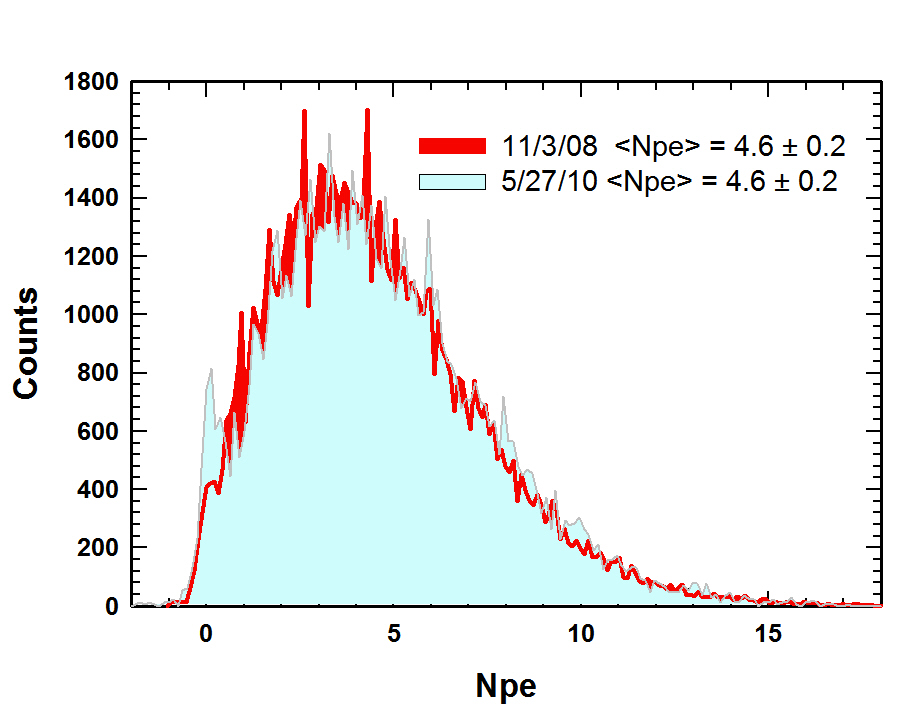}
   \caption{\label{fig:scint_cube_west}Photoelectron yield measured with the scintillation cube in
    the West detector in November of 2008 and again in May of 2010, showing no significant change
    over this 18 month period.}
 \end{center}
\end{figure}
yield. In addition, the $^{55}$Fe source mounted in the cube in approximately the same location provides a
means to determine the gas gain, which allows a determination of the peak of the Poisson distribution
in terms of photoelectrons. The number of photoelectrons determined using the gas gain can also
be compared to the number given by the shape of the Poisson distribution convoluted with the
fluctuations due to gas gain and electronic noise. The two methods generally agreed quite well.

Measuring the photoelectron
yield from the cube requires setting up a special readout configuration that is rather disruptive
to the normal operating mode of the detector, and consequently, the measurements of the photoelectron yield
from the cubes were not done very frequently (typically only once or twice per run). However, each time
the measurements were done, we observed no change in the photoelectron yield, and hence the quantum
efficiency of the photocathodes, from when the photocathodes were originally produced.
Fig.~\ref{fig:scint_cube_west} shows the photoelectron distribution measured with the
scintillation cube in the West detector in November of 2008 and in May of 2010, demonstrating that the photoelectron
yield of 4.6 $\pm$ 0.2 did not change over this period of more than 18 months.

\section{Performance}
\label{sec:Performance}   
\vspace{-0.3cm}
\subsection{Noise}
\label{subsec:Noise}
\begin{figure}[h!]
\begin{center}
\includegraphics[height=70mm, width=0.9\linewidth]{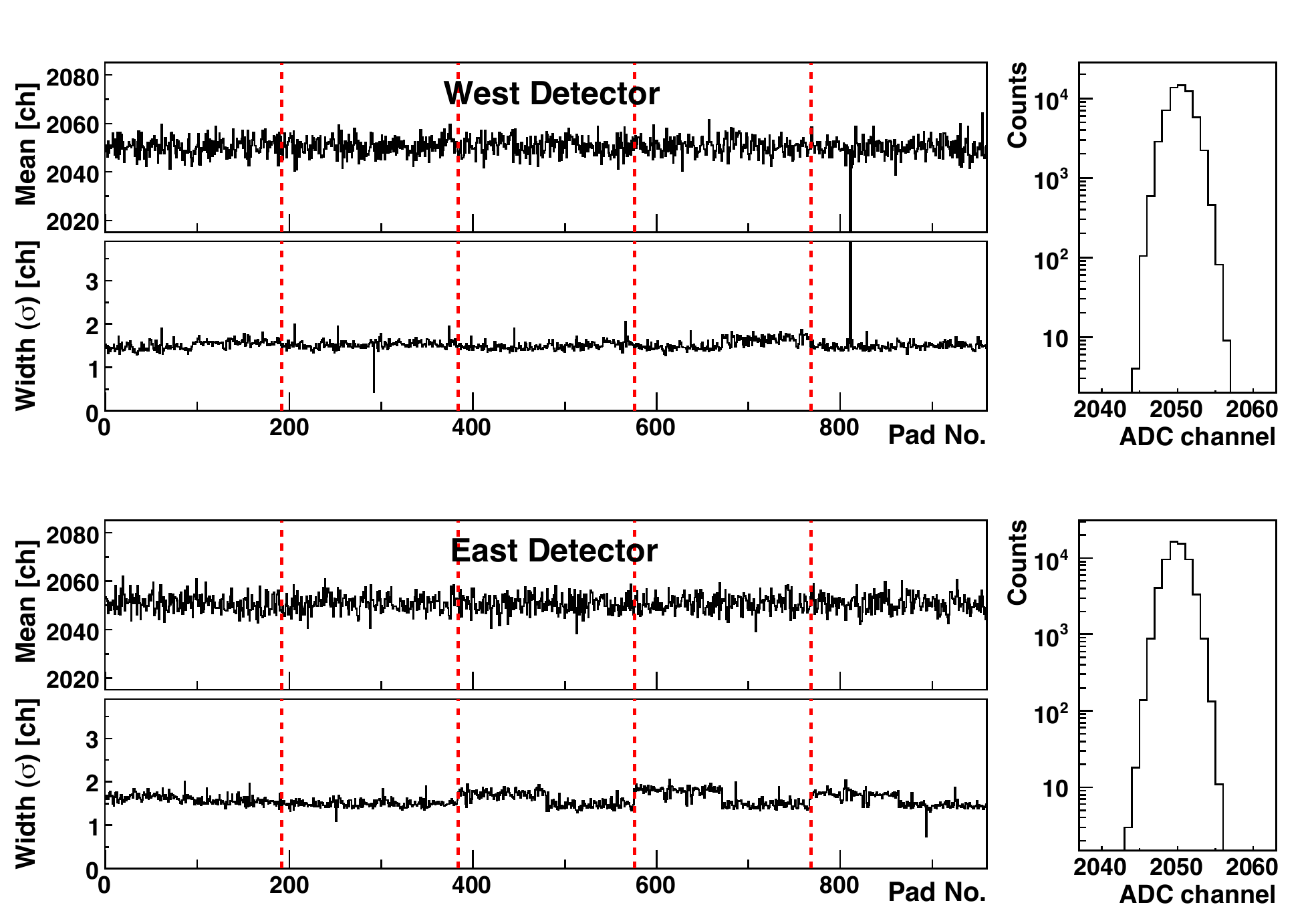}
\caption{ Left panels: pedestal mean and width ($\sigma$) as a function of pad number in the west and east detectors. Right panels: a typical pedestal distribution of a single pad
in the east and west detectors.}
\label{fig:PedestalVsPad}
\end{center}
\end{figure}

\vspace{-0.1cm}
The signals from the GEM readout system are amplified by pre-amplifiers installed
outside of the HBD detector, and sent to the front end module (FEM), $\sim$5\,m
away from the detector, as described in Section~\ref{sec:Elec}.
Fig.~\ref{fig:PedestalVsPad} shows the mean value and width ($\sigma$) of the
pedestal distributions for all pads in the West and East detectors.
The pedestal mean values sit close to the middle of the dynamic range
(4096 ADC counts).
As shown in the figure, the noise level is almost the same in all pads with a typical sigma value of $\sim$1.5\ ADC channels.
The histograms on the right of Fig.~\ref{fig:PedestalVsPad} show the pedestal distributions for
one single pad demonstrating a Gaussian distribution over more than three orders of magnitude.
To reduce the data volume, an online zero-suppression was applied requiring the pad signal to be
larger than 5 ADC channels i.e. $\sim 3 \sigma$ larger than the pedestal mean.

\begin{figure}[htbp]
\begin{center}
\includegraphics[width=0.9\linewidth]{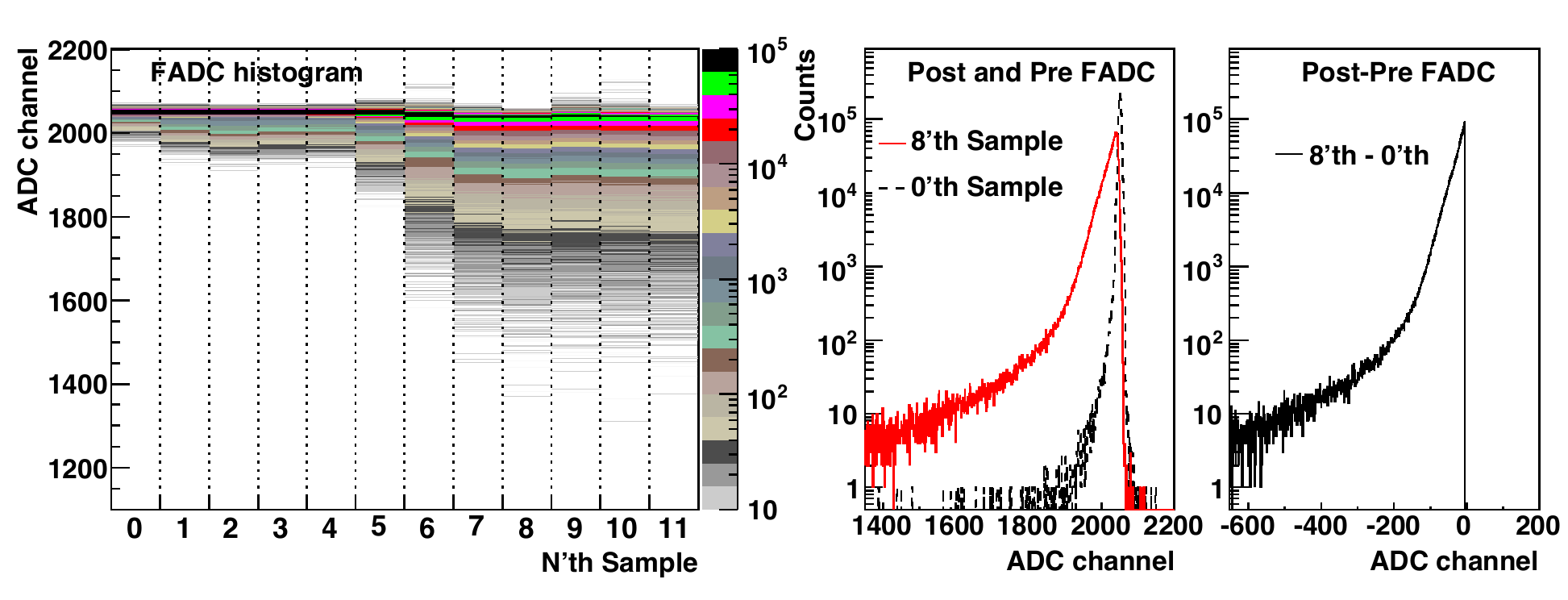}
\caption{Output signal from the front end electronics as a function of time slices (left panel),
histogram of the 8th and 0th time sample (middle panel), and difference of 8th and 0th sample (right panel).}
\label{fig:SignalFADC}
\end{center}
\end{figure}

The left panel of Fig.~\ref{fig:SignalFADC} shows a typical FADC histogram
for an electron signal in the detector.
Twelve samples are taken at a rate of 57.6 MHz, corresponding to $\sim$ 17.4 ns time bins, and
spanning an interval of $\sim$ 209 ns.
In the offline analysis, the signal is defined as the difference of the
samples (8+9+10) - (0+1+2). The zero-suppression is applied to the difference
of sample 0 and 8. In the right plot, the difference of sample 8 and 0 is shown.
The width of the distribution is narrower than the sample 8 itself. This is because
the low frequency component of the noise is eliminated by subtracting the  0$^{th}$ sample.
This subtraction also makes it possible to calculate the net charge when
signal pile-up occurs.

\subsection{Pattern recognition}
\label{sec:PatternRecognition}  
The pad size (hexagon shape with side a = 15.5 mm and area = 6.2~cm$^2$) was chosen
to be comparable but smaller than the blob size (area = 9.9 cm$^2$) such
that an incident electron produces a signal distributed over a small
cluster of a few pads. For single electrons the cluster size is
typically 2-3 pads whereas a somewhat larger cluster is produced by
close \ee pairs from $\gamma$ conversions or $\pi^0$ Dalitz decays.
On the other hand a hadron typically produces a single
pad hit.

A simple cluster finding algorithm is used to identify electron
candidates in the detector. Clusters are built around a seed pad
having a charge larger than a  selected threshold (typically 3-5
photoelectrons). In a first step, the fired pads among the first
six neighbors of the seed are added to the seed. A pad is
considered fired if it has a signal larger than typically one
photoelectron. In a second step clusters that have in common at least one
fired pad are merged together to form a single larger
cluster. The total charge of the cluster is determined as the sum of
charges of the pads assigned to the cluster. The center of gravity of
the cluster is taken as the hit position of the incident particle.

\subsection{Position resolution}
\label{sec:PositionResolution}
 
\begin{figure}
 \begin{center}
   \includegraphics[keepaspectratio=true,width=65mm]{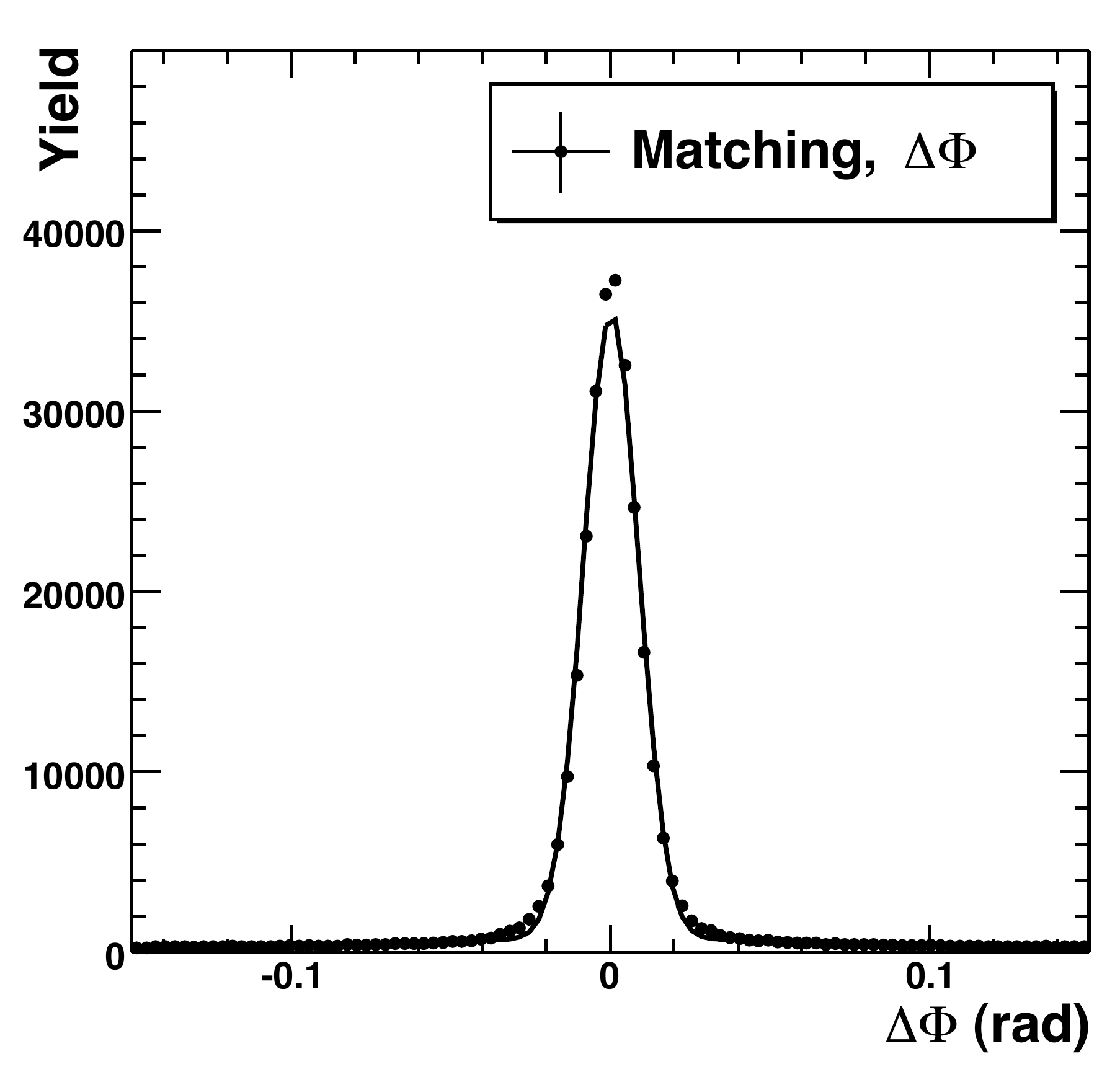}
   \includegraphics[keepaspectratio=true,width=65mm]{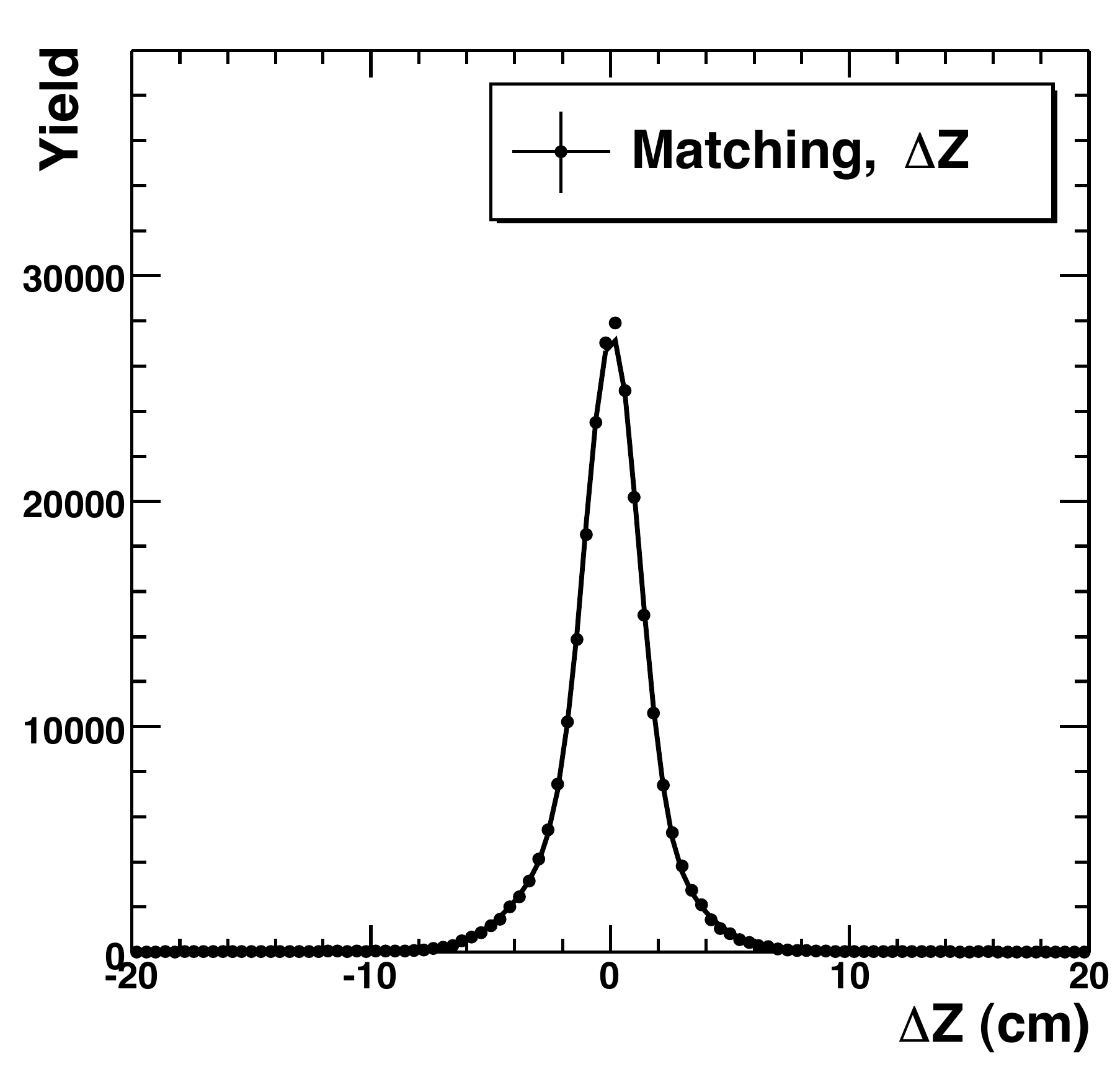}
   \caption{Matching of electron tracks in $\Phi$ (left panel)
   and Z (right panel) directions. The solid lines represent the fits
   to a Gaussian function.}
   \label{fig:matching}
 \end{center}
\end{figure}

The HBD position resolution is determined from the matching of
tracks defined in the PHENIX central arm detectors to HBD clusters. Fig.~\ref{fig:matching}
shows the matching distribution i.e. the distance between the track
projection point onto the HBD photocathode plane  and the closest HBD cluster,
in the $\Phi$ (left panel) and Z (right panel) directions. These
distributions were obtained from the p+p run of 2009 using a highly
pure sample of electron pairs originating from $\pi^0$ Dalitz decays (with
a mass \mee = 50-150~MeV/c$^2$) fully reconstructed in the PHENIX
central arms. The matching distributions have almost no background of
random matching as expected from a highly pure sample of electrons
and from a very efficient detector. The electron detection efficiency
will be discussed in Section~\ref{sec:ElecEff}.

\begin{figure}
 \begin{center}
   \includegraphics[keepaspectratio=true,width=65mm]{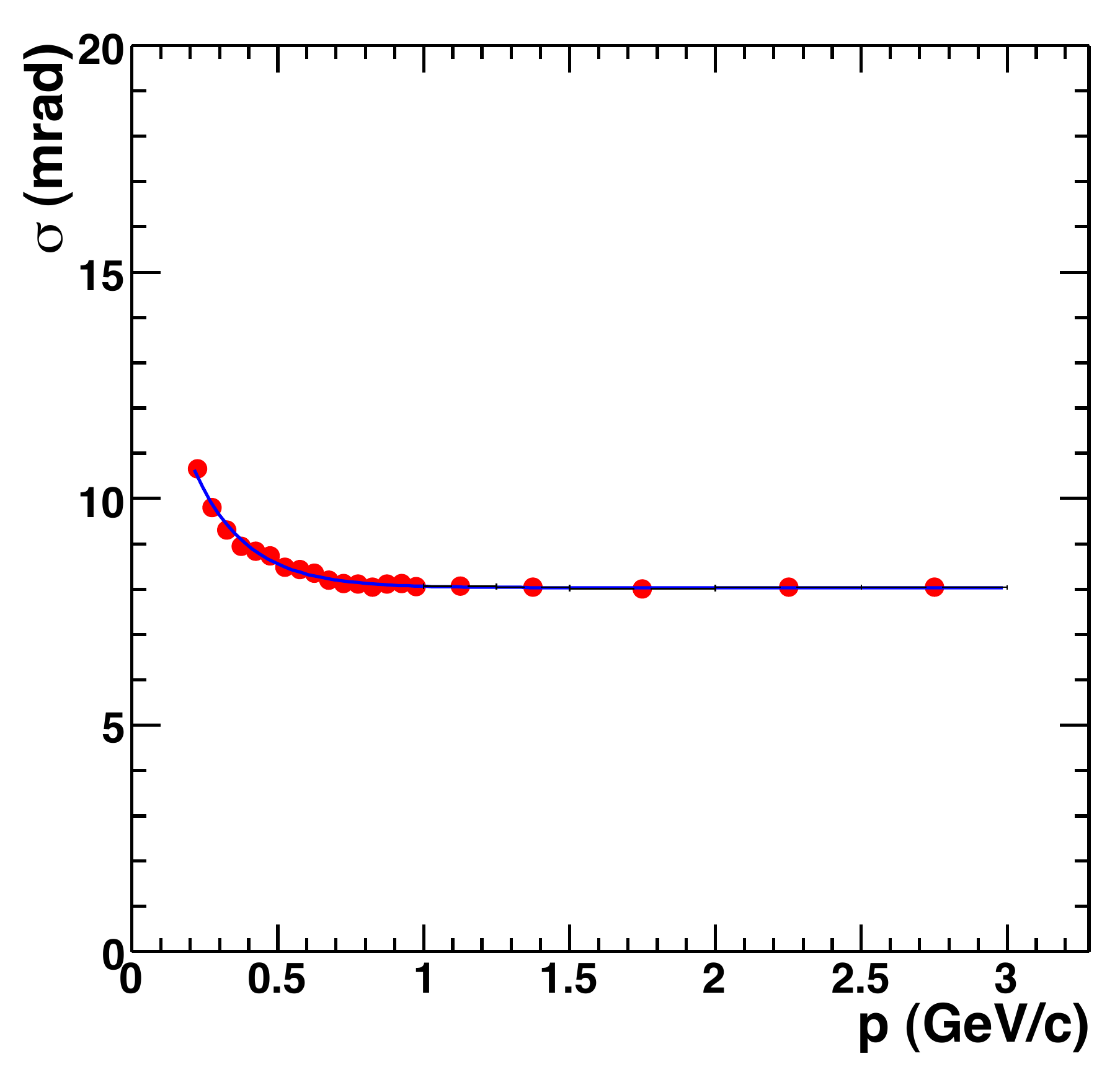}
   \includegraphics[keepaspectratio=true,width=65mm]{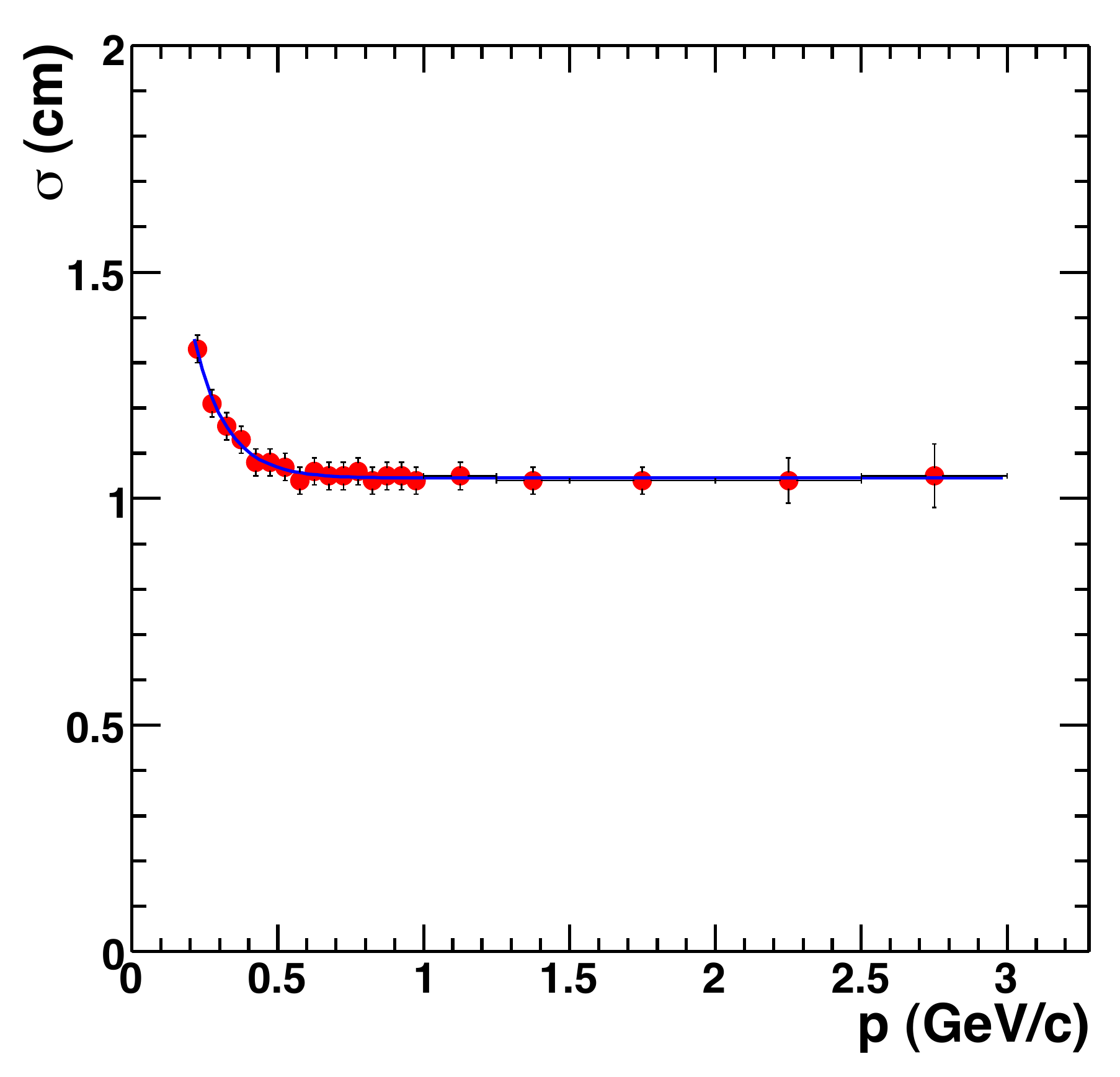}
   \caption{Matching resolution of electron tracks $\sigma_{\Phi}$
   (left panel) and $\sigma_{Z}$ (right panel) as a function of momentum.}
   \label{fig:sigma_matching}
 \end{center}
\end{figure}

The distributions exhibit a Gaussian shape as demonstrated by the
fits in the figure. They are used to align the HBD with respect to the central
arm detectors by requiring the centroid of the distributions to be
centered at 0. The $\sigma$ values of the fits are presented as
function of the track momentum in Fig.~\ref{fig:sigma_matching}. They show the
expected 1/p dependence at low momenta and a constant value at high
momenta representing the intrinsic detector resolution. The
latter is dominated by the size of the hexagonal pads. The position
resolution of single pad hits is expected to be given by 2a/$\sqrt{12}$= 0.9~cm.
For electron tracks, the center of gravity hit determination leads to
a better resolution resulting in the asymptotic $\sigma_{\Phi}$ value
of 8~mrad or 4.8~mm. This value is taken as the HBD intrinsic position resolution
since the central arm track is determined with much better precision.
In the Z direction the asymptotic resolution $\sigma_Z$ of $\sim$1.05~cm results
from the quadratic sum of the intrinsic detector resolution and the
Z resolution of the vertex position which is about 1~cm in p+p collisions.

\subsection{Hadron response and hadron rejection factor}
\label{sec:HadronResponse}

\begin{figure}
 \begin{center}
   \includegraphics[keepaspectratio=true,width=65mm]{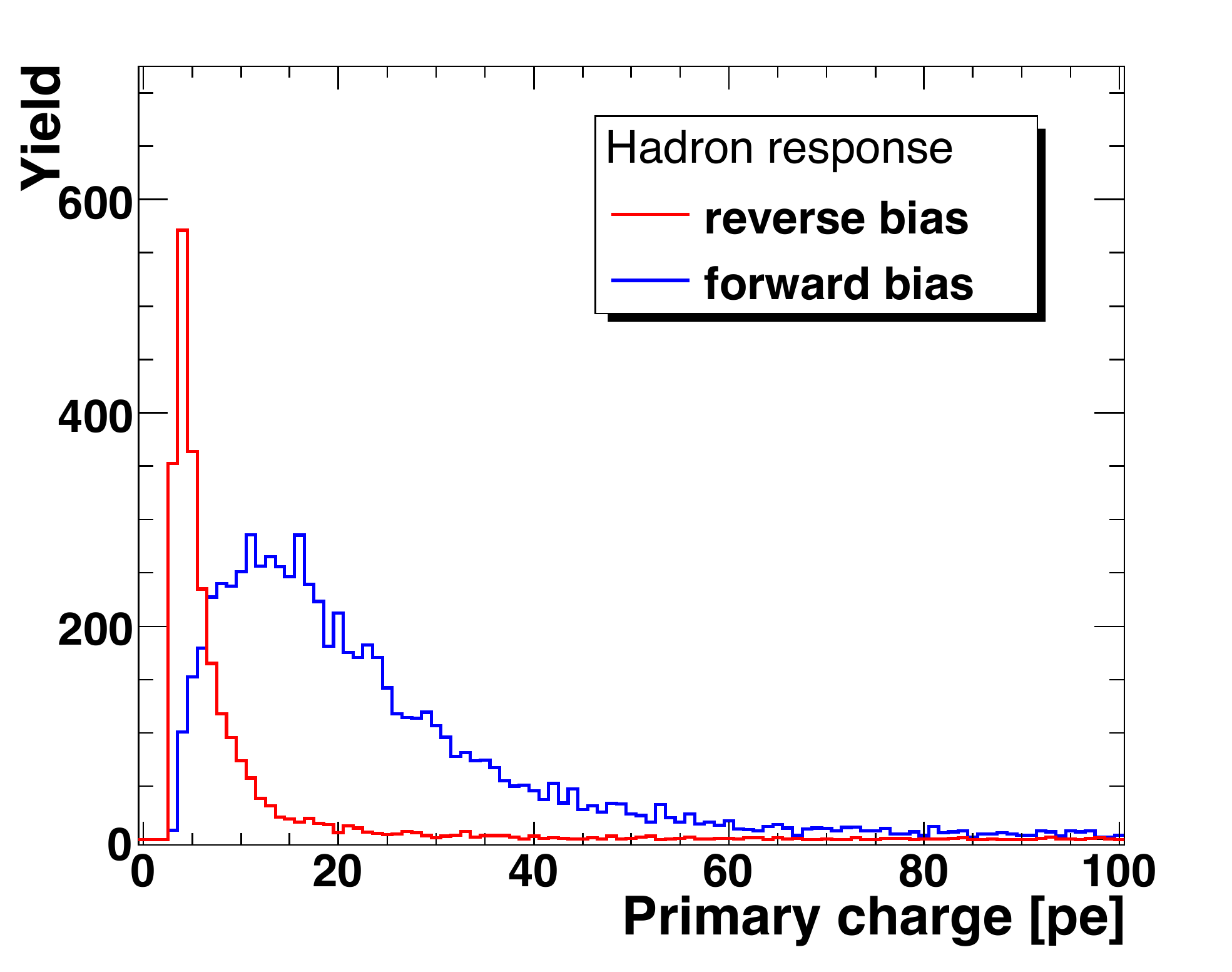}
   \includegraphics[keepaspectratio=true,width=65mm]{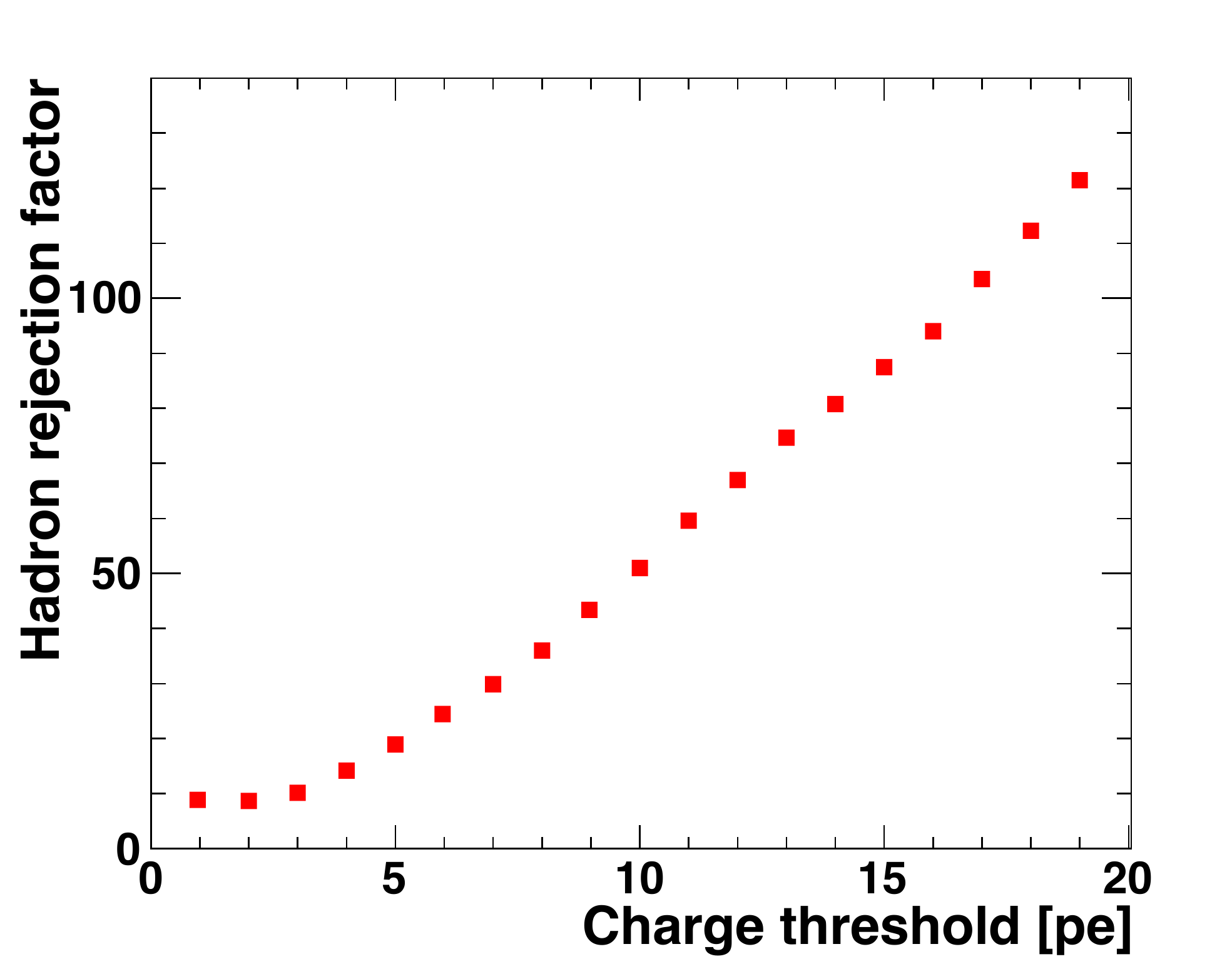}
   \caption{Left panel: HBD response to hadrons in FB and RB. Right panel: The hadron rejection factor derived from the hadron pulse-height distribution in RB as function of the signal amplitude threshold in units of the primary ionization charge. }
   \label{fig:hadrons}
 \end{center}
\end{figure}

The left panel of Fig.~\ref{fig:hadrons} shows the HBD response to hadrons in the FB
and RB modes. Hadrons identified in the central arm detectors are projected
into the HBD and the amplitude of the closest cluster within $\pm 3 \sigma$
matching windows in $\Phi$ and Z is plotted. The signal is expressed in terms of the
primary ionization charge, using the measured detector gain. The FB spectrum
is well reproduced by a Landau distribution characteristic of the
energy loss of minimum ionizing particles. The measured mean amplitude is
consistent with an energy loss of dE/dx = 7~keV/cm \cite{ref:CF4-energyloss}
and a primary ionization of $\sim$ 50~eV/ ion-pair.

In RB, there is a sharp drop in the pulse height as the
primary charges get repelled towards the mesh. In this mode, the pulse height distribution
results from the collection of (i) ionization charges from a thin
layer of about 100 $\mu$m above the first GEM surface and (ii) ionization charges  from the entire first
transfer gap, which are subject to a two-stage amplification \cite{ref:hbd2}.

We define the hadron rejection factor as the ratio of the number of hadron
tracks identified in the central arm detectors to the number of corresponding matched hits in the HBD with
a signal larger than a pre-determined charge threshold.
The hadron rejection factor derived
from the hadron spectra measured in RB is shown in the right panel of Fig.~\ref{fig:hadrons} as function of the
charge threshold. The rejection is limited by the long Landau tail and depends on
the charge threshold that can be applied without compromising too much the single
electron detection efficiency. Rejection factors of the order of 50 can be achieved
with an amplitude threshold of $\sim$ 10~e.

\subsection{Single versus double electron response}
\label{sec:ElectronResponse}
 
As mentioned in the Introduction, the combinatorial background originates
mainly from  $\pi^0$ Dalitz decays and $\gamma$ conversions. Most of
these pairs are reconstructed in the HBD as a double electron cluster
(overlapping electron and positron hits) due to their small opening
angle and the coarse granularity of the HBD pad readout. The HBD
exploits these two facts and  reduces the combinatorial background
by rejecting central arm electron tracks if the associated hit in
the HBD has a double hit response or if there is a nearby hit within
an opening angle of typically 200~mrad. This is done by two cuts, an
analog cut and a close hit cut, respectively. The analog cut requires
good separation between single and double electron hits, whereas the
close hit cut requires good hadron rejection in order not to veto
the signal with the overwhelming yield of hadrons.

\begin{figure}
 \begin{center}
   \includegraphics[keepaspectratio=true,width=65mm]{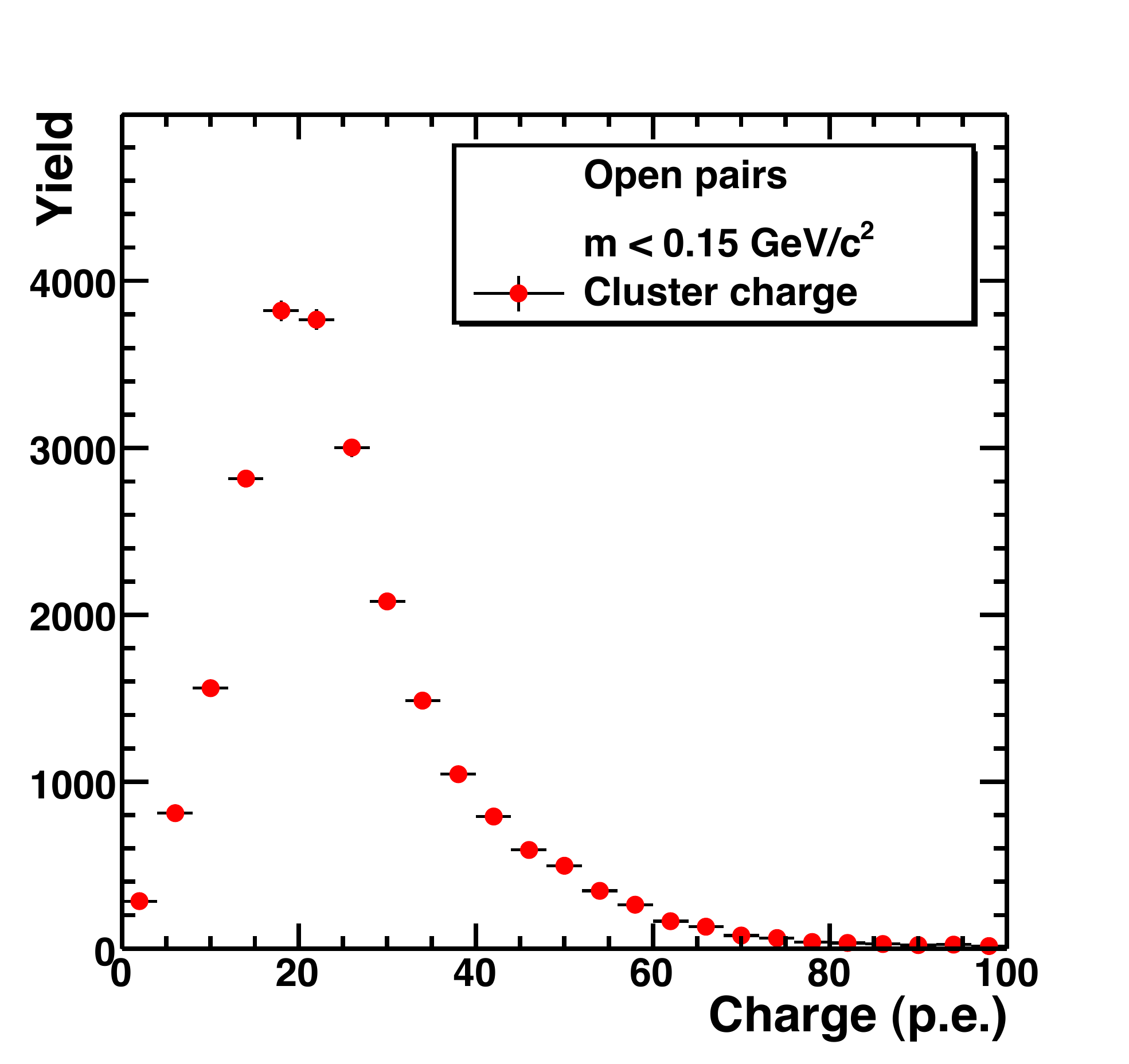}
   \includegraphics[keepaspectratio=true,width=65mm]{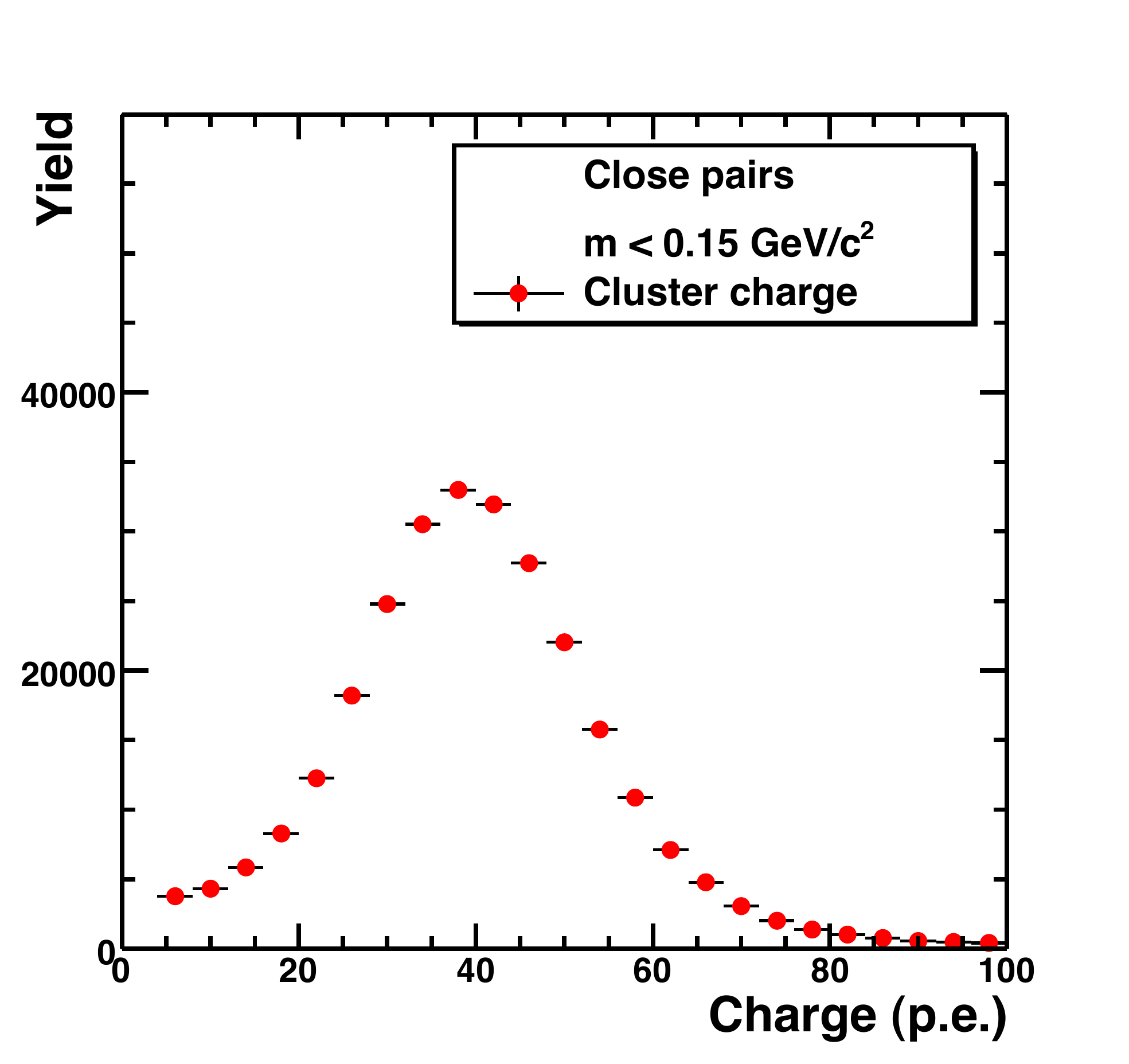}
   \caption{HBD response to single electrons (left panel)and to an unresolved double electron hit (right panel).}
   \label{fig:single-double-electron}
 \end{center}
\end{figure}

In order to study the HBD response to single and double electrons, we
select a sample of  pairs in the mass region below 0.15 GeV/c$^{2}$
where the combinatorial background is negligible. This sample is divided
into two categories: if both the electron and positron tracks reconstructed
in the PHENIX central arms are matched within 3$\sigma$ in both $\Phi$ and
Z directions to two separate HBD clusters we interpret this as the
response of the HBD to single electrons. If they are matched to the same
HBD cluster we interpret it as the HBD response to a double electron.
The HBD single electron response is shown in
the left panel of Fig.~\ref{fig:single-double-electron},
whereas the HBD double electron response is shown in
the right panel.
The former is peaked at around 20~photoelectrons, whereas the latter is
peaked at about twice that value, at $\sim$40 photoelectrons.
The mean value of the tagged single electrons is
significantly higher, probably reflecting the fact that this sample
contains a small fraction of double electron hits. We therefore take the
peak values of 20 and 40 photoelectrons to represent the mean HBD response to
single  and double electrons respectively.

The comparison of left panels of Figs.~\ref{fig:hadrons} and ~\ref{fig:single-double-electron} shows a very good separation
between single electrons and hadrons in RB. A large fraction of the hadrons can be rejected by applying a low amplitude
cut to the HBD signal.

\subsection{Figure of merit N$_0$ and photon yield}
\label{sec:Npe}
The average number of photoelectrons N$_{pe}$ in a Cherenkov counter
with a radiator of length L is given by:
\beq
   N_{pe} = N_0 \times L / \overline{\gamma}^2_{th}
\eeq
where $\overline{\gamma}_{th}$ is the average Cherenkov threshold over the
sensitive bandwidth of the detector and N$_0$ is the figure of merit
of the Cherenkov counter.

The ideal figure of merit, i.e. in the absence of any losses, is obtained by integrating the
CsI quantum efficiency (QE) times the CF$_4$ gas transmission (T$_G$) over the sensitive bandwidth of the detector. The HBD is sensitive to photons between the ionization threshold of the CsI photocathode ($\sim$6.2 eV) and the CF$_4$ cut-off (the 50\% cut-off point is at $\sim$11.1~eV  and the transmission drops to zero at $\sim$12.4~eV). For the CsI QE in CF$_4$ we use our measured values given in \cite{ref:hbd2} where it was shown that the QE increases linearly from 6.2 eV to 10.2 eV (corresponding to the highest energy where it was measured). We assume the same linear dependence to extrapolate the QE from 10.2 eV till the absolute cut-off at 12.4 eV. For T$_G$ we use our measured values (see below). We then obtain an ideal value for N$_0$ of:
\beq
     N_0^{ideal} = 370 \int_{6.2}^{12.4}QE(E) \cdot T_G \cdot dE = 714~cm^{-1}
\eeq

In the actual detector, this figure gets degraded by a number of factors that reduce the overall photoelectron yield. These include the transparency of the radiator gas, T$_G$, the optical transparency of the entrance
mesh, T$_M$, the optical transparency of the top GEM (which reduces the effective photocathode area), T$_{PC}$, the loss of photoelectrons due to the reverse bias mode of operation, $\epsilon_{RB}$,
the transport efficiency of the photoelectrons, once extracted from the photocathode, into the holes of the GEM, $\epsilon_{Tr}$, and the loss of signal due to the pad amplitude threshold that is applied to the readout, $\epsilon_{th}$.
Some of these efficiencies are wavelength independent and straightforward to measure or estimate, while others are wavelength dependent and may have greater uncertainties. In the following we discuss all these factors and quote their average  values in Table~\ref{tab:npe}.

The optical transparency of the mesh, T$_M$, is simply determined by the opacity of the wire mesh and was calculated to be 88.5\%.

The optical transparency of the photocathode, T$_{PC}$, gives the effective area of the photocathode and is determined
by the hole pattern in the GEM foil. However, the GEM holes are not perfectly cylindrical and have a tapered shape that consists of an outer hole in the copper layer and an inner hole in the kapton. By measuring the photocathode efficiency of a solid planar photocathode and comparing it with that of a photocathode deposited on a GEM foil, we determined that the effective photocathode area of the GEM is given by the average of the inner and outer hole diameters. This leads
to an average value for the optical transparency of the GEM foil of 81\% as given in Table~\ref{tab:npe}.

The radiator gas transparency T$_G$ consists of the intrinsic transmission of CF$_4$ combined with the absorption caused by any impurities such as oxygen and water. The intrinsic transmission of the HBD gas is essentially given by the transmission of input gas as shown in the top panel of Fig.~\ref{fig:Gas_Transmission_Spectra}, and the transmission including the oxygen and water impurities in the output gas is shown in the lower panels of Fig.~\ref{fig:Gas_Transmission_Spectra}.
The shape of the transmission spectrum at the shortest wavelengths is not measured with high precision due to limitations in the light output of the transmission monitor. In order to obtain the transmission down to the UV cutoff, it is assumed that the shape of the spectrum is a symmetric
S-shaped curve with a 50\% transmission point at 111 nm, which is then extrapolated down to an absolute cut-off of 100 nm. We estimate that this approximation leads to an uncertainty of $\sim$10 \% in our estimation of the photoelectron yield. Using typical transmission curves with 20 ppm of water and 3 ppm of oxygen, we obtain an average value of 89\% for the gas transmission as given in Table~\ref{tab:npe}.

The transport efficiency, $\epsilon_{Tr}$, for transferring photoelectrons produced on the photocathode to the holes in the GEMs was measured in \cite{ref:TNS1}. The value is independent of wavelength and is given as 80\% in Table~\ref{tab:npe}.

Finally, the losses due to reverse bias operation, $\epsilon_{RB}$, as described in \cite{ref:hbd2}, were minimized by optimizing the reverse bias operating point for each module, as described in Section~\ref{subsec:reverse-bias}. These losses, along with
the loss of signal  due to the amplitude threshold applied in the readout, $\epsilon_{th}$, is estimated to be 90\% as given in Table~\ref{tab:npe}.

With all of these losses, the expected figure of merit is computed to be
N$_0^{calc}$~=~328~cm$^{-1}$ with an estimated uncertainty of 14\%. The uncertainty comes primarily from the CF$_4$ transmission near its cut-off, and from the extrapolation of the CsI QE from from 10.2 eV  to the CF$_4$ cut-off. Using the calculated average $\overline{\gamma}_{th}$ = 28.8 \cite{ref:chromatic} and an average radiator length of L = 51.5 cm, the expected number of photoelectrons is 20.4$\pm$2.9. A more accurate calculation based on the convolution of the QE with the gas transmission and $\gamma_{th}$ (which varies with wavelength due to the chromatic aberration)  according to:
\beq
   N_{pe} = 370 \cdot L \cdot T_M \cdot T_{PC} \cdot \epsilon_{RB} \cdot \epsilon_{Tr} \cdot \epsilon_{th} \int_{6.2}^{12.4}QE \cdot T_G \cdot \cdot  dE / \gamma_{th}^2
   \label{eq:N_pe}
\eeq
 gives a very similar value of N$_{pe}$ = 20.3$\pm$2.8.

 \begin{table*}
 \caption{Figure of merit and Cherenkov photon yield.}
\label{tab:npe}
\begin{tabular}[]{l c}  \\
N$_0$ ideal value                     &  714 cm$^{-1}$   \\
\hline
Optical transparency of mesh          &  88.5\% \\
Optical transparency of photocathode  &  81\%   \\
Radiator gas transparency             &  89\%   \\
Transport efficiency                  &  80\%   \\
Reverse bias and pad threshold        &  90\%   \\
\hline
N$_0$ calculated value                & 328$\pm$ 46 cm$^{-1}$    \\
N$_{pe}$ expected                     &  20.4$\pm$ 2.9   \\
\hline
N$_{pe}$ measured                     & 20   \\
N$_0$ measured value                  &  322 cm$^{-1}$ \\
\hline
\end{tabular}
\end{table*}

The experimental number was obtained from a sample of resolved Dalitz pairs as defined in
Section~\ref{sec:ElectronResponse}, i.e. pairs reconstructed in the central arms with a mass
\mee $<$ 150 MeV/c$^2$ matched to resolved clusters in the HBD. As shown in
Fig.~\ref{fig:single-double-electron}, the HBD response to these single electrons gives a most
probable value of N$_{pe}^{meas} \sim$ 20 photoelectrons
corresponding to a measured figure of merit N$_0^{meas}$ = 322 cm$^{-1}$,
 in very good agreement with the calculated values. The observed N$_0$ value is  very
large compared to those achieved in any other gas Cherenkov counter \cite{ref:ceres94,ref:seguinot99,ref:phenix2003,ref:nappi2005}.
 
\subsection{Single electron efficiency}
\label{sec:ElecEff}
\begin{figure}
 \begin{center}
  \includegraphics[width=0.55\linewidth]{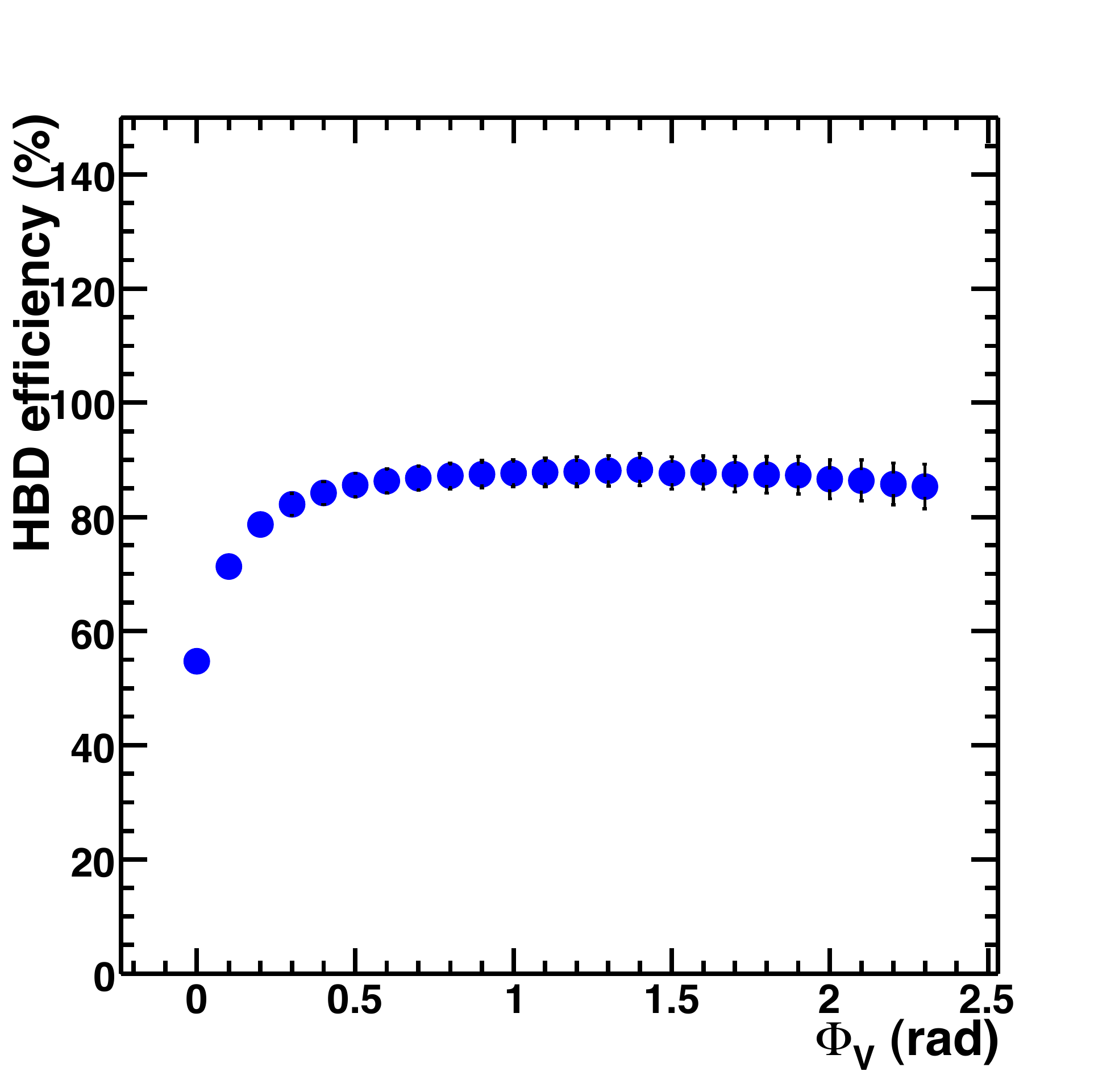}
  \caption{HBD single electron detection efficiency as function of the $\phi_V$ angle cut used
  to remove the conversion electrons (see \cite{ref:ppg088} for definition of the $\phi_V$ angle).}
  \label{fig:single_elec_efficiency}
 \end{center}
\end{figure}

The HBD single electron identification efficiency is a key factor for
the dilepton physics with the HBD. The electrons reconstructed in the PHENIX central arms
cannot be used to determine the HBD single electron efficiency since most of them do not
originate from the vertex but from downstream $\gamma$ conversions. We have used two methods
to determine the HBD electron efficiency.
In the first method, we select  a sample of reconstructed $\pi^0$ Dalitz open
pairs with low mass where the number of the conversions is
relatively small and where the combinatorial background is negligible,
namely 25 $<$ \mee $<$ 50~MeV/c$^{2}$. The conversions in this mass window are
effectively removed by applying a cut on the orientation $\phi_V$ of the pairs in the magnetic
field \cite{ref:ppg088}. The electrons in this sample are matched to hits in the HBD within
a 3$\sigma$ matching window and the ratio of the matched hits to the total number of electrons define the
HBD single electron efficiency. This ratio is plotted in Fig.~\ref{fig:single_elec_efficiency}
as function of the  $\phi_V$ cut. The figure demonstrates
that the efficiency averaged over the entire detector is close to $\sim90\%$.
Most of the losses occur near the edges of the detector modules. Excluding the boundaries results
in an efficiency close to 100 \%.

\begin{figure}
 \begin{center}
   \includegraphics[keepaspectratio=true,width=65mm]{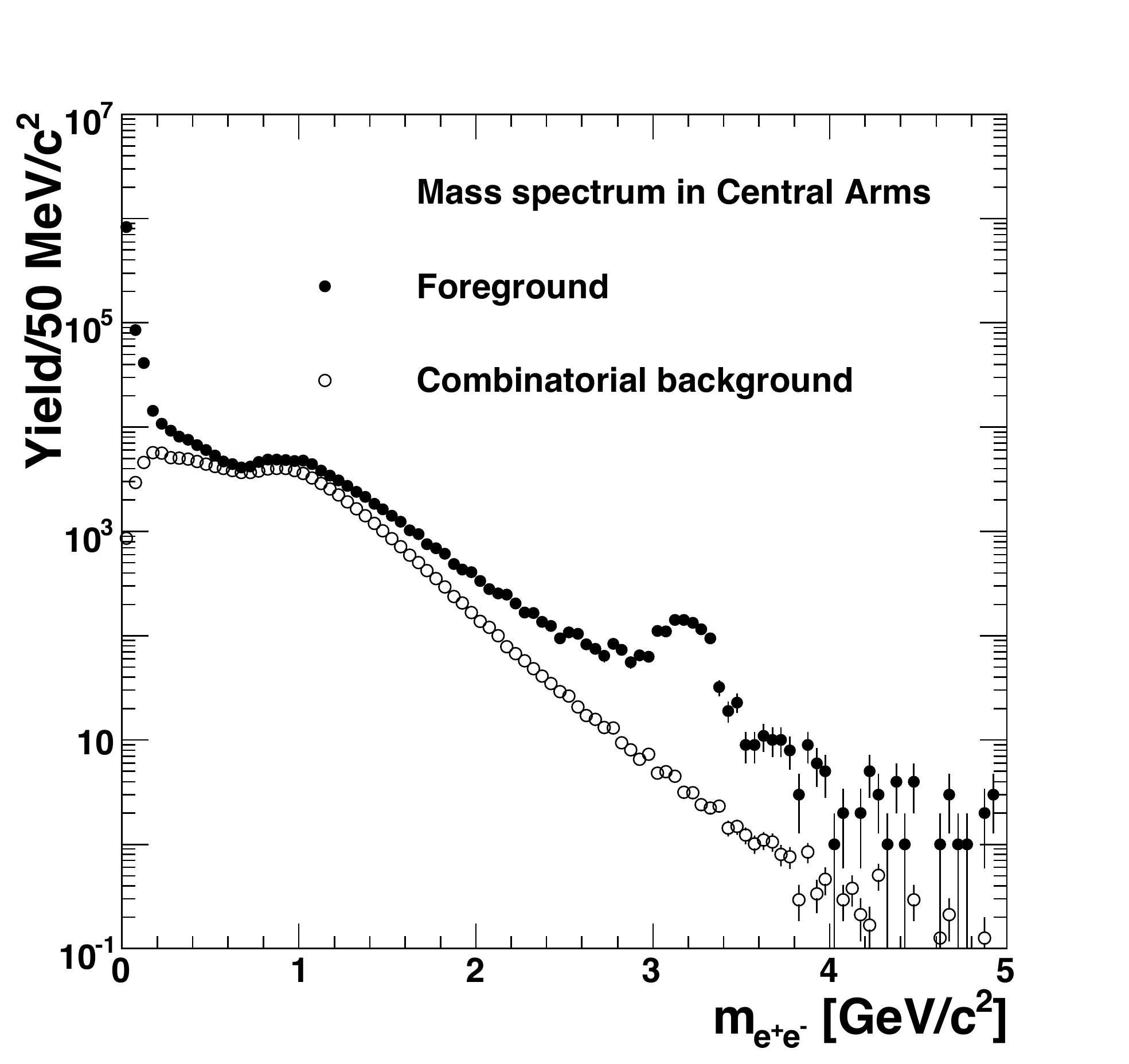}
   \includegraphics[keepaspectratio=true,width=65mm]{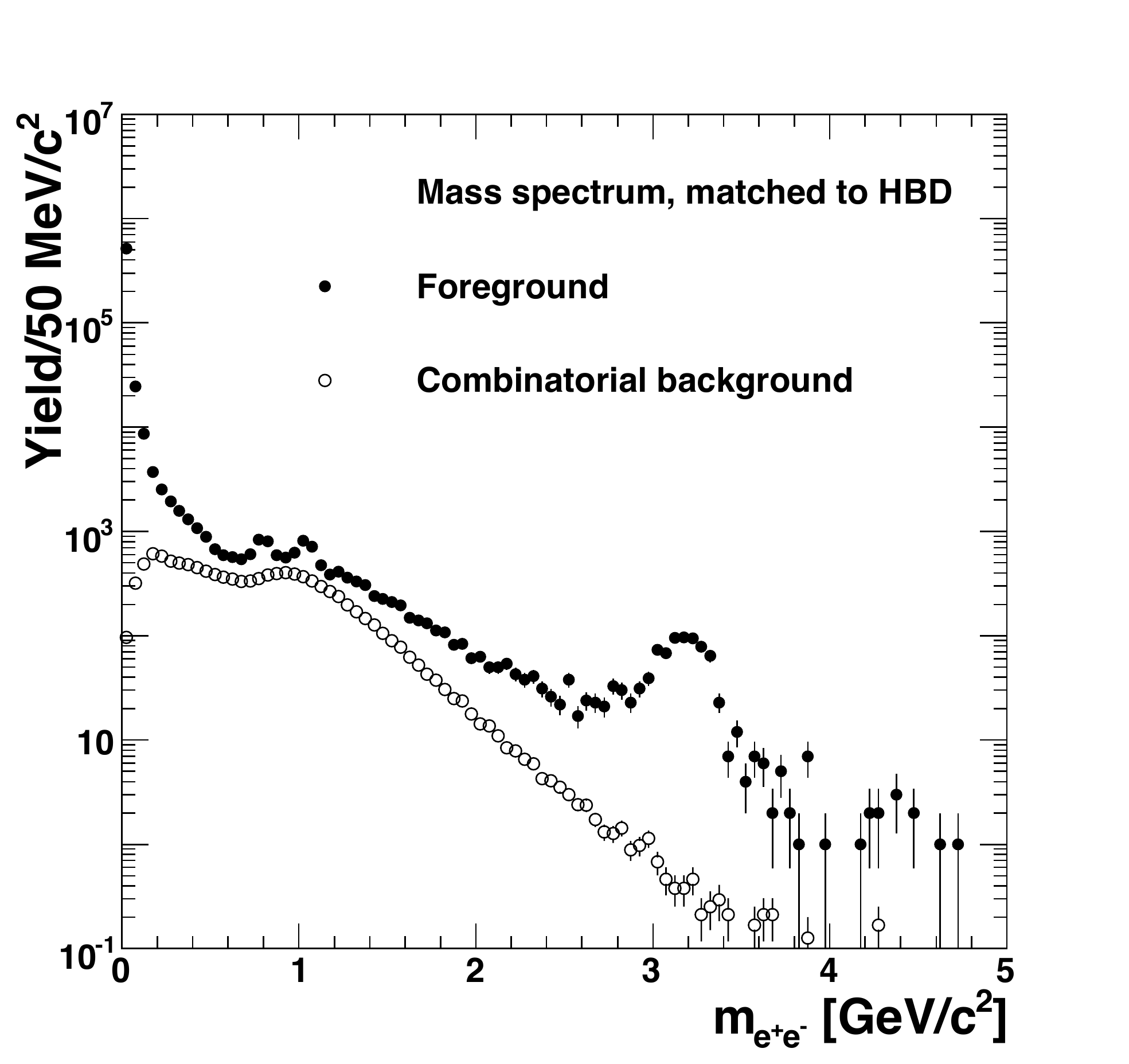}
  \caption{Invariant \ee mass spectrum measured (solid dots) in pp collisions at \sq = 200 GeV. The left panel shows the mass spectrum reconstructed using the central arms of the PHENIX detector only and the right panel shows the same mass spectrum after requiring matching hits in the HBD. The combinatorial background is shown in both panels by the open circles.}
  \label{fig:mass-spectrum}
 \end{center}
\end{figure}

In the second method, we use the sample of fully reconstructed \ee pairs in the  high mass region
(\mee $>$ 2.5~GeV/c$^{2}$) from the p+p run of 2009. This sample is dominated by $J/\psi$ decay into
\ee pairs with a relatively low background, consisting of combinatorial pairs and correlated pairs from the semileptonic decays
of charmed mesons. The left panel of Fig.~\ref{fig:mass-spectrum} shows the invariant \ee mass spectrum measured (solid dots) in \sq = 200 GeV p+p collisions
using the  PHENIX central arms only. The combinatorial background evaluated by a mixed event technique is shown by the open circles.  The right panel shows the same mass spectrum after requiring a matching of the electron and positron tracks to hits in the HBD. The matching to the HBD effectively removes conversions occurring downstream of the HBD and misidentified electrons in the central arms and consequently the combinatorial background is considerably reduced as demonstrated in the right panel. On the other hand, the  $J/\psi$ yield is almost preserved. A proper evaluation of the  $J/\psi$ yield can be obtained by fitting the mass spectrum (after subtraction of the combinatorial background) in the vicinity of the $J/\psi$ peak with a Gaussian function (for the  $J/\psi$) plus an exponential function for the open charm contribution. Comparing the so extracted $J/\psi$ yield before and after matching to the HBD,
gives also a single electron efficiency of $\sim90\%$ for the entire HBD detector.

 \section{Summary}
\label{sec:Conclusions}  
We described the concept, construction, operation and performance of the Hadron Blind Detector that was developed for the PHENIX experiment at RHIC. The HBD is a Cherenkov detector with a 50 cm long CF$_4$ radiator connected in a windowless configuration to a triple GEM coupled to a pad readout and with a CsI photocathode layer evaporated on the top face of the GEM stack. The detector was successfully operated in the 2009 and 2010 RHIC runs where large samples of p+p and Au+Au collisions, respectively, were recorded. The detector showed very good performance in
terms of noise, stability, position resolution, hadron rejection, single vs. double hit recognition  and single electron detection efficiency. The novel concept of using CF$_4$ in a windowless configuration results in an unprecedented bandwidth of sensitivity from 6.2 eV (the threshold of the CsI photocathode) up to  11.1 eV (the CF$_4$ cut-off). This translated in a measured figure of merit N$_0$ of $\sim$330 cm$^{-1}$, much higher that in any other existing gas Cherenkov detector.

\section{Acknowledgements}

We are grateful to the PHENIX Collaboration for their support and help during the various phases of the HBD upgrade
project. We are grateful to Franco Garibaldi and his group from INFN, Roma for loaning to us their CsI evaporation facility. We are grateful for the technical support of  Mr. Richard Hutter,
Mr. Richard Lefferts and Mrs. Lilia Goffer.
We acknowledge support of the work
at Brookhaven National Lab by the U.S. Department of Energy, Division of Nuclear Physics, under Prime Contract No. DE-AC02-98CH10886,
at Columbia University's Nevis Labs by the U.S. Department of Energy, Division of Nuclear Physics, under Prime Contract No. DE-FG02-86ER40281,
at Stony Brook University by the U.S. National Science Foundation under contract PHY-0521536 and by the U.S. Departement of Energy, Division of Nuclear Physics under contract DEFG - 0296ER40980,
and at the Weizmann Institute of Science by the Israeli Science Foundation, the Minerva Foundation with funding from the Federal German Ministry for Education and Research and
the Leon and Nella Benoziyo Center for High Energy Physics.


\end{document}